## *Data Management Plans: the Importance of Data Management in the BIG-MAP Project*

Ivano E. Castelli,[1,*] Daniel J. Arismendi-Arrieta,[2] Arghya Bhowmik,[1] Isidora Cekic-Laskovic,[3] Simon Clark,[4] Robert Dominko,[5] Eibar Flores,[1] Jackson Flowers,[6] Karina Ulvskov Frederiksen,[1] Jesper Friis,[7] Alexis Grimaud,[8,9] Karin Vels Hansen,[1] Laurence J. Hardwick,[10] Kersti Hermansson,[2] Lukas Königer,[11] Hanne Lauritzen,[1] Frédéric Le Cras,[12] Hongjiao Li,[13] Sandrine Lyonnard,[14] Henning Lorrmann,[15] Nicola Marzari,[16] Leszek Niedzicki,[17] Giovanni Pizzi,[16] Fuzhan Rahmanian,[6] Helge Stein,[6] Martin Uhrin,[1] Wolfgang Wenzel,[13] Martin Winter,[3,18] Christian Wölke,[3] and Tejs Vegge[1,*]

[1] Department of Energy Conversion and Storage, Technical University of Denmark, DK-2800 Kgs. Lyngby, Denmark
[2] Department of Chemistry-Ångström Laboratory, Uppsala University, Box 538, S-75121, Uppsala, Sweden
[3] Helmholtz Institute Münster, IEK-12, Forschungszentrum Jülich GmbH, DE-48149 Münster, Germany
[4] SINTEF Industry, New Energy Solutions, 7034 Trondheim, Norway
[5] National institute of chemistry, Hajdrihova 19, 1000 Ljubljana, Slovenia
[6] Department of physical chemistry, Karlsruhe Institute of Technology (KIT), Helmholtz Institute Ulm (HIU), Lise-Meitner Str. 16, 89081 Ulm, Germany
[7] SINTEF Industry, Materials and Nanotechnology, 7034 Trondheim, Norway
[8] Chimie du Solide et de l'Energie, Collège de France, UMR 8260, 75231 Paris Cedex 05
[9] Réseau sur le Stockage Electrochimique de l'Energie (RS2E), CNRS FR3459, 33 rue Saint Leu, 80039 Amiens Cedex, France
[10] Stephenson Institute for Renewable Energy, Department of Chemistry, University of Liverpool, Liverpool, L69 7ZF UK
[11] Lab automation and bio-reactor technology, Fraunhofer Institute for Silicate Research ISC, Neunerplatz 2, 97082 Würzburg, Germany
[12] University of Grenoble Alpes, CEA, LITEN, DEHT, 38000 Grenoble, France
[13] Institute of Nanotechnology (INT), Karlsruhe Institute of Technology (KIT), Hermann-von-Helmholtz Platz-1, 76344, Eggenstein-Leopoldshafen, Germany
[14] University of Grenoble Alpes, CEA, CNRS, IRIG, SyMMES, Grenoble 38000, France
[15] Fraunhofer R&D Center Electromobility, Fraunhofer Institute for Silicate Research ISC, Neunerplatz 2, 97082 Würzburg, Germany
[16] Theory and Simulation of Materials (THEOS), and National Centre for Computational Design and Discovery of Novel Materials (MARVEL), École Polytechnique Fédérale de Lausanne, CH-1015 Lausanne, Switzerland
[17] Faculty of Chemistry, Warsaw University of Technology, Noakowskiego 3, 00-664 Warszawa, Poland
[18] MEET Battery Research Center, University of Münster DE-48149 Münster, Germany
* Corresponding authors: ivca@dtu.dk (IEC, data management responsible); teve@dtu.dk (TV, project coordinator)

**Abstract**: Open access to research data is increasingly important for accelerating research. Grant authorities therefore request detailed plans for how data is managed in the projects they finance. We have recently developed such a plan for the EU-H2020 BIG-MAP project – a cross-disciplinary project targeting disruptive battery-material discoveries. Essential for reaching the goal is extensive sharing of research data across scales, disciplines and stakeholders, not limited to BIG-MAP and the European BATTERY 2030+ initiative but within the entire battery community. The key challenges faced in developing the data management plan for such a large and complex project were to generate an overview of the enormous amount of data that will be produced, to build an understanding of the data flow within the project and to agree on a roadmap for making all data FAIR. This paper describes the process we followed and how we structured the plan.

## Introduction

Consistent data management is of paramount importance for the acceleration of science and attracts steadily increasing attention from all entities throughout the entire research value chain, from the ones funding the research, to the researchers and the users of the data, which include industrial stakeholders. The attention is reasoned in a need for not only storing data related to published discoveries, but also to build new research on previously generated research data from any trustworthy source, which contributes to accelerating research. For us, the term research data includes all data generated during the research project, e.g., experiments and simulations, as well as any material underpinning the data such as laboratory notebooks, standards and protocols, pictures and videos, scripts and codes. All these should be open and accessible for reuse well beyond the duration of the specific project, if there is no passable reason for restricting the openness.

With the integration of automated workflows[1–3] and Artificial Intelligence (AI) in our everyday research,[4] streamlined access to data becomes even more needed than before as AI models depend on the availability of large quantities of scientific data typically not found in a single publication, a single dataset or produced by an individual research group or institution. In support of this trend, several databases designed for curation and easy sharing of experimental and computational research data have been established. Examples from the world of materials research are: Inorganic Crystal Structure Database (ICSD),[5] Springer Materials,[6] the Cambridge Crystallographic Data Centre,[7] Materials Project,[8,9] Materials Cloud,[10] Open Quantum Materials Database (OQMD),[11] the AFLOW consortium,[12] the Open Materials Database,[13] and the Novel Materials Discovery Laboratory (NOMAD).[14] Another initiative designed for increasing awareness on research data (store, share, reuse), which is requested by many grant authorities and academic institutions, is the mandatory submission of a Data Management Plan (DMP). An example is the European Commission, who requests submission of a DMP within the first six months for their ERC (European Research Council) and H2020 projects. The purpose of the DMP is to promote curation, storage, and sharing of the generated data by describing the research data generated within the project, by stating for whom these data might be useful and by unfolding how the project ensures that the storing, sharing, publication and preservation of the generated data comply with the FAIR (Findable, Accessible, Interoperable, Reusable) principles[15] to the greatest possible extent. Whereas this set compliance rules might be seen as a burden, there are measurable positive impacts for science as a whole as the fundamental way of pursuing research is rethought towards a sharing community.

This article describes how we developed a DMP for the H2020-LC-BAT-12 project Battery Interface Genome – Materials Acceleration Platform (BIG-MAP, www.BIG-MAP.eu). BIG-MAP[16] is a large-scale European research initiative, spanning 34 leading partners from academia, research organizations and large-scale R&D facilities as well as leading enterprises in the field. BIG-MAP is one of the six research projects belonging to the BATTERY 2030+ initiative,[17] aiming at discovering the battery of the future through disruptive development of new technologies. The goal of BIG-MAP is to achieve a five to ten fold acceleration of the discovery of new battery materials. To achieve this goal, it is necessary to develop a modular, closed-loop platform able to bridge physical insights and data-driven approaches. The BIG-MAP strategy is to cohesively integrate machine learning, computer simulations and AI-orchestrated experiments covering materials synthesis plus characterization and testing of materials and components, in order to accelerate the discovery and optimisation of battery materials. Optimal utilization of research data is crucial for achieving this goal and, in broader terms, to set a standard for future large-scale, data-centric projects aiming at accelerated discovery of materials for use in clean-energy technologies. Pivotal for this is the implementation of the BIG-MAP shared infrastructure, designed to support interoperability of the generated data, the BIG-MAP code repository[18] for shared development of computational tools and the BIG-MAP App Store[19] designed for easy public access to user-friendly versions of the developed tools. Tools will be made available in the GitHub organization and in the App Store. The tools will be able to access specific data needed for performing their task irrespectively of where the data is stored, e.g., the OPTIMADE API.[20–22] The data generated in the project will be archived in recognized repositories such as the open-access repository Materials Cloud[10,23] for computational materials science recognized by Open Research Europe,[24] large-scale facility repositories for synchrotron data, and the project partners' repositories for electrochemical measurements. Gradually, as the BIG-MAP infrastructure develops, the data will also be made available for the project partners via this shared infrastructure.

This article comprises four main parts: i) an overview of BIG-MAP and its link with the BATTERY 2030+ family of European battery projects, ii) the strategy and process we have followed when working out the DMP, iii) an overall description of the research data generated in the project and iv) a description of how the project methodologies and open-data policy facilitate the storage of data of high FAIR-ness. The detailed description of the data generated within each work package (WP) is reported in the Supplementary Information (SI). Included in the SI, there is also a detailed description of the tools and precautions implemented in order to ensure high FAIR-ness of the research data. The DMP published here is a snapshot of a living document, which will be constantly updated for the whole duration of the project.

## BIG-MAP and the BATTERY 2030+ Large-scale Research Initiative

Today, energy production and transport are evolving fast to meet challenging environmental targets and growing demand. Energy storage is the Achilles heel of the accelerating efforts for sustainable energy production and use. The search for both low-cost and high-performance materials and devices cannot rely on incremental improvements of conventional technologies, but rather require the accelerated discovery of disruptive technologies and of ultra-performing storage materials. To answer this need, the European Union has funded several battery initiatives under the Horizon 2020 scheme. These projects are organized under the umbrella of the BATTERY 2030+ consortium.[17] This consortium is composed of a coordination and support action (CSA) and six research innovation projects, which address multiple key challenges related to batteries. Following a chemistry neutral approach, BATTERY 2030+ and its member projects aim at "*reinventing the way we invent batteries*" by generating a toolbox of versatile infrastructures, common data frameworks and transferable knowledge that can be translated into design principles for discovering and developing new battery materials, as described in the BATTERY 2030+ Roadmap.[25] The vision of BIG-MAP is fully aligned with the goals of BATTERY 2030+. BIG-MAP focuses specifically on the challenge of *accelerating the discovery of new battery materials and interfaces that can only be achieved by understanding and predicting how the battery interfaces evolve in time and space*, which is a theme shared by all projects under the BATTERY 2030+ umbrella. In BIG-MAP, this will be achieved by creating an autonomous infrastructure able to design, synthesize, and test new materials across all domains of the battery development cycle, as well as orchestrate experiments and simulations on-the-fly. The project will be a lever to create the infrastructural backbone of a versatile and chemistry-neutral battery Materials Acceleration Platform in Europe, focusing in particular on the role of the interface and on how to achieve a five to ten fold increase in the rate of discovery of novel battery materials, interfaces and cells. This will set the stage for the European battery research community to efficiently develop and proliferate new battery chemistries in the coming decade.

## Our strategy and process for working out an operational and consistent DMP

Our ambition has been to layout a DMP that serves multiple purposes:

- An operational tool promoting the interaction and flow of data between the 10 scientific WPs of the project.
- A practical guide for the project partners for how to fulfil the obligation set by the H2020 Open Research Data Pilot.
- A demonstrator for how a DMP can facilitate active interaction with related R&D projects, here with the BATTERY 2030+ projects as a case.

We have chosen to work out the DMP in a collaborative effort crossing the entire project. The template we used follows to a large extent the standard DMP template for H2020 projects. This consists of two sections: a data summary describing the generated data and the FAIR section describing the incentives taken for making the data FAIR. However, we have introduced some modifications to the template.

In order to arrive at a tangible description of the generated data, we have chosen to work out separate data summaries for each WP – summaries that can serve as read-alone documents. The summaries have been supplemented by a general description of the main data categories generated in the project, which gives an overview of the data. Standard tables describe the main properties of the data, while flow diagrams illustrate in/out data streams of each WP. An overview showing the main routes for data exchange – the data highways – has also been included. See examples of the tables and data flow diagrams in the next section.

For the FAIR section, the process has been somewhat different. Here, the starting point has been a draft for each WP worked out by the WP leaders and following the H2020 template. Subsequently, the drafts have been condensed into one unified description of the incentives that shall be taken for ensuring that the generated data is FAIR. The description is done at three levels: the agreed general principles to be followed, an elaboration of these when needed, and a listing of reasoned exceptions from the general principles.

The WP leaders managed the work related to their WPs with the help of key participants from the WP, and the task was solved concurrently for all WPs and under the supervision of the BIG-MAP data manager. The concurrent process is regarded as important as it allowed for regular coordination between the WPs, for efficient tightening of possible open ends, and for developing a shared understanding of the complex structure of the generated data that crosses many disciplines. The result is an operational plan with shared ownership by all WP leads. The plan concretely describes the exchange of data inside the project and the outflow of data to the research community. Moreover, the plan serves the purpose of guiding the project work towards trustworthy data that can be readily accessed and used also from outside the project.

The strategy and process used can readily be adapted to other projects and be used for mapping and promoting collaboration between related initiatives. The key issues here will be to agree on a unified way of describing the data in order to ensure transparency and promote collaborative work. Presently, we are assisting the BATTERY 2030+ community in implementing this concept. We hope that the present publication will encourage other projects (within and outside BATTERY 2030+) to follow our approach for the preparation of their DMP plans, as this would be the first step towards an interoperability (and ultimately standardization) of research data.

One of the tasks of BATTERY 2030+ is to standardize data and protocols. Workshops and a task force have been organized to achieve this. The first step towards this is to homogenize the DMPs of the projects under the BATTERY 2030+ umbrella. This will allow us to reach a higher level of data interoperability between the projects than without such homogenization. By pioneering the definition of DMPs in large (battery) projects, BIG-MAP has the ambition of proposing a DMP template for the other projects dealing with petabytes of data and with data stemming from a broad range of disciplines and sources. It is important to note that writing a DMP is a team effort, which involves the expertise of many scientists. Our approach has been to involve all WP leaders, each one responsible for the writing of their WP DMP followed by a homogenization of the WP DMPs. This has created a sense of ownership and awareness of both the DMP and the need for data sharing across the project.

## The DMP Data Summary and the Highway of Data

The research data generated in BIG-MAP is essential for creating a modular platform that future large-scale battery projects can build upon. The data originate from multiple sources: from physical experiments, simulations, and artificial intelligence. The data comes in many different forms: data can be processed or unprocessed, and data can be structured (e.g., files holding labelled rows and columns with numbers) but also unstructured (e.g., models, metadata, code, documents in natural language). Efficient exploitation of such heterogeneous datasets calls for the project-wide guidelines for generating, storing and querying of data described in a DMP.

BIG-MAP is organized in one management work package (WP1) and ten scientific work packages, spanning from the generation of data describing battery materials by means of computational methods and experiments (WP2-WP6), to the design and development of data management tools facilitating data exchange within the WPs (WP7, WP8, WP9), and to the development of AI-accelerated materials discovery and deep learning models for inverse materials design (WP10-WP11), trained using the data generated in WP2-WP6. An outline of the generation and flow of research data within the project is shown in Fig. 1.

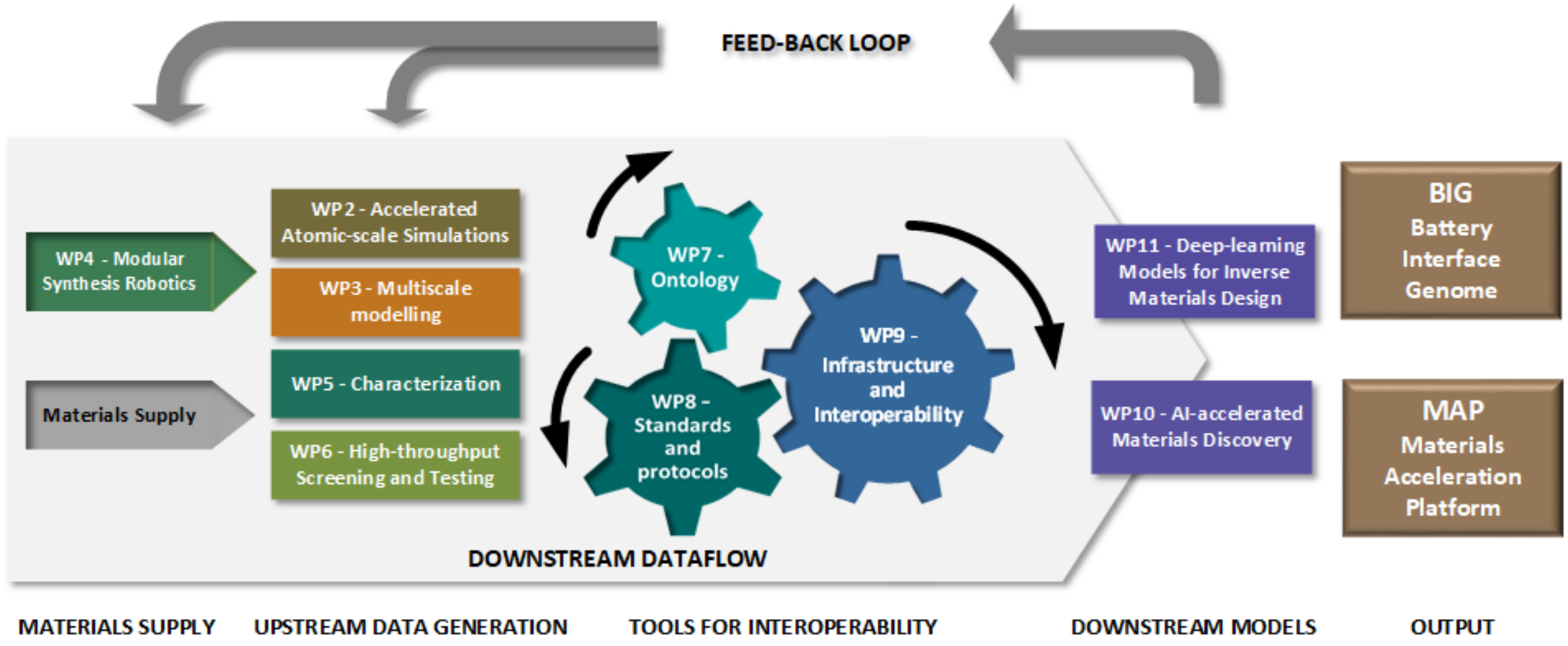

*Fig. 1 Schematic view of the types of data generated in the various WPs and the downstream data flow from WP2-WP6 via the tools developed in WP7-WP9 ensuring that the data can readily be used in WP10 and WP11.*

The starting point of the work is the knowledge and data already generated by the project partners combined with any relevant open data, i.e. data available in the research literature and in public repositories, e.g. topical repositories such as Inorganic Crystal Structure Database (ICSD),[5] Materials Cloud,[10] Materials Project,[8,9] and the Computational Materials Repository,[26] as well as general purpose repositories such as GitHub, Zenodo and FigShare.

Following the flow of the data within the project is crucial. This does not only allow keeping track of how the work progresses, but also minimizes redundant research activities and, when needed, enables requests for new data across the WPs. Fig. 2 shows the data and information flow in the project in more details than Fig. 1. The developed infrastructure (WP9) is the hub that will ensure interoperable communication and transfer of data across the project. Included in the infrastructure will be a shared data storage and processing facility accessible for all BIG-MAP partners. The infrastructure will manage automated requests for data across the project, typically requests from AI-based models (WP10-11) for new data and the reporting back of the requested data to the models. The infrastructure will also include apps that can perform autonomously analyses of datasets on-the-fly, e.g., for spectroscopic or structural datasets. The tools supporting interoperability of the data and automation of the request-replay system are the battery interface ontology BattINFO (WP7) and the standards and protocols for data acquisition, for data processing and for reporting data and metadata (WP8). A typical automated workflow could be an acquisition request for new training data for the AI-models. The request will be sent to the infrastructure that in response to this, will search for existing data or request new data from the data-generating WPs (WP2-6) or automatically launch calculations that can generated the missing data. The ability of the system to perform such autonomous processes hinges on the built-in ability to describe the requested data and metadata explicitly. Here the ontology comes into play, as it allows for an unambiguous description of the data, e.g., a specific physio-chemical property of a specific part of the battery cell in a specific spatial domain. Based upon this, the infrastructure can autonomously and accurately search for existing data, or alternatively identify the standards and protocols to be included in a request for new data to be forwarded to the data-generating WPs. All these workflows will be made available in the App Store.

WP2 and WP3 deal with AI-tools embedded atomistic and multiscale simulations of materials and interfaces. Overall, the synergy between these two WPs allows development of a multiscale knowledge of the battery using input from the experimental WPs, i.e., WP4, WP5, and WP6 (Fig. 2) that cover respectively modular robotics experiments, characterization, and high-throughput synthesis and testing. The three experimental WPs work together on the automation of experiments following the agreed experimental protocols and standards that have been defined in WP8 and following the WP7 ontology (BattINFO), as well as on automated integration of the results in the simulation WPs which is essential for verifying the predictions and completing the models. Within the WP10 and WP11 a closed loop is created where all experimental and computational data are utilized for building a deep-learning model for spatio-temporal evolution of battery interfaces that will be used for effective exploration of the chemical and structural design space of batteries.

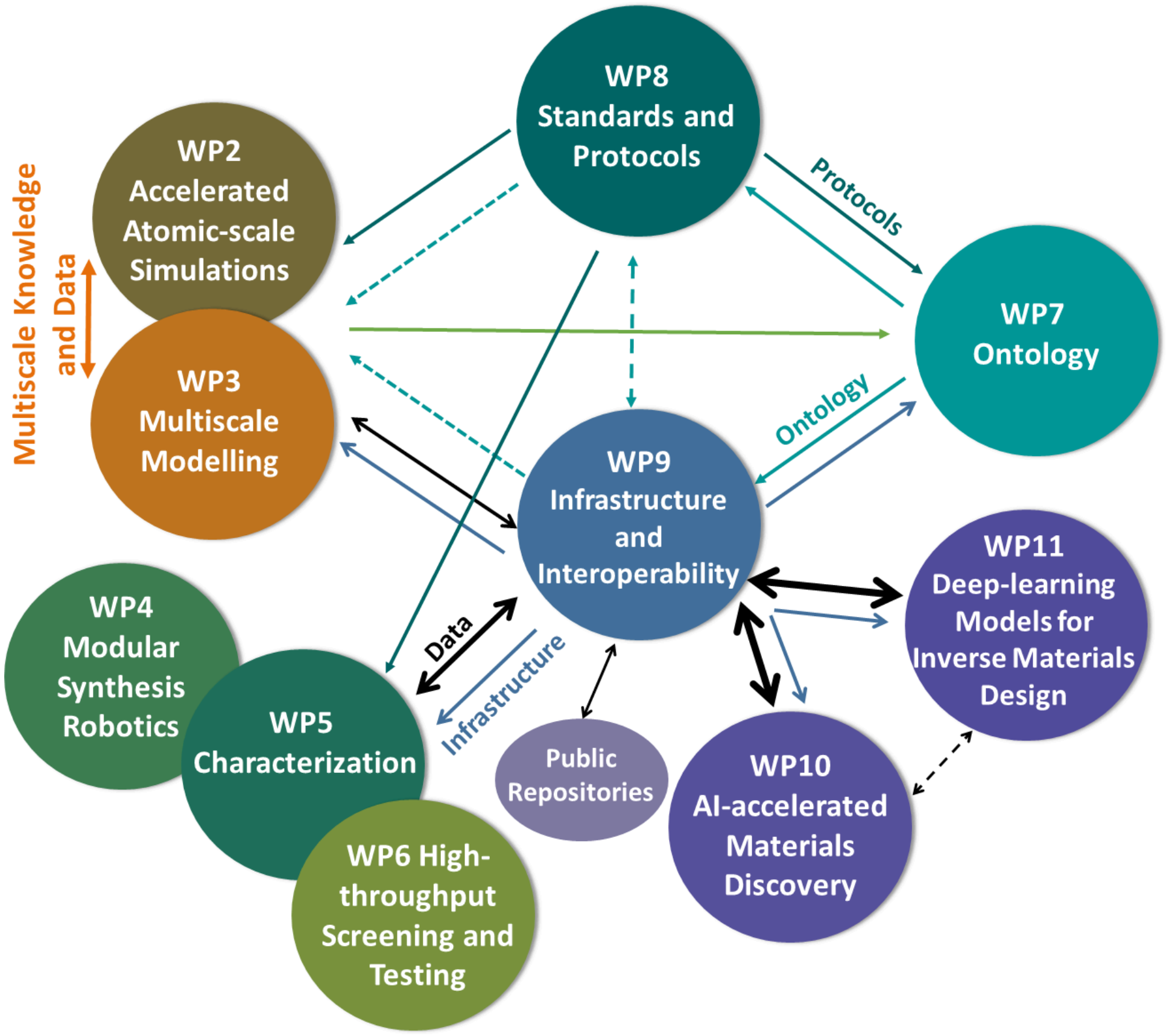

*Fig. 2 Highway of data – Overall information and data flow within BIG-MAP. The arrows show the main flows of information. The thickness of the arrows gives a visual indication of the amount of data transferred, which span from*

*few KB and MB of text files and small datasets to TB of synchrotron data, as explained in more details in the data tables in the SI. The TB-PB dataset size puts serious constraints on where such data are stored, transferred and how the user can access them. The WPs have been grouped according to their affinity.*

As the success of BIG-MAP hinges on easy and efficient sharing of data across the WPs and the scientific disciplines involved, it is crucial that all partners and WPs are fully aware of which data they shall deliver to other WPs and which data they shall receive from other WPs. Data format, data size and timing are essential information to convey between the involved partners in order to make this orchestra play. For this reason, we have chosen to highlight the data exchange within the project in the DMP. This has been done by including tables describing the inflow and outflow of data from each WP (see table 2 and 3 for examples and find all tables in SI) and diagrams showing how each WP exchange data with the other WPs (see SI). Fig. 2 represents an integration of the data-flow diagrams developed for each WP. The diagram shows only the most important data flows (the data highways), whereas the less significant but still important data flows have been omitted in order to arrive at a readable overview. Even though the data flow diagrams indicate that data is delivered and collected from a specific WP, all data passes through WP9, which ensures that data is transferred in a standardized manner between the WPs.

*Table 1. Example of types and formats of the data generated by WP2*

| Datatype | Description | Data sets | Type | Format | Size |
|---|---|---|---|---|---|
| Electronic Structure: WFT, DFT, QMC | Structures, energy-related data, wave functions & electronic properties, ab-initio molecular dynamics (AIMD) trajectories, different types of spectra | Data generated by different tools: Engines (molecular): GAUSSIAN, ORCA, MOLPRO, TURBOMOLE, NWChem, QChem, ADF, PSI4, MRCC, NECI Engines (periodic): CP2K, VASP, QUANTUM ESPRESSO, Yambo, Castep, GPAW, QuantumATK, Crystal, NECI | Tarball files can be created from the calculation folder, including relevant inputs and output raw data | .tar.gz (an archive of input and output text, XML, netctdf, hdf5, or any other machine-readable file) | TB |

*Table 2. Example of types and formats of the data collected by WP2*

| WP | What | To be used for | Suggested | | |
|---|---|---|---|---|---|
| | | | type | format | size |
| From WP5 | Parameters from structural and chemical characterizations at the local scale, including e.g., spectra (vibrational, absorption, …) | Model refinement, cross-analysis, validation, and design of simulated experiments | Tarball files can be created with post-processed data and parameters of interest | .tar.gz (with databases and reproducibl e- data sheets) | MB |

*Table 3. Example of types and formats of the data delivered by WP2*

| WP | What | Usable for | Suggested |
|---|---|---|---|

| | | | type | format | size |
|---|---|---|---|---|---|
| To WP5 | Computational predictions for bulk/interfacial structures, "chemical environments", spectra, transport properties, or any experimental characterization requiring atomistic or electronic interpretation | Guiding the characterization effort and supporting the interpretation of the results | Tarball files can be created with post-processed data and parameters of interest | .tar.gz (with databases and reproducible-data sheets) | TB |

**FAIR-ification of the BIG-MAP Data**

In addition to the description of the generated research data given in the Data Summary section, the DMP shall also describe how the project ensures that the generated research data is FAIR, and thereby giving the highest possible value for use within the project and reuse outside the project. BIG-MAP is part of the H2020 Open Research Data Pilot Programme of the European Commission. This programme provides Guidelines for FAIR Data Management, which serve therefore as guidelines for BIG-MAP's handling of data. BIG-MAP will utilize the Open Research Data Pilot opening for restricting the access to specific data. For BIG-MAP this applies for specific data underpinning business secrets, data for which disclosure can obstruct protection of intellectual property, data collected from propriety databases and some data generated by propriety software. Any disclosure of BIG-MAP research data shall respect the Consortium Agreement.

Due to the complex relationships between the generated research data and the need for a high-level interoperability, it is seen as crucial to develop a project-intern data infrastructure and project-intern data-handling procedures that ensure high interoperability and thereby high data FAIR-ness for data throughout the entire data life cycle, i.e., already from the data acquisition. The exact procedure for handling the data and the incentives for FAIR-ification depend on the nature of the research data. The BIG-MAP data falls into two categories: basic data (tabulated data and images) and computational tools (code, software, scripts and apps).

The basic data will be documented, equipped with a unique identifier and transferred to the BIG-MAP shared data storage facility where the data will become available and usable for the consortium. Transfer to the shared storage facility shall happen as soon as practically possible after data generation. The basic data shall appear findable, accessible, interoperable and reusable for the project partners as soon as it becomes available via the consortium-wide storage facility. This serves as a test in a restricted forum of how FAIR the data is. Passing the test means that the datasets and their documentation are in form that allows them to be indexed in a data repository and subsequently published provided that there are no restrictions on the data openness. An exception from this standard data handling procedure applies for datasets generated at large-scale synchrotron or neutron facilities, as such datasets are typically in the TB scale. Due to their size, such data will remain in their raw form in the repositories of large-scale facilities, and only processed datasets of manageable size will be transferred from there and made directly available for a broader audience. The preferred site for publishing basic data will be repositories that are recognized within the battery research community, e.g., the ones from which BIG-MAP also harvests open data, and the institutional data repositories of the partners. All basic data shall, to the largest degree possible, be made available in open, non-propriety formats. The software used should ideally be open source. If not possible, the software and tools needed for accessing the data shall be named. Published data shall be licensed in the least restrictive manner, e.g., Creative Commons Attribution (CC BY 4.0). Upon publication a persistent identifier (a DOI or handle) is assigned to the dataset.

In addition to the basic data, computational tools supporting the development of novel battery chemistries are highly valuable research outputs. The BIG-MAP GitHub organization serves as a shared space for developing the computational tools (code, software and scripts). GitHub version control, documentation and licensing will be utilized to their full extent. The preferred licences will be MIT, BSD, or GPL for software, codes and scripts, and Creative Commons Attribution (CC BY 4.0) for other content, e.g., the ontology. For further exposition of the tools of high usability for a broader audience, the GitHub organization has been linked with the public-facing BIG-MAP App Store where any user can access user-friendly versions of the tools, e.g., the "Quantum ESPRESSO AiiDAlab" app[27] for computing band structures and other structure properties. The App Store is equipped with a GUI (Graphic User Interface) that allows search for content, i.e., apps, via standard web browsers, and the apps are equipped with APIs (Application Programming Interfaces) allowing software-independent communication and usage of the app, see Fig. 3. The App Store is interlinked with both the BIG-MAP GitHub organization where the related source codes are deposited and with data storage facilities (the shared BIG-MAP data storage and public repositories) where the code/apps can access the data needed for performing their tasks. This architecture ensures maximum exploitation of all datasets, while complying with the legal constraints attached to the use of sensitive data.

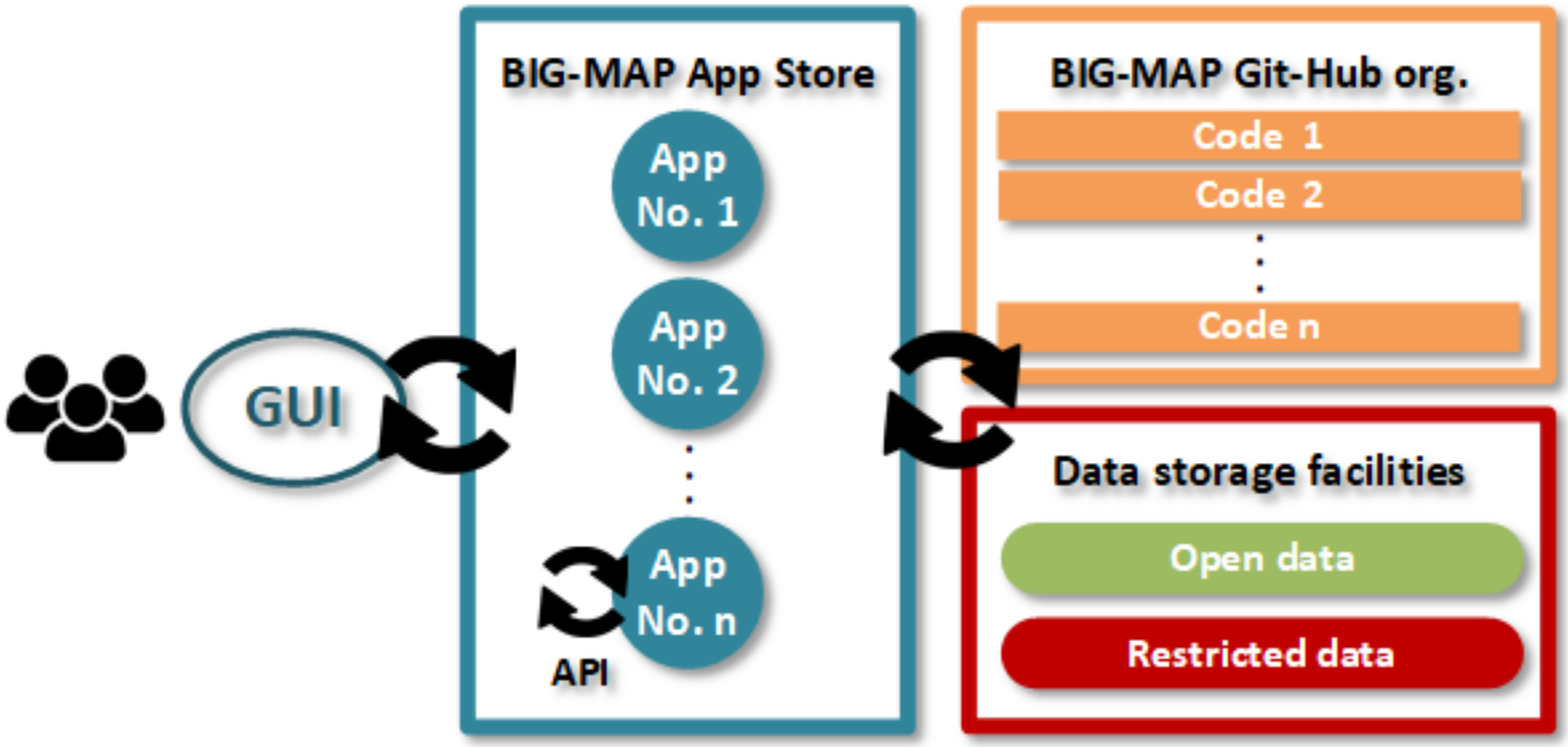

*Fig. 3. The BIG-MAP App Store – conceptual view*

All research data, irrespective of their form (basic data or computational tools), will be documented to the extent needed for relevant users to efficiently reuse the data. The documentation shall hold information on why, how and when the data was generated and who generated the data. For experimental data the documentation could take the following form: the specific aspect of the battery interface described by the data, the method used to generate the data with reference to standard protocols and standard operating procedures, and information on data fidelity and provenance. Essential for documentation is the battery interface ontology (BattINFO) and the BIG-MAP standards and protocols. BattINFO provides explicit naming of the key aspects related to the battery materials in general and especially the battery interfaces, see Fig. 4, whereas the standards and protocols allow for precise references to how the data was generated and processed.

The documentation to follow the research data shall be in the form of metadata and keywords, preferably supplemented with written guidelines, e.g., Readme text files, to improve the reusability of the data. Metadata and keywords shall be machine-readable. The metadata shall, as far as possible, be generated through automatic routines. The automatically generated metadata can be manually complemented with custom metadata, if needed for efficient query and reuse of the data. Data and metadata need to be stored together or connected via persistent links and indexed with machine-readable search keywords.

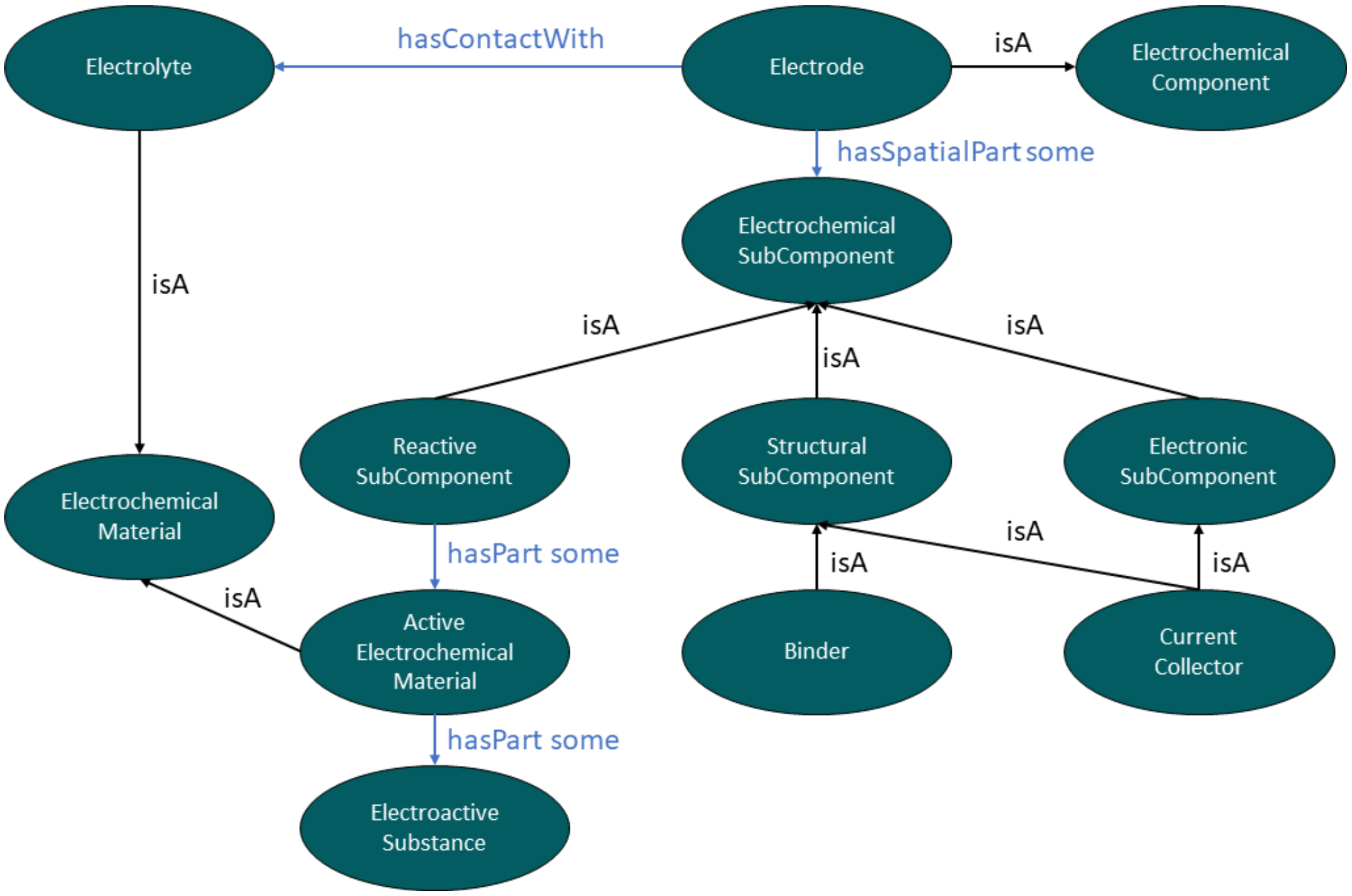

*Fig 4. Example of one branch of the BattINFO ontology show the classes and relationships that describe an electrode*

All research data shall be available for at least 10 years after the project's end date.

The BIG-MAP management structures have placed the responsibility for handling the research data securely and in compliance with the DMP per default with the partner who generated the data, but other models for sharing the responsibility can be agreed if more convenient and more partners are involved. A data manager whose task it is to guide the partners in any aspect related to data management has been appointed. All partners are responsible for complying the European Code of Conduct for Research Integrity[28] when performing their tasks in the project. More details regarding our approach to FAIR data are provided in the SI.

## Conclusions

In this work, we have highlighted the importance of data management in large and multi-domain data-centric research projects, where interoperability of data is fundamental for connecting many different tasks reliant on effective interactions between WPs and scientific disciplines.

Our take-home message is that a well worked-through DMP, holding concrete information on how to structure and handle the data, will serve as a valuable tool for guiding the project work towards trustworthy data that can be readily accessed and used also from outside the project. Another important point is that it takes efforts to concisely describe the data generated in large multi-disciplinary projects like BIG-MAP, however the effort pays back as a clear picture of how the many different datasets generated can and shall play together for maximizing the project output. This is not only valuable for managing the project, but also for providing a map that will give the project members an understanding of how their data fits into to the overall puzzle – an understanding that is essential for targeting the data generation to the precise need of the destination WPs.

We have also learned that the process used for working out the DMP is important. By working out the plan in a collaborative effort crossing the entire project, the DMP will appear not only as a document describing the generated data and how these are stored, documented and shared, but will also appear as an effective tool for planning and execution of the project and for keeping track of the work done by various partners in different WPs and their role within the overall project. The collaborative process itself is also important, as it raises the awareness in the team members of consistent data management and facilitates the project development of methodologies that support FAIR data handling. The DMP is thus a living document, which will evolve during the project and be constantly updated. In this respect, the DMP will benefit from a deeper use of the developed ontology, from the general implementation, which would help in better organizing the data tables into more operational categories, to the tools, such as Protégé,[29] which will contribute to visualize the connections between data in a more dynamic way. With these updates, we aim at increasing the connectivity between the data, not only of the BIG-MAP project, but, more ambitiously, among the whole BATTERY 2030+ community, establishing for the first time a common platform for sharing battery data.

Thanks to its connection with the BATTERY 2030+ initiative, BIG-MAP is defining new standards for data management across the whole battery community. The DMP described here could be used as a template by other initiatives aiming at a deeper integration of data and projects under a unified umbrella, which will ultimately contribute to accelerate battery discovery.

**Acknowledgements**
This project has received funding from the European Union's Horizon 2020 research and innovation programme under grants agreement No 957189 (BIG-MAP) and No 957213 (BATTERY 2030+). Sandrine Lyonnard would like to acknowledge Poul Norby, Aleksandar Matic, Marnix Wagemaker, Ennio Capria, Duncan Atkins, and Stéphanie Belin for the fruitful discussions during the preparation of the section describing WP5.

## References

1 Bölle, F. T.; Mathiesen, N. R.; Nielsen, A. J.; Vegge, T.; Garcia-Lastra, J. M.; Castelli, I. E. Autonomous Discovery of Materials for Intercalation Electrodes. *Batter. Supercaps* **2020**, *3* (6), 488–498.
2 Kahle, L.; Marcolongo, A.; Marzari, N. High-Throughput Computational Screening for Solid-State Li-Ion Conductors. *Energy Environ. Sci.* **2020**, *13* (3), 928–948.
3 Blau, S. M.; Patel, H. D.; Spotte-Smith, E. W. C.; Xie, X.; Dwaraknath, S.; Persson, K. A. A Chemically Consistent Graph Architecture for Massive Reaction Networks Applied to Solid-Electrolyte Interphase Formation. *Chem. Sci.* **2021**, *12* (13), 4931–4939.
4 Bhowmik, A.; Castelli, I. E.; Garcia-Lastra, J. M.; Jørgensen, P. B.; Winther, O.; Vegge, T. A Perspective on Inverse Design of Battery Interphases Using Multi-Scale Modelling, Experiments and Generative Deep Learning. *Energy Storage Mater.* **2019**, *21*, 446–456.

5 Inorganic Crystal Structure Database https://icsd.fiz-karlsruhe.de.
6 Springer Materials https://materials.springer.com/.
7 The Cambridge Crystallographic Date Centre https://www.ccdc.cam.ac.uk.
8 The Materials Project https://materialsproject.org/.
9 Jain, A.; Ong, S. P.; Hautier, G.; Chen, W.; Richards, W. D.; Dacek, S.; Cholia, S.; Gunter, D.; Skinner, D.; Ceder, G.; Persson, K. A. Commentary: The Materials Project: A Materials Genome Approach to Accelerating Materials Innovation. *APL Mater.* **2013**, *1* (1).
10 Materials Cloud https://www.materialscloud.org/.
11 The Open Quantum Materials Database https://oqmd.org.
12 Automatic - FLOW for Materials Discovery http://aflowlib.org/.
13 The Open Materials Database http://openmaterialsdb.se.
14 NOMAD Centre of Excellence https://nomad-coe.eu.
15 Wilkinson, M. D.; Dumontier, M.; Aalbersberg, Ij. J.; Appleton, G.; Axton, M.; Baak, A.; Blomberg, N.; Boiten, J.-W.; da Silva Santos, L. B.; Bourne, P. E.; Bouwman, J.; Brookes, A. J.; Clark, T.; Crosas, M.; Dillo, I.; Dumon, O.; Edmunds, S.; Evelo, C. T.; Finkers, R.; Gonzalez-Beltran, A.; Gray, A. J. G.; Groth, P.; Goble, C.; Grethe, J. S.; Heringa, J.; ’t Hoen, P. A. .; Hooft, R.; Kuhn, T.; Kok, R.; Kok, J.; Lusher, S. J.; Martone, M. E.; Mons, A.; Packer, A. L.; Persson, B.; Rocca-Serra, P.; Roos, M.; van Schaik, R.; Sansone, S.-A.; Schultes, E.; Sengstag, T.; Slater, T.; Strawn, G.; Swertz, M. A.; Thompson, M.; van der Lei, J.; van Mulligen, E.; Velterop, J.; Waagmeester, A.; Wittenburg, P.; Wolstencroft, K.; Zhao, J.; Mons, B. The FAIR Guiding Principles for Scientific Data Management and Stewardship. *Sci. Data* **2016**, *3* (1), 160018.
16 The Battery Interface Genome – Materials Acceleration Platform (BIG-MAP) project https://www.big-map.eu/.
17 BATTERY 2030+ https://battery2030.eu/.
18 BIG-MAP GitHub Organization https://github.com/BIG-MAP.
19 BIG-MAP App Store https://big-map.github.io/big-map-registry/.
20 Andersen, C. W.; Armiento, R.; Blokhin, E.; Conduit, G. J.; Dwaraknath, S.; Evans, M. L.; Fekete, Á.; Gopakumar, A.; Gražulis, S.; Merkys, A.; Mohamed, F.; Oses, C.; Pizzi, G.; Rignanese, G.-M.; Scheidgen, M.; Talirz, L.; Toher, C.; Winston, D.; Aversa, R.; Choudhary, K.; Colinet, P.; Curtarolo, S.; Di Stefano, D.; Draxl, C.; Er, S.; Esters, M.; Fornari, M.; Giantomassi, M.; Govoni, M.; Hautier, G.; Hegde, V.; Horton, M. K.; Huck, P.; Huhs, G.; Hummelshøj, J.; Kariryaa, A.; Kozinsky, B.; Kumbhar, S.; Liu, M.; Marzari, N.; Morris, A. J.; Mostofi, A.; Persson, K. A.; Petretto, G.; Purcell, T.; Ricci, F.; Rose, F.; Scheffler, M.; Speckhard, D.; Uhrin, M.; Vaitkus, A.; Villars, P.; Waroquiers, D.; Wolverton, C.; Wu, M.; Yang, X. OPTIMADE: An API for Exchanging Materials Data. **2021**.
21 OPTIMADE - Open Databases Integration for Materials Design https://www.optimade.org/.
22 OPTIMADE Web Client https://big-map.github.io/big-map-registry/apps/optimade-web.html.
23 Talirz, L.; Kumbhar, S.; Passaro, E.; Yakutovich, A. V.; Granata, V.; Gargiulo, F.; Borelli, M.; Uhrin, M.; Huber, S. P.; Zoupanos, S.; Adorf, C. S.; Andersen, C. W.; Schütt, O.; Pignedoli, C. A.; Passerone, D.; VandeVondele, J.; Schulthess, T. C.; Smit, B.; Pizzi, G.; Marzari, N. Materials Cloud, a Platform for Open Computational Science. *Sci. Data* **2020**, *7* (1), 299.
24 Open Research Europe https://open-research-europe.ec.europa.eu/.
25 BATTERY 2030+, Inventing the sustainable batteries of the future https://battery2030.eu/research/roadmap/.
26 The Computational Materials Repository (CMR) https://cmr.fysik.dtu.dk/.
27 Quantum ESPRESSO AiiDAlab App https://big-map.github.io/big-map-registry/apps/aiidalab-qe.html.
28 The European Code of Conduct for Research Integrity https://allea.org/code-of-conduct/.
29 Protégé - A free, open-source ontology editor and framework for building intelligent systems https://protege.stanford.edu.

# Supplementary Information

## Structure of the Data Management Plan

BIG-MAP's Data Management Plan (DMP) consists of two main sections:

- Data Summary starting from page 3
- FAIR Data starting from page 8787

The Data Summary section is further divided into 11 subsections. The first subsection describes BIG-MAP's overall data structure and the "highways" for exchange of research data between the project's ten scientific work packages (WP2-WP11). The following ten subsections give a detailed description of the research data generated in each WP and the exchange of research data with the other WPs. The FAIR Data section describes the general principles agreed within the project for FAIR-ification of BIG-MAP's research data, i.e. the principles for making the data Findable, Accessible, Interoperable and Reusable. This section also describes any exceptions from the general principles applying for specific data sets. Both the Data Summary section and the FAIR Data section have been written to be readable as independent documents, i.e. there are a few repetitions of central information that is essential for the readability of each sections. All the abbreviations are listed and explained at the end of the document.

# Table of contents

# 1. Data Summary

## 1.1 BIG-MAP's overall data structure

The multi-disciplinarity of the BIG-MAP project makes research data one of its central components. The generated research data is essential for creating a platform that future large-scale battery projects can build upon. This might be projects targeted at discovery and optimization of new battery materials and battery technologies, projects within the BATTERY 2030+ family and projects serving the European battery community.

BIG-MAP's research data is generated from different sources (multi-sourced): experiments, simulations, testing and artificial intelligence and comes in many different forms: files holding rows and columns with numbers, i.e. what we normally consider as data, but also in the forms of models, code, scripts, samples, etc. The priming and aggregation of data into a larger interoperable datasets and the flow of data between the work packages (WPs) are fundamental for the success of the project. For this reason data flow diagrams are seen as important information to include in this DMP.

This document describes the first version of BIG-MAP's DMP. The DMP will be updated in connection with the project's mid-term review.

### 1.1.1 Purpose of data collection/generation

The purpose of BIG-MAP's data generation is to facilitate accelerated discovery and optimization of new sustainable battery materials and interfaces. This is to be done by applying computational and experimental methods in an orchestrated effort designed to achieve full integration of the two worlds and thereby take full benefit from both worlds. The structure of the project can be seen in Fig. 1.

BIG-MAP's overall goal is unfolded in the 10 scientific WPs' data generation having the following sub-goals:

- WP2-WP6 to generate data describing battery materials by means of computational methods and experiments.
- WP7-WP9 to design and develop tools that allow the WPs to talk efficiently with each other. The tools are a battery interface ontology (WP7), standards and protocols (WP8) and an IT infrastructure designed for data interoperability across the project (WP9).
- WP10-WP11 to develop AI-accelerated materials discovery models and deep learning models for inverse materials design. The data generated in WP2-WP6 makes the fundament upon which the models are built.

Fig. 2 outlines how the generation of research output (data, tools and models) is shared between the WPs.

### 1.1.2 Explain the relation to the objectives of the project

BIG-MAP aims at creating a paradigm shift in the traditional sequential battery development approach, which proceeds from theoretical materials prediction, via synthesis and characterization, to cell and pack-level testing and modelling. BIG-MAP relies on the development of a unique data infrastructure, which combined with multi-scale simulations, robotic experiments, and artificial intelligence accelerates the discovery of novel battery materials.

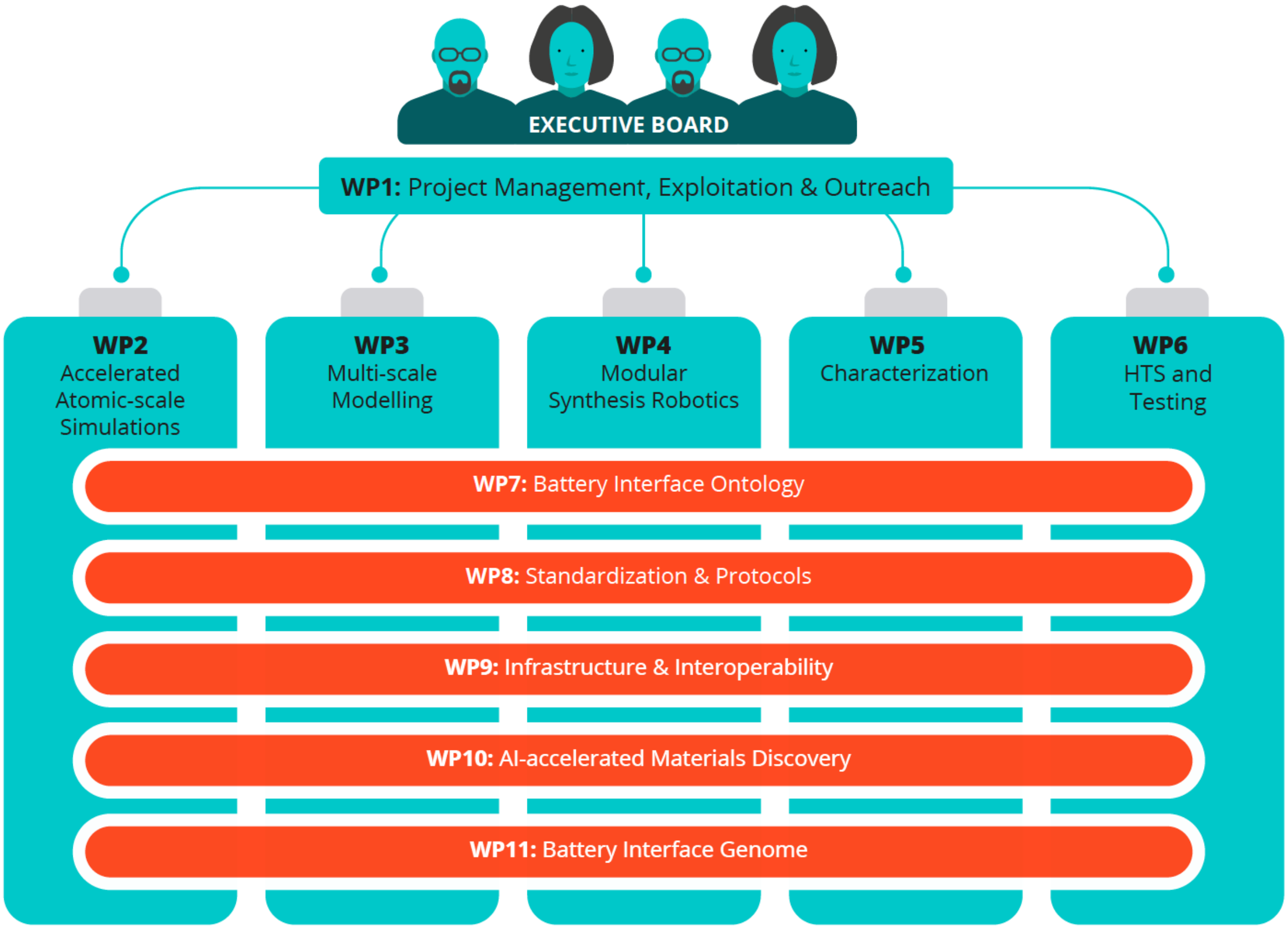

*Fig. 1. The WP structure of BIG-MAP*

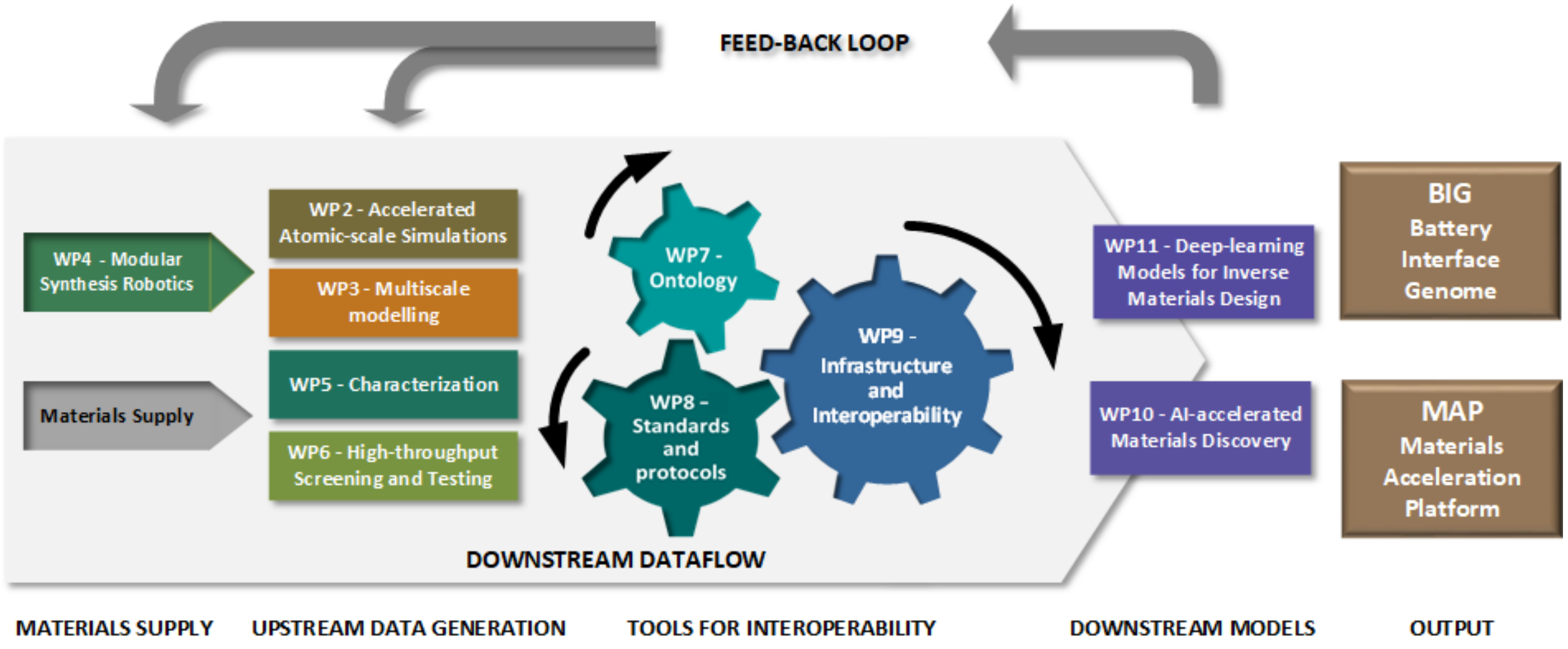

*Fig. 2 Schematic view of the types of data generated in the various WPs and the downstream data flow from WP2-WP6 via the tools developed in WP7-WP9 ensuring that the data can readily be used in WP10 and WP11.*

### 1.1.3 Specify the types and formats of data generated/collected

The generated and collected data come from different sources (experiments, simulations, and artificial intelligence). The data do not only include results obtained by a specific technique, but also scripts and workflows, laboratory notebooks and standard operating procedures, videos, pictures, samples. The infrastructure built in BIG-MAP is designed to store all this information and to ensure provenance, long-term storage, and simplicity in sharing and collecting the data. BIG-MAP's App Store and GitHub organization will play an important role in sharing and for publishing computational output.

The types of data generated in the WPs are as follows:

WP2 Accelerated Atomic-scale Simulations
- AI-accelerated atomic-scale simulation techniques for battery materials and interfaces.
- Data from simulations describing the electronic structure and atomic structure of battery materials and interfaces.

WP3 Multi-scale Modelling
- Predictive scale-bridging models for battery interphase/interfaces.
- Data describing the evolution of battery interfaces in time and space generated by the developed models.

WP4 Modular Synthesis Robotics
- An experimental setup for automated formulation and synthesis of battery-relevant protection materials.
- Synthesized materials that potentially can be used for protection of the electrodes.
- Experimentally generated data describing the chemistry, structure and electrochemical properties of the synthesized protective materials.

WP5 Characterization
- Methodology for accelerated characterization workflows including automated acquisition, curation and analysis of experimental data sets.
- Experimentally generated data describing the electrolyte-electrode interface structure and the evolution of interphases at relevant length and time scales.

WP6 HTS and Testing
- An experimental setup for chemistry-neutral, high-throughput formulation-characterization-performance testing of electrolytes and compatible electrodes across time and length scales.
- Experimental data describing the structure and properties of the electrolyte/electrode interfaces and interphases over the materials lifecycle.

WP7 Battery Interface Ontology
- A common representational system for describing battery interfaces that enables interoperability within and between BIG-MAP's WPs and that can be consumed by both humans and machines.

WP8 Standardization and Protocols
- Standards for data acquisition, battery performance evaluation and reporting that shall ensure interoperability across BIG-MAP's WPs.

WP9 Infrastructure and Interoperability
- An infrastructure allowing for automated requests for new simulations and experiments and allowing for communication between simulations and experiments.

WP10 AI-accelerated Materials Discovery

- AI/ML algorithms for effective sampling of the chemical search space and selection of the optimal subsequent experiments or simulation to run in WP2-WP6.
- An automated data analysis framework able to perform routine analysis of battery related research data and generated figures of merit (FOM).
- A software interface for the interlinking of latent spaces across simulations and experiments into the active-learning algorithms

WP11 Battery Interface Genome

- A unified description of chemical and spatio-temporal evolution of the battery interface based upon multi-fidelity data from simulations, experimental characterization and testing.
- Methods for autonomous identification of the descriptors, i.e. the genes that encode the spatio-temporal evolution of battery interfaces and interphases.
- The critical components of a deep-learning model for predicting the evolution of the interface over multiple time - and length scales, including autonomous uncertainty quantification and error correction

For a comprehensive description of these data and their formats, see sections 0-1.11.

Following the flow of all these data is crucial. This does not only allow keeping track of the progresses, but also to request new data across WPs, if necessary. An overview of the data flow is shown in *Fig. 3*.

### 1.1.4 **Specify if existing data are being re-used (if any)**

The starting point of the project is data generated and available from BIG-MAP's partners and data available from open sources, such as databases holding computationally and experimentally generated data, GitHub repositories, and research data repositories (e.g. Materials Cloud, Zenodo, FigShare). Data published in the literature will also be used.

### 1.1.5 **Specify the origin of the data**

The generated data stems from multiple sources:

- Experiments (synthesis robots, autonomous characterization and testing)
- Large-scale facilities and the partners' laboratories
- Simulations (from atomic-scale up to the continuum level)
- Artificial intelligence.

### 1.1.6 **State the expected size of the data (if known)**

The generated datasets are very different in size, ranging from few KB (kiloByte) for text files to TB-sized (TeraByte-sized) data files typically generated at large-scale facilities.

The total amount of data generated in BIG-MAP is expected to be in the PetaByte (PB) range.

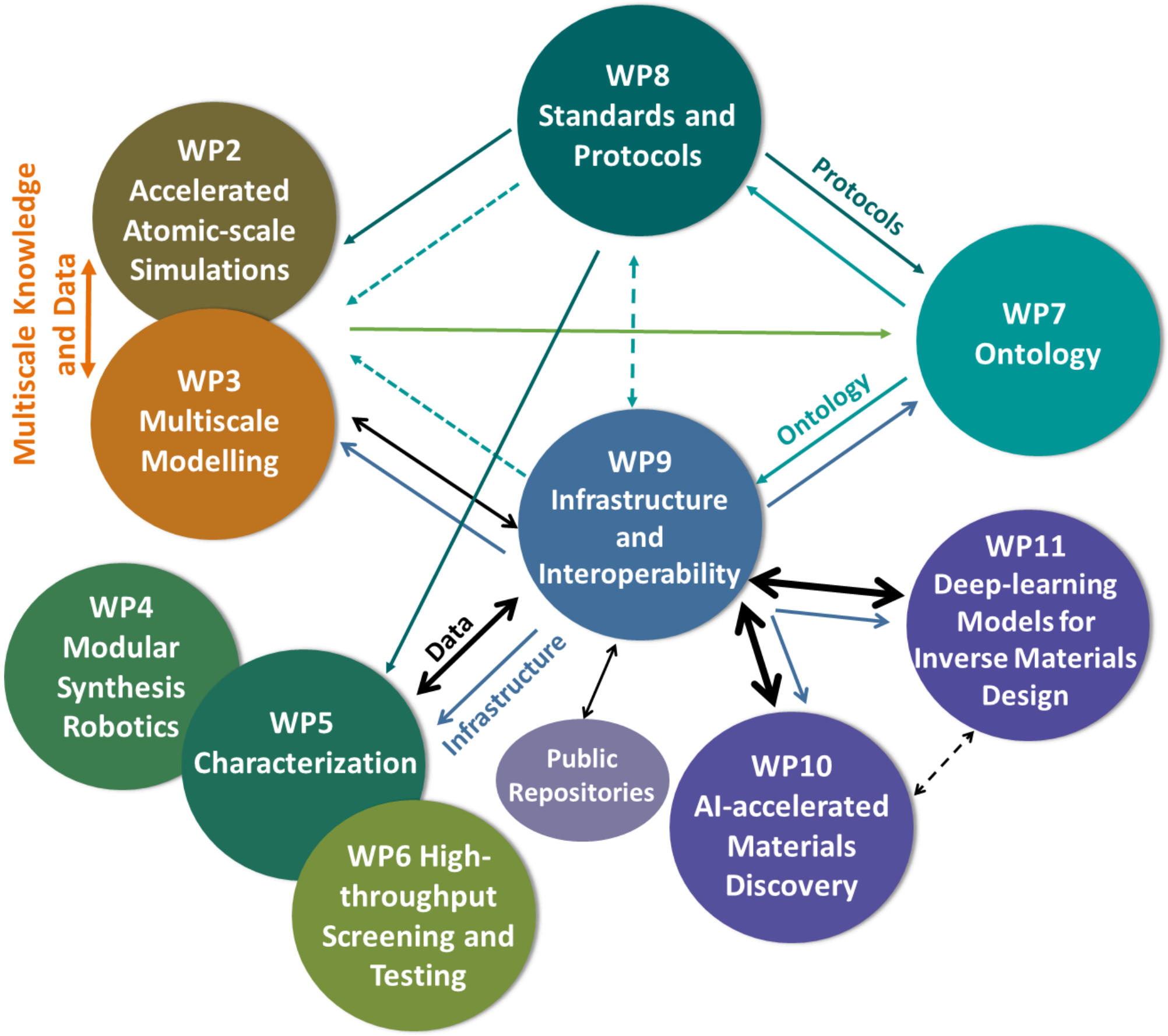

*Fig. 3 Highway of data – Overall information and data flow within BIG-MAP. The arrows show the main flows of information. The thickness of the arrows give a visual indication of the amount of data transferred, which span from few KB and MB of text files and small datasets to TB of synchrotron data, as explained in details in the data tables in the SI. The TB size puts serious constraints on where such data are stored and how the user can access them. The WPs have been grouped according to their affinity.*

### 1.1.7 **Outline the data utility: to whom will it be useful**

The generated research data are essential for achieving BIG-MAP's objective. In addition, the data will provide the backbone needed for achieving a five to ten fold acceleration in the materials-discovery process, for which BIG-MAP is the three-year demonstrator.

The data will be useful for the research community involved in:

- Development of new battery materials and battery technologies in silico and in experiments and especially those working on post-lithium technologies.
- Utilization of artificial intelligence and machine learning in materials discovery and optimization, i.e. MAP initiatives.
- Development of tools ensuring data operability across scientific disciplines.

The data will also be useful for any stakeholder in the battery industry.

## 1.2 WP2 – Accelerated atomic-scale simulations

### 1.2.1 Purpose of data collection/generation

The focus of WP2 is in modelling battery systems at the electronic and atomistic scales. This WP serves as a technology demonstrator, and will use example chemistries that are generally simpler than those in commercial batteries, but nevertheless sufficiently complex to be relevant and push the boundaries of what is possible to simulate computationally. The internal work and data flows of WP2 are summarized in Fig. WP 1.1.

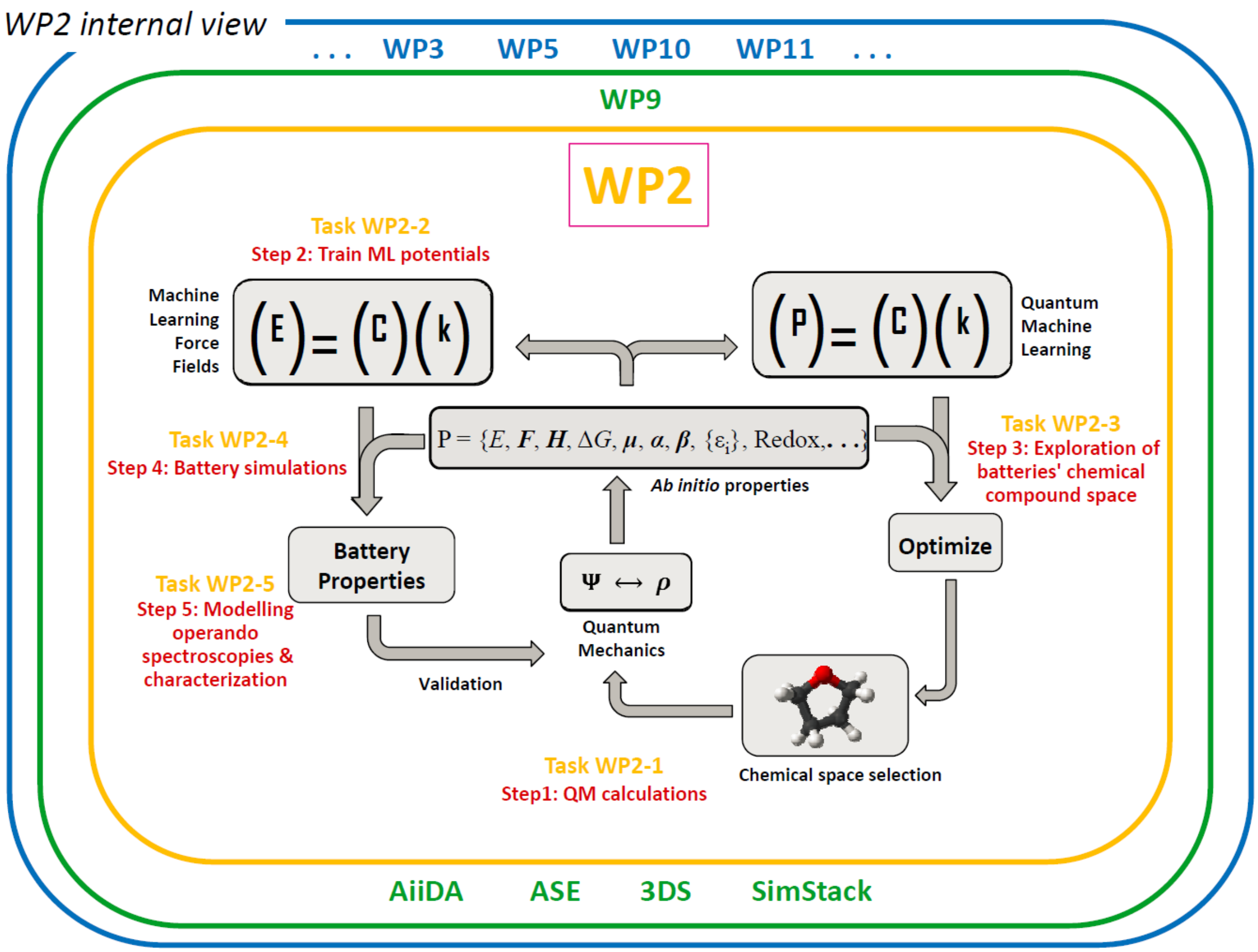

*Fig. WP 1.1. Internal workflows and data flows of WP2.*

In WP2, we will develop an automated closed-loop workflow that demonstrates two strands:

<u>A (*left strand in the figure*)</u>: (i) the calculation of reliable quantum mechanics (QM) data, (ii) fitting of machine learning force fields (ML-FF) models to the QM data and (iii) application of the new ML-FFs to provide new insights and identify relevant features of the simulated systems.

<u>B (*right strand*)</u>: (i) the calculation of reliable QM data, (ii) fitting of property models to the QM data and (iii) prediction and optimization of relevant properties throughout compound space in search for the most promising new material candidates. WP2 will also incorporate automated workflows for prediction of spectra, as well as interpretation of other materials characterization techniques. The details of the internal WP2 data flows are given in Table WP 1.1. The details of the data flows between WP2 and other WPs are given in Table WP 1.2 and Table WP 1.3 and are illustrated in Fig. WP 1.2.

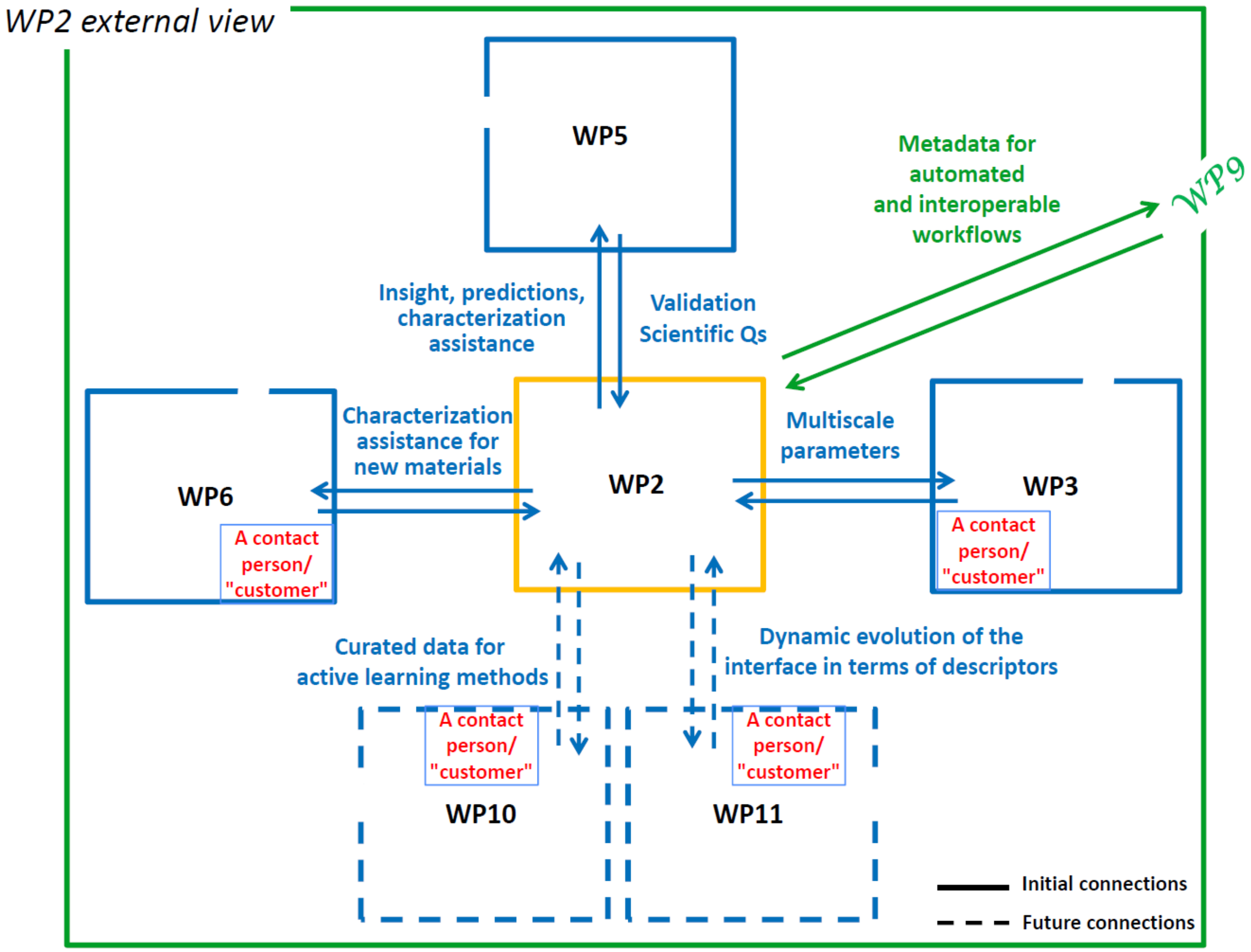

*Fig. WP 1.2. Work and data flows between WP2 and other WPs.*

### 1.2.2 Explain the relation to the objectives of the project

The work relates to BIG-MAP's Specific Objective O1: "Develop the scientific and technological building blocks and models for accelerated battery discovery" and especially the two sub-objectives:

(i) AI-accelerated atomic-scale simulation techniques for battery materials and interfaces.

(ii) Multi-scale modelling by coupling AI with multi-fidelity data from simulations and experimentation.

This translates to the following objectives for WP2:

- Benchmark and generate QM data of adequately high quality for the selected demonstrator cases.
- Develop ML force fields for simulations of electrolytes and battery interfaces.

- Develop alchemical ML models for property optimization to explore battery compound space.
- Validate and apply the developed ML models to selected demonstrator cases (systems and properties).
- Design best protocols for modelling/predicting "experimental" spectroscopies (and other characterization techniques).
- Develop a semi-automated workflow that integrates many of the previous stages.

### 1.2.3 Specify the types and formats of data generated/collected

WP2 will generate large volumes of data from e.g. electronic structure calculations and trajectories from atomistic simulations. For the atomistic configurations, for example, WP2 envisions a hierarchy of data formats (Table WP 1.1). Thus:

1. There are simple human readable formats, such as XYZ, that everyone can interpret, but the cost of making them interoperable and part of workflows is medium high because interpreters need to be written if they do not exist already for a given workflow system.
2. At the other end, there are highly specified machine readable data formats that work well for the particular purpose they were designed for, e.g. as part of a workflow system. These require little human intervention, but may not be applicable in a new situation.
3. Data in WP2 will be made available in a variety of formats along this hierarchy, depending on needs of the particular task that generates the data and also others in other WPs.

Types and formats of the data that will be generated by WP2 is shown in Table WP 1.1.

*Table WP 1.1. Types and formats of the data that will be generated by WP2*

| Datatype | Description | Data sets | Type | Format | Size |
|---|---|---|---|---|---|
| Electronic Structure: WFT, DFT, QMC | Structures, energy-related data, wave functions & electronic properties, ab-initio molecular dynamics (AIMD) trajectories, different types of spectra | Data generated by different tools: Engines (molecular): GAUSSIAN, ORCA, MOLPRO, TURBOMOLE, NWChem, QChem, ADF, PSI4, MRCC, NECI Engines (periodic): CP2K, VASP, QUANTUM ESPRESSO, Yambo, Castep, GPAW, QuantumATK, Crystal, NECI | Tarball files can be created from the calculation folder, including relevant inputs and output raw data | .tar.gz (an archive of input and output text, XML, netctdf, hdf5, or any other machine readable file) | TB |
| ML-FF (representation & regression) | SOAP/ACE + GAP Parameters | Engines: QUIP & GAP codes | As above | As above | MB |
| Alchemical Exploration and Optimization | Property relationships & Compound space search | Engines: QML & APDFT codes | As above | As above | GB |
| Atomistic simulations | Atomic trajectories, and associated transport, spectral etc. properties | Engines: LAMMPS, GROMACS,QUIP | As above | As above | TB |

WP2 will collect data to the following WPs (Table WP 1.2).

*Table WP 1.2. WP2 will collect data to the following WPs.*

| WP | What | To be used for | Suggested type | format | size |
|---|---|---|---|---|---|
| WP3 | Top-down data and knowledge | Model refinement and validation | Tarball files can be created with post-processed data and parameters of interest | .tar.gz (with databases and reproducible-data sheets) | GB |
| WP4 | Experimental data (materials, input parameters for models) | Model refinement and validation | As above | As above | MB |
| WP5 | Parameters from structural and chemical characterizations at the local scale, including e.g. spectra (vibrational, absorption, …) | Model refinement, cross-analysis, validation, and design of simulated experiments | As above | As above | MB |
| WP6 | Parameters from electrochemical and spectroscopic characterizations. New candidates from High Throughput Screening (HTS) | Model refinement, validation and design of simulated experiments | As above | As above | MB |
| WP7 | Domain knowledge on battery active materials, coating materials, characterization of interfaces, characterization of bulk and surface reactions in electrochemical systems | Ontology design | Text, excel, images | .pdf, .doc, .xls, .ppt, .png | MB - GB |
| WP8 | List of metadata that must be filled to identify the key information for modelling/simulations data | Reporting and sharing results generating in WP2 and WP3 | Text and web interface | .txt, html or java | kB and MB |
| WP10 | Suggestion for the next optimal candidates | Model refinement and validation Autonomous feedback loops | Tarball files can be created with post- | .tar.gz (with databases and | MB |
| | | Simulated experiments | processed data and parameters of interest | reproducible-data sheets) | |
| WP11 | ML based descriptor search methods at atomic scale and uncertainty estimation of ML models | Optimal chemical space and interface composition exploration | As above | As above | MB |

WP2 will deliver data to the WPs listed in Table WP 1.3.

*Table WP 1.3. Data delivery from WP2 to other WPs.*

| WP | What | Usable for | Suggested type | format | size |
|---|---|---|---|---|---|
| WP3 | Atomic-scale structures, energetics and transport coefficients required for bottom-up multiscale models Decomposition pathways and reaction rates, and more | Multiscale modelling | Tarball files can be created with post-processed data and parameters of interest | .tar.gz (with databases and reproducible-data sheets) | GB |
| WP5 | Computational predictions for bulk/interfacial structures, "chemical environments", spectra, transport properties, or any experimental characterization requiring atomistic or electronic interpretation | Guiding the characterization effort and supporting the interpretation of the results | As above | As above | TB |
| WP6 | Simulated experiments/trajectories on materials from HTS | Characterization assistance Insight | As above | As above | GB |
| WP9 | Metadata for automated and interoperable workflows | Platform and workflow engines (AiiDA, ASE/Myqueue, 3Dexperience/Pipeline pilot, SimStack) | See WP9 | See WP9 | GB |
| WP10 | Curated atomistic data | Active learning and joint in-silico/experimental feedback loops | Tarball files can be created with post-processed data and parameters of interest. | .tar.gz (with databases and reproducible-data sheets) | GB |
| WP11 | Atomistic-scale dynamics data for interfaces, related properties and features for descriptor search | Building spatio-temporal generative model and ML based descriptor search models | As above | As above | GB |

### 1.2.4 Specify if existing data are being re-used (if any)

Relevant publicly available data collected from data repositories (e.g. NOMAD, Materials Cloud, CMR, etc.) will be used when appropriate. WP2 teams will mainly generate new data.

### 1.2.5 Specify the origin of the data

The data is generated/collected from computational work performed by the involved organizations:

Task 2.1: Electronic Structure; UU, UCAM, SINTEF, UNIVIE, WWU, CTH;

Task 2.2: ML-FF (representation & regression); UCAM, UU, DTU, UNIVIE, WWU;

Task 2.3: Alchemical Exploration and Optimization; UNIVIE, UCAM, UU, DTU;

Task 2.4: Electronic and Atomistic simulations; CTH, WWU, DTU, UU, UCAM, 3DS, CEA, UMI, SAFT;

Task 2.5: Electronic and Atomistic simulations (spectroscopy etc.); CNR, EPFL, UU, CTH;

### 1.2.6 State the expected size of the data (if known)

The total size of the generated data will be in the TB+ range (see Table WP 1.1).

### 1.2.7 Outline the data utility: to whom will it be useful

- Curated reference data for repositories or interactive databases. Useful for the consortium and wider.
- New interatomic potentials. Useful for wide modelling communities.
- Alchemical data sets. Useful for the consortium and wider.
- MD trajectories. Useful for the consortium and wider.
- Property and spectral predictions. Useful for wide experimental and computational communities.
- Computational protocols and workflows. Useful for wide modelling communities.

## 1.3 **WP3 – Multiscale Modelling**

### 1.3.1 **Purpose of data collection/generation**

WP3 deals with the development of predictive scale-bridging models for battery interphase/interfaces on the basis of the atomic constituents of the materials involved. Using AI to enhance the scale-bridging and validating the models against first-principles calculations and experimental data. Physics-based models that lack predictive power, will be augments with data to generated predictive models that can be integrated into the multiscale modelling chain. This WP serves as a technology demonstrator, and will use model chemistries that are much simpler than those in commercial batteries, but nevertheless sufficiently complex to be relevant and push the boundaries of what is possible to simulate computationally. In WP3 we will develop and validate multiscale simulations that utilize a number of different models and conceptual approaches that will significantly accelerate research regarding the battery interface genome (WP11). Mesoscale and continuum modelling can be used to explore rare events, cell function and long-term degradation at the device level. A comprehensive model of the Solid Electrolyte Interphase (SEI) / Cathode Electrolyte Interface (CEI) remains a significant challenge that will be addressed by the development of new scale-bridging models by the partners in this WP. In particular, the development of mesoscale models that incorporate material specific information is key to describe the complexities of the SEI/CEI accurately. Given the long time-scales involved, the development of advanced coarse-graining simulation techniques is a key challenge for this WP, as well as their harmonization with continuum modelling.

### 1.3.2 **Explain the relation to the objectives of the project**

The work relates to BIG-MAP's Specific Objective 1: "Develop the scientific and technological building blocks and models for accelerated battery discovery" and especially the two sub-objectives:

1. AI-accelerated atomic-scale simulation techniques for battery materials and interfaces.
2. Multiscale modelling by coupling AI with multi-fidelity data from simulations and experimentation.

In WP3 we will develop and validate multiscale simulations that utilize a number of different models and conceptual approaches to be used in WP11. Mesoscale and continuum modelling can be used to explore rare events, cell function and long-term degradation at the device level. We will develop a comprehensive model of the SEI/CEI, which remains a significant challenge that will be addressed by the development of new scale bridging models by the partners in this WP. In particular, the development of mesoscale models that incorporate material specific information is key to describe the complexities of the SEI/CEI accurately. Given the long-time scales involved, the development of advanced coarse-graining simulation techniques is a key challenge for this WP, as well as their harmonization with continuum modelling.

### 1.3.3 **Specify the types and formats of data generated/collected**

WP3 will generate large volumes of data from e.g. electronic structure calculations and trajectories from atomistic simulations. For the atomistic configurations, for example, WP3 envisions a hierarchy of data formats (see Table WP 1.4). Thus:

- There are simple human readable formats, such as XYZ, that everyone can interpret, but the cost of making them interoperable and making them part of work flows is medium high because interpreters need to be written if they don't exist already for a given workflow system.

- At the other end, there are highly specified machine readable data formats that work well for the particular purpose they were designed for, e.g. as part of a workflow system. These require little human intervention, but may not be applicable in a new situation.

Data in WP3 will be made available in a variety for formats along this hierarchy, depending on the needs of the particular task that generates the data and also others in other WPs. We note that data generated in WP3 may slightly differ from the data generated in other WPs as some data is generated as intermediate data in scale-bridging workflows. Such data, in particular from lower scales (atomistic, electronic structure) is then passed to higher scales or vice versa. This data is reported in tables below with the tag (intermediate data in the column datasets) and is not the primary output of the workflow, but required for completeness, according to our data standards. Types and formats of the data that will be generated by WP3 are shown in Table WP 1.4**.**

*Table WP 1.4. Types and formats of the data that will be generated by WP3*

| Datatype | Description | Data sets | Type | Format | Size |
|---|---|---|---|---|---|
| Workflows (Simstack, AiiDA) | Computational protocols for workflows and data access | AiiDA, ASE, CLEASE and SimStack Apps and their components | Uncompres sed raw data | GitHub | MB |
| Category: CGMD, DEM; Engines: LAMMPS, LIGGGTHS | Description of the meso-structure of electrodes | Trajectories, structures, engines: LAMMPS, HOMOMD, LIGGGTHS | Tarball | .tar,gz | GB |
| Category: 4D-resolved continuum electrochemica l simulations Engine: Comsol Multiphysics | Prediction of electrochemical observables and reaction/transport heterogeneities | Trajectories, Structural data | Tarball | .tar,gz | GB |
| Electronic Structure: DFT, SCC-DFTB; Engines: VASP, CP2K, DFTB+ | Structure, energies, wave function, electronic properties, spectra | Intermediate Data in Workflows | Uncompres sed raw data, tarball, database | .tar,gz. Databas e | TB |
| Atomistic simulations: FF and ML-FF, Engines: LAMMPS | Structures, energies, trajectories, spectra | Intermediate Data in Workflows | Uncompres sed raw data, tarball, database | .tar,gz, Databas e | TB |
| Platform for calculations and workflows, | Curated data | Access protocols, ontology components | Database, semantic access, web | Text | GB |
| Engines: ASE, AiiDA | | | access, API access | | |
| (Category of data):<br>1) SEI (CEI) molecular structures;<br>2) Solution trajectories from kinetic Monte Carlo (KMC) algorithms;<br>3) solution trajectories from reactive FF molecular dynamics simulations.<br>(Engines): LAMMPS, SCM, SCIPY, NUMPY, MATLAB, in house codes; | 1) An ensemble of SEI (CEI) structures corresponding to different aging and operating conditions;<br>2) the growth of the SEI (CEI) layer during charging (discharging) phases (KMC);<br>3) MD solution trajectories for a number of SEI (CEI) configurations with the main reaction products forming the SEI under imposed conditions. Other relevant quantities will be also investigated (e.g. exchange currents, free-energies for Li across the interface); | Mathematical models for SEI, CEI parameters, Structural data, Trajectories | Uncompressed raw data, tarball | .tar,gz, Database | TB |
| P2D+Population balance equation model | Lithium concentration profiles in electrolyte/electrode particles, SEI layer thickness distribution on electrode particles, state of charge, voltage, capacity fading curves | Trajectories and time-lines for concentrations and reactions in batteries with spatial resolution and aggregated | Uncompressed raw data | Code-Specific | GB |
| Category: Electronic Structure: WFT, DFT<br>Engines (molecular): GAUSSIAN, ORCA, TURBOMOLE<br>Engines (periodic): CP2K, VASP | Structures, energy-related data, wave functions & electronic properties, AIMD trajectories | Intermediate Data in Workflows | Uncompressed raw data, tarball | .tar,gz, Database | TB |
| Category: ML-FF (representation & regression) | Force field models for battery materials and compositions | SOAP/ACE + GAP Parameters | Uncompressed raw data, tarball | .tar,gz, Database | MB |
| Engines: QUIP & GAP codes | | | | | |
| Category: Hybrid MD/MC (rs@md) Engines: GROMACS + in-house wrapper | The in-house code regularly interrupts the GROMACS simulation and changes the topology (e.g. creation/deletion of chemical bonds). The output is the same as for GROMACS. | Intermediate Data in Workflows | Uncompres sed raw data, tarball | .tar,gz, Databas e | TB |
| Electronic structure: DFT, NEGF. Engines (molecular and periodic): SIESTA (also with TranSIESTA module) | We will be using the SIESTA code (https://siesta.icmab.es/sie sta), that implements both standard DFT and the NEGF formalism (TranSIESTA) for calculations under potential bias. SIESTA is interfaced to the AiiDA platform, so some of the data might be made available as collections of AiiDA database nodes. Part of the data can also be generated in netCDF form | Intermediate Data in Workflows | Tarball, database, netCDF | .tar,gz, Databas e NETCD F | GB |
| Category: P4D Continuum modelling simulation, cell-level variables evolution. Engine: in-house code with FEniCS (Python) | Cell-level parameters i.e.: voltage curves, performance evolution, up to 3D detailed variables (potentials, concentrations, exchange-currents, SEI layer thickness) | Extended P2D to 3D with thermal, mechanical and electrochemical (CEI, SEI) degradation models accounting inhomogeneities within the cell and particle level. Generates datasets of evolution of internal variables and cell performance across long cycle life tests in hdf5 or csv format. Both formats can be easily iterated using open libraries in almost any language (i.e Python) | Uncompres sed raw data, tarball, HDF5 | HDF5, .tar,gz, Databas e | MB |
| GCMD + Electronics Structure: LAMMPS | Trajectories and Snapshots for the morphology of nanoparticles (e.g. Si) intercalated with Li | Intermediate Data in Workflows | Tarball, SimStack, ASE | .tar,gz, Databas e | TB |
| KMC: in-house code, mesoscopic | Structure and morphology (snapshots and trajectories) of SEI formation at model electrodes | In-house KMC code orginally used for catalysis which has been adapted to SEI formation | Tarball | .tar,gz, Databas e | TB |

More details about the workflows and calculation platforms mentioned in the table above can be found in the section dedicated to WP9.

WP3 will collect data to the WPs listed in **Table WP 1.5**.

*Table WP 1.5. WP3 will collect data to the following WPs*

| WP | What | To be used for | Suggested type | format | size |
|---|---|---|---|---|---|
| WP2 | Force fields Ab-initio MD trajectories | Input to MD Models | Tarball, an archive of input and output text, XML, netctdf, hdf5 | .tar,gz, Database | GB |
| WP4 | Experimental Data (materials, input parameters for models) | Validation of models for battery interfaces/interphases, battery models in general, training of AI | Output data, parameter sets | .doc/ .csv | KB-MB |
| WP5 | Electrochemical characterization Scattering data (XRD, SAS, XRR, NR,..) Spectroscopy data (NMR, XPS, IR/Raman,..) Imaging data (TEM/SEM, FIB-SEM, XRT, XRD-CT, NI, NCT, NanoTomo, NDP,..) | Validation of models for battery interfaces/interphases, battery models in general, training of AI | Uncompressed raw data Data Sheets | Database | TB |
| WP6 | Electrochemical characterization, XPS, IR/Raman, TEM/SEM, FIB-SEM, XRD | Validation of models for battery interfaces/interphases, battery models in general | Uncompressed raw data | Database | TB |
| WP7 | Semantic data representation, Scale bridging data representations | Workflow generation, interaction with data-driven approaches, data-exchange with WP10, WP11 | Uncompressed raw data | Database | TB |
| WP8 | Data standards and list of metadata | Contribution to materials data bases and workflows in a community accessible format | Uncompressed raw data | Database | TB |
| WP9 | Workflow tools, Databases | Validation of multiscale models, training of AI | Tarball | .tar,gz | TB |
| WP10 | AI models for scale-bridging and surrogate models | Models that may bridge different scales and supply surrogate modelling | hdf5 or json | Hdf5 | TB |
| WP11 | Uncertainty propagation scheme between scales | Efficient bound on uncertainty for multiscale simulation | Tarball | Python code | TB |

WP3 will deliver data to the WPs listed in Table WP 1.6.

*Table WP 1.6. Data delivery from WP3 to other WPs.*

| WP | What | Usable for | Suggested type | format | size |
|---|---|---|---|---|---|
| WP2 | Top-down data and knowledge | Accelerated atomic scale simulations | Tarball, an archive of input and output text, | XML, netctdf, hdf5, | TB |
| WP5 | PxD model output, MD trajectories, KMC output | Benchmark workflow; Acquisition analysis | Tarball, uncompressed raw data | .tar,gz | TB |
| WP6 | PxD model output, MD trajectories, KMC output | HTS and testing Analysis | Tarball, uncompressed raw data | .tar,gz | TB |
| WP7 | Domain knowledge | Battery interface ontology | Tarball, an archive of input and output text, | .tar,gz, XML, netctdf, hdf5 | TB |
| WP7 | Domain knowledge on battery active materials, coating materials, characterization of interfaces, characterization of bulk and surface reactions in electrochemical systems | Ontology design | Text, excel, images | .pdf, .doc, .xls, .ppt, .png | MB-GB |
| WP8 | Domain knowledge | Standardization and protocols | Tarball, an archive of input and output text, XML, netctdf, hdf5 | XML, netctdf, hdf5, .tar,gz | TB |
| WP9 | Domain knowledge; Workflows | Infrastructure and interoperability | Tarball, an archive of input and output text, XML, netctdf, hdf5 | XML, netctdf, hdf5, .tar,gz | TB |
| WP10 | Domain Knowledge; Workflows; Training Data | Workflow interface between WP2 and WP10, training data for WP10 | Tarball, uncompressed raw data | Hdf5, json | TB |
| WP11 | Multiscale simulation data | Model building and verification and validation of the performance of generative model | Tarball, uncompressed raw data | Text | TByte |

The flow of data between WP3 and its tasks and to the other WPs is shown in Fig. WP 1.3 and Fig. WP 1.4.

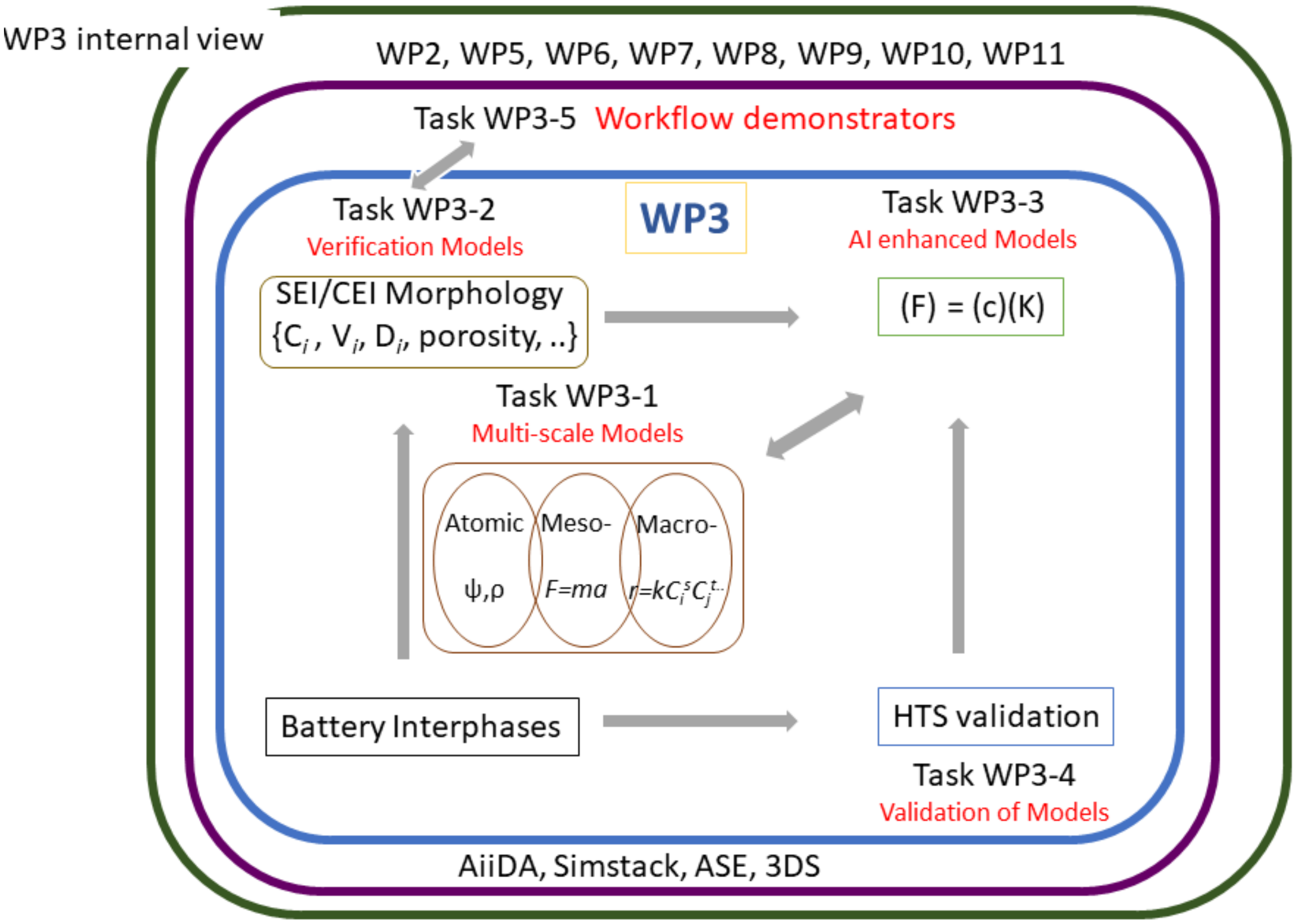

*Fig. WP 1.3. Data flow between WP3 and its tasks.*

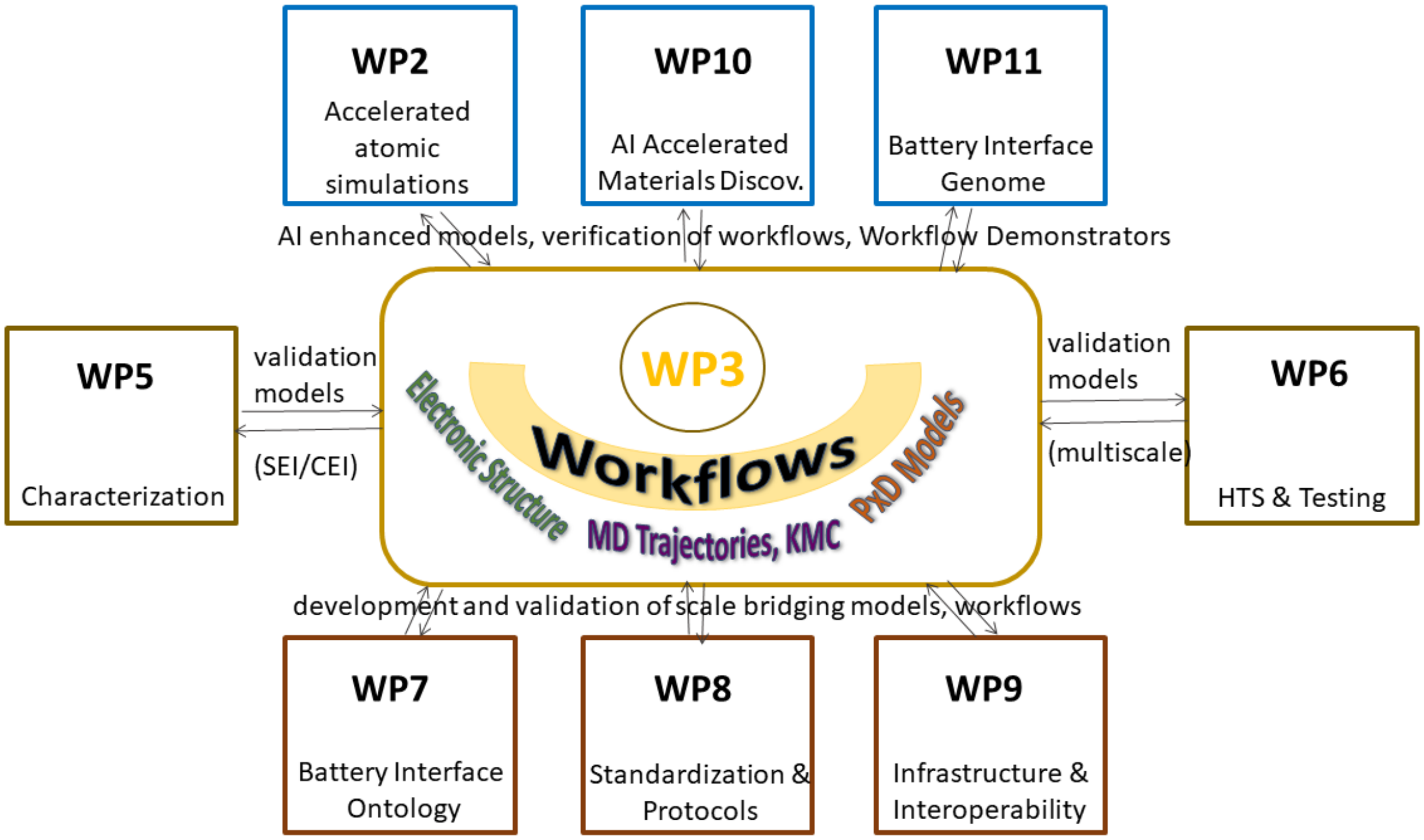

*Fig. WP 1.4. Data flow between WP3 and the other WPs.*

### 1.3.4 Specify if existing data are being re-used (if any)

Relevant publicly available data collected from data repositories (e.g. NOMAD, Materials Cloud, CMR, etc.) will be used when appropriate. WP3 teams will mainly generate new data

### 1.3.5 Specify the origin of the data

The data is generated/collected from computational work performed by the involved organizations:

Task 3.1: Development and validation of multiscale modelling frameworks (PDT, DTU, UU, KIT, CNRS, CEA, NIC, CNR, CID, FZJ): electronic structure method, atomistic modelling, coarse grained & mesoscopic models, continuum models, PxD, workflow systems (AiiDA, ASE, Simstack)

Task 3.2: Verification and validation of scale-bridging protocols for SEI/CEI (ALL): workflows developed in task 3.1

Task 3.3: AI enhanced models (KIT, DTU, UU, CNRS, CEA, PDT): AI-frameworks, workflows developed in task 3.1 and 3.2

Task 3.4: Verification and validation of models for HTS experiments (WWU, DTU, NIC, KIT, SAFT): workflows developed in Task 3.1 and 3.2, ontologies, databases, access protocol to robotic experiments

Task 3.5: Workflow demonstrators (KIT, DTU, UU, CNRS, NIC, PDT, CNR, CID, FZJ), AiiDA, ASE, Simstack): workflows developed in Task 3.1, 3.2

### 1.3.6 State the expected size of the data (if known)

The total size of the generated data will be in the TB+ range.

### 1.3.7 Outline the data utility: to whom will it be useful

- Curated reference data for repositories or interactive databases. Useful for the consortium and wider.
- New interatomic potentials. Useful for wide modelling communities.
- Alchemical data sets. Useful for the consortium and wider.
- MD trajectories. Useful for the consortium and wider.
- Property and spectral predictions. Useful for wide experimental and computational communities.
- Computational protocols & workflows. Useful for wide modelling communities.
- Long term cell degradation capacity fade prediction and internal variables, including SEI resistance and thickness evolution. Useful for the consortium and wider.

## 1.4 **WP4 – Modular Synthesis Robotics**

### 1.4.1 **Purpose of data collection/generation**

The scope of WP4 is:

- construction and integration of a modular and automated formulation and synthesis framework of hard- and software
- synthesis and test of inorganic and organic materials acting as ex-situ (inorganic or organic protective layers) or in-situ (SEI-forming additives and salts) coatings on substrates (anode and cathode particles or planar surfaces).

The data generated within this work packages falls in three general categories:

1. Robotic hard- and software
2. Material properties and synthesis procedures
3. Test and characterization data

The data in category 1 will mostly be used for WP-internal as well as for the preparation of interfaces to the work packages WP6 and WP10. Category 2 and category 3 data will be shared with the theoretical WPs 2, 3 and 11 as well as the experimental WPs 5, 6, 9 and 10.
The interoperability of the experimental data (raw, processed and analysed) will be ensured by applying the ontology developed and defined within the WP7 and the standards for data and metadata defined in WP8. The generated output data will be specified in the form defined by WP9 and labelled in terms of fidelity level. Obtained experimental data will be transferred to WP10 and WP11 being in charge of handling, storage and AI-training.

### 1.4.2 **Explain the relation to the objectives of the project**

The work in WP4 is directly related to the sub-objective "Demonstration of an open-source framework for seamless integration of software, hardware and robotics modules to enable autonomous synthesis processes for advanced battery materials and cells" within the objective O1 "Develop the scientific and technological building blocks and models for accelerated battery discovery".

For the Demonstration of autonomous synthesis robotics of protective electrode coatings (BIG-MAP's key demonstrator 3) the three abovementioned categories 1 to 3 must be considered, more specifically:

- Documentation of the soft- and hardware specifications, machine communication etc.
- Description of the basic synthesis workflows, including mapping to the hardware utilized, material properties and test data
- Generation and transfer of test data to other WPs, especially to WP5 (in-depth characterisation of protective coatings), WP6 (HTS assay of protective coatings).

### 1.4.3 **Specify the types and formats of data generated/collected**

Data generated/collected in the frame of the WP4 will be mostly text-based and can be supplied in different text-based formats depending on individual requirements (JSON, CSV, XML). The experimental data will belong to 4 main categories: raw data (2D, 3D data sets), processed data (corrected, normalized, averaged, etc.), analysed data (e.g. fitted, for instance) and output data (e.g. parameters from fits, for instance). The access to data as well as transfer to other WPs will be extremely dependent on the technique used.

Types and formats of the data that will be generated by WP4 are listed in Table WP 1.7.

*Table WP 1.7. Types and formats of the data that will be generated by WP4*

| Datatype | Description | Data sets | Type | Format | Size |
|---|---|---|---|---|---|
| Material specifications | Selection of materials to be used for inorganic and organic protective layers | Specification workshop, discussion, documentation | Word files | .doc | MB |
| **Robotic hardware and software** | | | | | |
| Technical documentation of robotic system | Machine specifications and product data sheets, Technical description of the software and hardware developed | Documentation | Word files | .doc | MB |
| | Specification of the communication interface | Documentation, Software | Word | .doc | KB-MB |
| Process documentation of robotic system | Run conditions (amounts, source of base materials, order of addition, mixing times, mixing speed, date of run, identity of run instigator) | Data available through "FLEX" access to data via internet login | | | KB-MB |
| | | to be specified (D4.1), e.g. Wiki via internet login | e. g. text, images | | KB-MB |
| | Standard operation procedure SOP; Parameters set and materials required to run a suggested experiment; automatically or manually generated | Documentation | Excel file | .csv | KB |
| | Materials request list from robot to perform requested experiment | Documentation | Excel file | .csv | KB |
| Software middle layer | Enabling the communication (e. g. via JSON files) of the different robotic systems and AI (WP10) | Software | tbd. | | KB-MB |
| **Material characterization: Precursors and polymers, Salts, SEI Additives** | | | | | |
| Electrochemical data | Properties of polymers, salts and additives<br>- electrochemical stability and redox properties<br>- conductivity and conductivity mechanisms | Potentiostatic amperometry, Galvanostatic potentiometry, Electrochemical | Spectra<br><br>Key performance indicators versus time | .mpt / .csv | MB-GB |
| | - transference number<br>- small scale electrolyte salts/additive synthesis quality check and basic materials data | impedance spectroscopy | | | |
| Spectroscopic data | Electronic structures, chemical state of bulk and surface, phase purity, local chemical environment, degree of polymerization, synthesis quality check and basic materials data | NMR, IR, FTIR, Raman, XPS | Spectra | ASCII | KB-MB |
| Physicochemical data | Titration data; check for residual water | Karl-Fischer-Titration | Table | .csv | KB |
| | Viscosity | Rheology | Table | .csv | KB |
| **Component characterization: Coated particles, Electrolytes, coated and uncoated thin film electrodes** | | | | | |
| Electrochemical data: | Properties of coated particles, coated thin film electrodes, electrolytes<br>- electrochemical stability and redox properties<br>- conductivity and conductivity mechanisms | Potentiostatic amperometry, Galvanostatic potentiometry, Electrochemical impedance spectroscopy | Spectra<br><br>Key performance indicators versus time | .mpt, .csv | MB-GB |
| Structural data | Crystal structure of particles and thin films | (GI)XRD, powder XRD, STEM-HAADF, HR-TEM | XRD spectra | .ASCII | KB-MB |
| | Morphology, roughness and thickness | SEM, FIB-SEM | Images | .tiff, .jpg | MB |
| | Geometrical/Microstructural data of thin films | Stylus profilometry | Tables | .csv | KB |
| Chemical data | Local chemical environment, elements, | XPS, FIB depth profile | Spectra | .csv | KB-MB |
| | Elemental analysis of thin films | ICP-OES, Atomic Absorption Spectroscopy, EDX, RBS, LIBS | Tables<br><br>Spectra | .csv | KB |
| | Elemental mapping | EDS/EDX, EELS, FIB-SEM depth profile | Images, spectra, | .jpg, .png, .tiff, .csv | MB |
| | Interface characterization; Local chemical environment at the interphase (organic and inorganic), crystal structure, | XPS, XRD | Images, spectra, measurements | .jpg, .png, .tiff | MB-GB |
| | Characterization of inorganic coatings on particles; Local chemical environment – film and electrode material | Raman, PXRD, | Spectra, images | .csv .jpg, .png, .tiff | MB-GB |
| **Materials and components for characterization and tests in other WPs** | | | | | |
| Materials for organic protective coatings and electrolytes | Molecules for organic protective coatings<br>Conductive salts<br>SEI forming additives | Sample material, product data sheets, meta data | Physical samples, Text, tables, JSON, links | .doc, .xls, .csv | g-kg / KB-MB |
| Protective coatings | Cathode powder with organic coatings<br>A library of inorganic electrode coatings | Sample material, product data sheets, meta data | Physical samples, Text, tables, JSON, links | .doc, .xls, .csv | g-kg / KB-MB |

WP4 will collect data to the following WPs listed in Table WP 1.8.

*Table WP 1.8. WP4 will collect data to the following WPs*

| WP | What | To be used for | Suggested type | format | size |
|---|---|---|---|---|---|
| WP5 | In-depth characterization of selected additives and coatings ; high-fidelity and high-throughput data ; | Evaluation of property-performance relationship Suggest next optimal screening & formulations | Output data & parameters sets | .doc | KB |
| WP6 | High throughput materials screening and performance evaluation | Suggest next optimal formulations and syntheses | Output data & parameters sets | .doc | KB |
| WP8 | Templates for standards and protocols for test data | Parameter sets for own experiments, reporting and data handling | Documentation | .doc, .xls | KB |
| WP10 | Suggested next experiments (e. g. via JSON file) | Perform synthesis experiment based on WP10 input | Parameter sets | JSON | KB |
| WP10 | Suggested candidate molecules to be synthesized. Molecules to be used for protective coatings from WP10's materials funnel | For synthesis, characterization and testing of the actual molecules as protective coatings on cathode powders | Documentation | .doc | KB-MB |
| WP10 | Raw data<br>Analysed data | Active learning runs and for automated data analysis of selected data | Output data<br>Raw data | .json<br>.hdf5<br>.csv | KB-MB |

WP4 will deliver data to the WPs listed in Table WP 1.9.

*Table WP 1.9. Data delivery from WP3 to other WPs.*

| WP | What | Usable for | Suggested | | |
|---|---|---|---|---|---|
| | | | type | format | size |
| WP2, WP3 | Experimental data (materials, input parameters for models) | Model building and simulation | Output data, parameter sets | .doc, .csv | KB-MB |
| WP5 | Selected electrolyte additives and salts, coated anode and cathode powder | High fidelity characterization | Sample sheets, data sheets | .doc samples | KB-MB |
| | | | Samples for characterization | g | |
| WP6 | Selection of electrolytes additives and salts; coated anode and cathode powders; basic characterizations | Design experiments for advanced characterization | Sample sheets, data sheets | Docs, samples | KB-MB |
| | | | Samples for tests | g-kg | |
| WP6 | Gradient coatings or combinatorial coatings | Design experiments for advanced characterization | Sample sheets, data sheets | Docs, samples | KB-MB |
| | | | Material for test | g-kg $cm^2$ | |
| WP7 | Domain knowledge | Ontology design | Text, excel, images | .pdf, .doc, .xls, .ppt, .png | KB-MB |
| WP10 | Data to run autonomous feedback loop | Synthesis parameters, synthesis control data, materials characterization | Data sheets | .csv, .json (tbd) | KB |
| WP10 | Communication protocol | Enable interaction with WP10 materials funnel / experiment suggestions (e. g. via JSON) | | .json | KB |
| WP10 | Experimentally determined performance of selected candidate molecules for protective coatings, (performance of molecules applied as protective coatings on cathode powder) | Feed back to the materials funnel | Data sheets | .csv | KB-MB |

The detailed data flow between the tasks of WP4 and between WP4 and the other WPs are shown in Fig. WP 1.5 and Fig. WP 1.6.

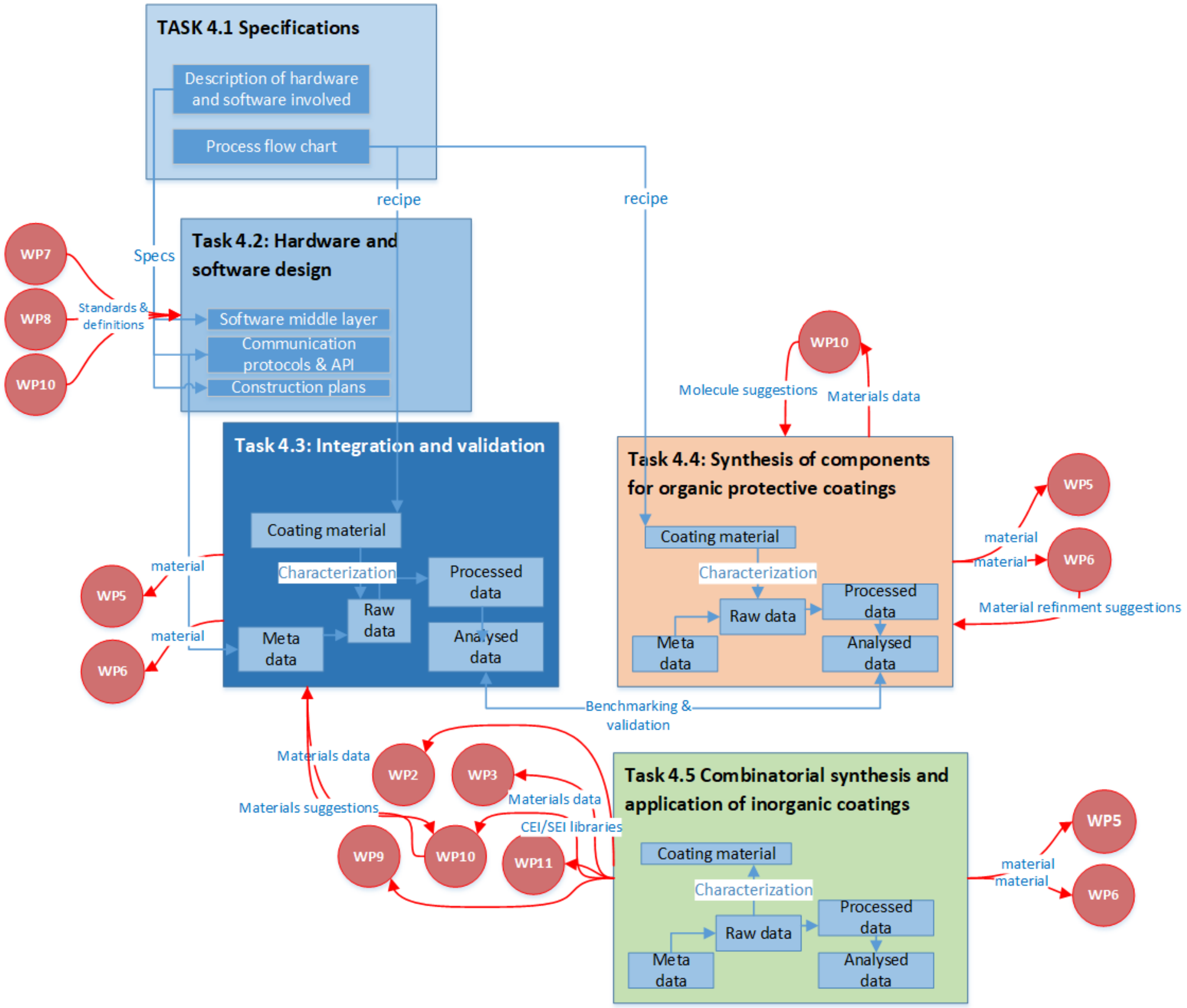

*Fig. WP 1.5. Data flow between the tasks of WP4 and relations for the other WPs.*

## 1.4.4 Specify if existing data are being re-used (if any)

Existing data, e. g. from material and crystal structure databases, peer-reviewed publications as well as own data will be used in order to verify manual and automated syntheses in this work package. Technical documentation of the machinery utilized will be used for the hard- and software construction and the definition of communication protocols.

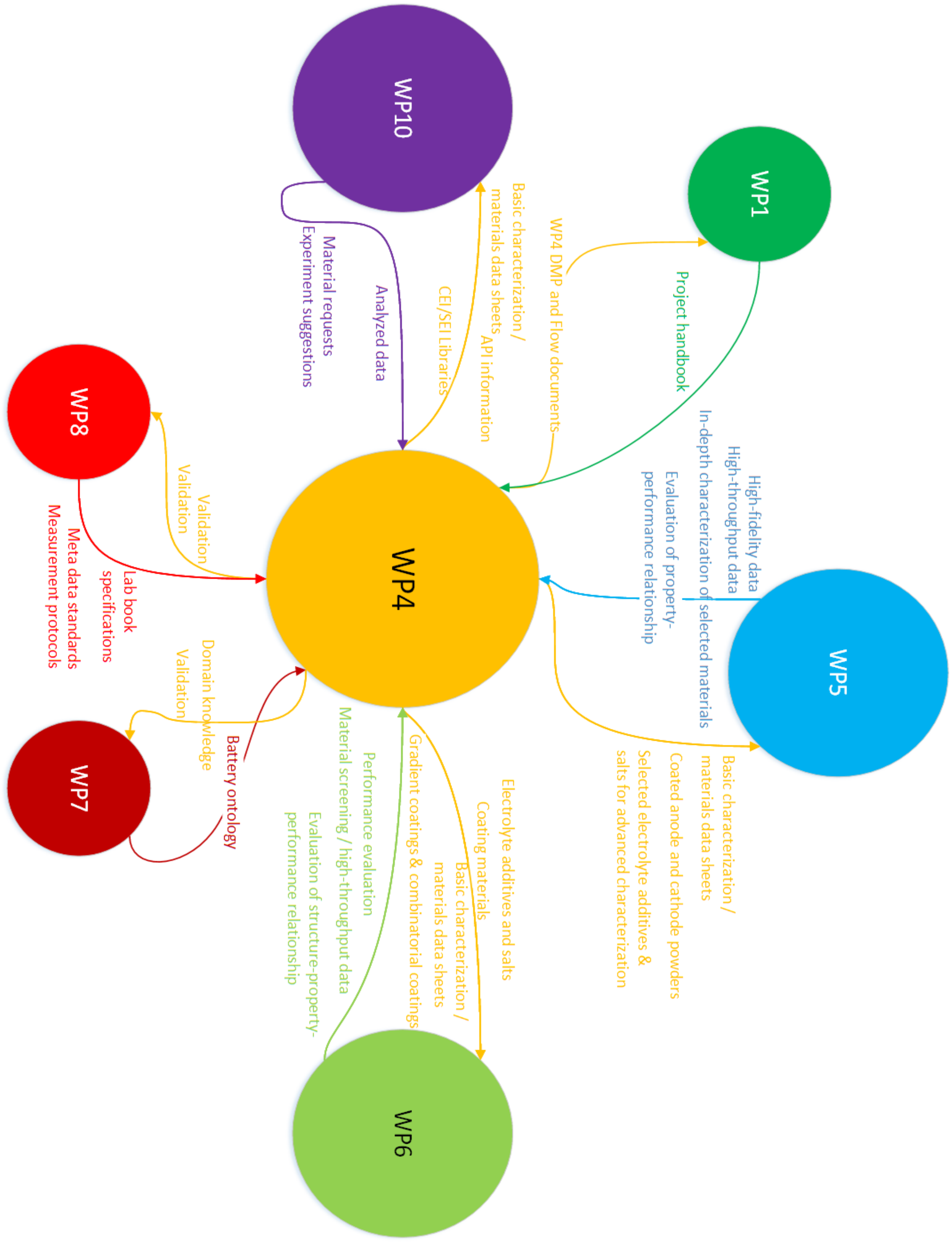

*Fig. WP 1.6. Data flow between WP4 and the WPs of BIG-MAP*

### 1.4.5 Specify the origin of the data

Task 4.1: Machine specifications (ITU, Fraunhofer, ULIV, FZJ, CEA, UT): documentation of soft- and hardware to be developed in WP4, technical meetings and workshops, technical documentation on-site

Task 4.2: Hardware and software design (UT, Fraunhofer, ULIV, ITU): soft- and hardware construction details

Task 4.3: Integration and validation (Fraunhofer, ULIV, UT, ITU, NIC): test protocols, hardware technical documentation and handbooks

Task 4.4: Synthesis of coating materials (NIC, WUT, Fraunhofer, ULIV, SOLB, UMI, FZJ): synthesis protocols, material data sheets and experimental data generated in the labs

Task 4.5: Combinatorial synthesis of protective coatings (CEA, FZJ): synthesis protocols and machine settings, material data sheets and experimental data generated in the labs.

### 1.4.6 State the expected size of the data (if known)

The expected size of the generated data is listed in the table above. The majority of data is expected to be in the MB range (images and electrochemical data), smaller file sizes (KB for spectra) and larger files in the GB scale can also be generated.

### 1.4.7 Outline the data utility: to whom will it be useful

The data created might be useful to the following groups:

- BIG-MAP consortium: especially WP2, WP5, WP8, WP10, WP11
- European Commission and European Agencies
- Stakeholders involved in the field of Li-ion batteries and materials
- Scientific community
- EU national bodies

## 1.5 WP5 – Characterization

### 1.5.1 Purpose of data collection/generation

The goal of WP5 is to collect and analyse experimental data describing the interface structure and the evolution of interphases at all relevant length- and time-scales, using a wide range of complementary in-lab and large scale facilities techniques. The generated data will be used for model development & validation and AI-training. In particular, experimental results are needed to identify realistic input parameters for the development of new computational models, and modelling results need to be validated against experimental results.

The interoperability of the experimental data (raw data, processed data, analysed data) will be ensured by applying the ontology developed and defined in WP7 "Battery Interface Ontology" and the standards for data and metadata defined in WP8 "Standardization and Protocols". The generated output data will be specified in the form defined by WP9 "Infrastructure & Interoperability" and labelled in terms of fidelity level. They will be transferred to central BIG-MAP, e.g. delivered to the WPs taking care of handling, storage, database, AI-training (WP10 and WP11) and modelling (WP2 and WP3).

### 1.5.2 Explain the relation to the objectives of the project

The work in WP5 relates to BIG-MAP's Specific Objective 1: "Develop the scientific and technological building blocks and modes for accelerated battery discovery" and especially the sub-objective: "Development of a European multi-modal characterization platform for battery interfaces and interphases bridging lab- and large-scale facilities (LSF) and techniques".

The key demonstrator of WP5 is the capability to run coordinated multi-techniques experiment to acquire multi-scale data, with specifically the following challenges and objectives:

- Develop a methodology for an accelerated experimental workflow
- Automate the acquisition, curation and analysis of experimental data sets
- Make advances in experimental techniques across multiple time- and length-scales with a linkage between in- and ex operando conditions
- Test the implementation of a European multi-modal experimental platform using standardized cells/protocols/metadata/data collection, treatment and analysis.

### 1.5.3 Specify the types and formats of data generated/collected

The types and formats generated and collected in WP5 will be very diverse, because of the variety of lab techniques, large-scale facility techniques, and measurement conditions (ex situ, in situ, operando). The experimental data will belong to 4 main categories: raw data (1D, 2D, 3D, 4D data sets), processed data (corrected, normalized, averaged), analysed data (e.g. fitted, for instance) and output data (e.g. parameters from fits, for instance). The access to data as well as transfer to other WPs will be extremely dependent on the technique used.

Types and formats of the data that will be generated by WP5 are listed in Table WP 1.10.

*Table WP 1.10. Types and formats of the data that will be generated by WP5*

| Datatype | Description | Data sets | Type | Format | Size |
|---|---|---|---|---|---|
| Methodology data | | | | | |
| Experimental logbooks | Conditions of realization of an experiment including sample description, technical set-up, type of measurement and output parameters, localization of raw data | Filling the on-line tool | Document | .doc | KB |
| Experimental matrix | Classification of techniques for battery interface characterization, criteria & indicators | Inputs from all partners and compilation of information using matrix template | Table | .doc | KB |
| Competence matrix | Summary of technical and experimental capabilities ; mappings of competence organized according to criteria (length-scales, type of information, operando cells availability, etc) | Inputs from all partners and compilation of information using mappings | Tables and graphical mappings | .doc | MB |
| Cluster organization | Description of techniques grouped in clusters | Diffraction cluster, bulk spectroscopy cluster, surface spectroscopy cluster, imaging cluster | Tables and graphical mappings | .doc | MB |
| Selection of Tier1 and Tier 2 techniques | Identification of key techniques based on competence and experimental matrixes | Compilation of available capabilities and selection according to other WPs targets | Tables | .doc | KB |
| Experimental workflow | Tier 1 and Tier 2 workflows describing the implementation of the experiments (timeline, sites, interconnections, cells) | Analysis of the clusters and tiers techniques | Tables and graphical mappings | .doc | MB |
| WP5 workflow interfacing with other WPs | Connections of WP5 with other WPs to develop the experimental program | Feed-back loops | Tables and graphical mappings | .doc | KB |
| Data acquisition methods | New modes of acquiring high-fidelity data, high throughput data, inter-operable data | Feed-back loops with modelling and AI<br>Depends on the techniques, to be adapted for each cluster family | Tables and graphical mappings | .doc | MB |
| Data analysis methods | Codes and algorithms for data analysis, interfaces for data representation | All partners' internal tools, mutualisation of best practices<br>Depends on the techniques, to be adapted for each cluster family | Many types depending on data | .py<br>Or others | KB<br>MB |
| Data sheets elaboration | Transforming analysed data into data transferable to other WPs | Depends on the techniques, to be adapted for each cluster family, guidelines to users to be able to extract, plot, represent the output parameters | Sheets | To be defined | KB |
| Cell harmonization | List of available operando cells classified by cluster, benchmark, development of new compatible cells | Identification of limits and needs for operando measurements (depending on techniques), selection of key cell designs | Summary of cell designs | .doc | MB |
| Materials characterization data | | | | | |
| Experimental logbooks | Conditions of realization of an | Filling the on-line tool | Document | .doc | KB |
| | experiment including sample description, technical set-up, type of measurement and output parameters, localization of raw data | | | | |
| Performance degradation | The electrochemical performance of the batteries versus operational time | Potentiostatic amperometry, Galvanostatic potentiometry, Electrochemical impedance spectroscopy | Spectra<br><br>Key performance indicators versus time | .mpt | KB |
| Structural data (lab.) | Crystal structure of the cathode and the anode before and after cycling | X-ray diffraction | XRD diffractograms | .xy | MB |
| | Phase distribution and phase transformations, strain & stress | Data analysis | To be defined | | MB |
| Structural data (large-scale facilities) | Relation between the phase transformations of the crystal structures in the electrodes, and the electrochemical activity, i.e. characterising the cells in real time. | X-ray diffraction<br>Neutron diffraction | To be defined | .xy | MB |
| | Relation between the nanoscale structural organisation in the electrodes, and the electrochemical activity, i.e. | SAXS<br>SANS | 2D patterns | .nxs<br>.dat | MB |
| | characterising the cells in real time. | | | | |
| Spatially-resolved microscopy-based structural/morphological data (lab.) | 2D/3D distribution of phases, nano- and microstructures including porosity, cracks and defects, SEI | FIB-SEM<br>STEM-HAADF<br>HR-TEM | Images, statistical analysis (profiles, distributions) | Many formats depending on technique | GB<br>TB |
| Spatially-resolved structural/morphological data (large scale facilities) | 2D or 3D distribution of phases, crystalline structures, nano- and microstructures including porosity mapping / cracks & defects | XRD-CT/ SAXS-CT<br>X-ray micro-tomography<br>X-ray nano-tomography<br>X-ray ptycho-graphy<br>Neutron Imaging<br>Neutron Depth Profiling | Images or mappings, statistical analysis (profiles, distributions, spectral information) | Many formats depending on technique | GB<br>TB |
| Spectroscopic data (lab.) | Local chemical environment, electronic structures, oxidation states, SEI compounds | XPS, IR, Raman<br>NMR<br>Auger<br>EDS/EDX,<br>EELS | Spectra | | MB |
| Spectroscopic data (large scale facilities) | Local chemical environment, electronic structures, oxidation states, SEI compounds | Synchrotron XPS<br>X-ray Raman Scattering<br>XAS (EXAFS / XANES) | Spectra | | MB |
| Inelastic Scattering data (large scale facilities) | Collective excitations, transport mechanisms | QENS, INS<br>RIXS | Spectra | | MB |
| Surface data (large scale facilities) | SEI composition and morphology | Neutron Reflectometry<br>X-rays Reflectivity | Spectra | | KB |
| Other lab data | SEI growth and evolution | EQCM, OEMS, ICP MS | Variations of parameter with time or SOC | | KB |

WP5 will collect data to the WPs listed in Table WP 1.11.

*Table WP 1.11. WP5 will collect data to the following WPs*

| WP | What | To be used for | Suggested type | format | size |
|---|---|---|---|---|---|
| WP2 | Structures, energy-related data, wave functions & electronic properties, AIMD trajectories, Atomic trajectories and associated transport, spectral etc. properties | Understand phenomena and mechanisms Cross-analysis of experimental data using the modelling predictions Designing new experiments to validate the modelling predictions | Parameters, data, spectra | .tar.gz | TB |
| WP3 | Predictions of structures and morphology (SEI, electrolytes, particles); Electronic structure and reactions of molecules; Atomic structures, vibrational properties; trajectories and representative configurations; topology/connectivity; Prediction of electrochemical observables and reaction/transport heterogeneities | Cross-analysis of experimental data using the modelling predictions Designing new experiments to validate the modelling predictions | Parameters, data, spectra | .tar.gz | TB |
| WP4 | Selection of electrolytes additives and salts; coated anode and cathode powders; basic characterizations | Design experiments for advanced characterization | Sample sheets, data sheets | g, docs | KB |
| WP6 | Selection of electrolytes; materials screening; performance evaluation; basic characterizations | Design experiments for advanced characterization | Sample sheets, data sheets | docs | KB |
| WP8 | Standardization for experimental protocols and logbooks, metadata and data formats | Perform experiments according to standards, generate metadata and data according to formats | Logbooks, metadata and data | Docs and many types depending on techniques | KB |
| WP9 | Standardization for data assimilation and storage | Analyse and store data according to standards | Archived data in FAIR, indexed, identifiable format | DOI | KB |
| WP10 | Data analysis for selected methods and data for | Data processing, Spectral deconvolution, | Parameters, fitted data | .json | KB |
| | experiment suggestions in active learning runs | peak fitting, experiment planning (suggestion) | & analysed spectra | | |
| WP11 | Data analysis; ML algorithms for segmentation of 3D volumes and statistical analysis of 3D microstructures | Extract quantitative parameters and their spatial distribution; design new experiments | Computer codes | Python code | MB |

WP5 will deliver data to the WPs listed in Table WP 1.12.

*Table WP 1.12. Data delivery from WP5 to other WPs*

| WP | What | Usable for | Suggested format | | |
|---|---|---|---|---|---|
| | | | type | format | size |
| WP2 | Parameters from structural and chemical characterizations at local scale; including e.g. spectra (vibrational, absorption, …) | Validate and refine models, design new experiments | Data analysis sheets | Parameters sheets | MB |
| WP3 | Raw data generated by the toolbox (in-lab/large scale facilities)<br>Parameters from multi-scale characterizations | Validate and refine models, design new experiments | Raw data, data analysis sheets | Many different formats depending on techniques<br><br>parameters sheets | KB-TB<br><br>KB |
| WP4 | In-depth characterization of selected additives and coatings; high-fidelity and high-throughput data | Evaluation of property-performance relationship<br>Suggest next optimal screening & formulations | Output data and parameters sheets | docs | KB |
| WP6 | In-depth characterization of selected electrolytes ; high-fidelity and high-throughput data | Evaluation of property-performance relationship<br>Suggest next optimal screening & formulations | Ouput data and parameters sheets | docs | KB |
| WP7 | Domain knowledge | Ontology design | Text, excel, images | .pdf, .doc, .xls, .ppt, .png | MB-GB |
| WP8 | Logbooks and metadata for all in-lab and LSF experiments; best practices | Evaluation/optimization of standards and protocols; | Sheets | docs | KB |
| | | optimization of formats | | | |
| WP9 | Curated experimental data or properties (chemical environment/structure/ morphology) evolution | Autonomous feed-back loops | Ouput data and parameters sheets | Docs and many types depending on techniques | KB-TB |
| WP10 | Curated experimental data or properties (chemical environment/structure/ morphology) evolution | Autonomous feed-back loops | Ouput data and parameters sheets | Docs and many types depending on techniques | KB-TB |
| WP11 | Curated experimental data or properties (chemical environment/structure/morphology) evolution; 3D images of microstructures | Generative models at microstructure level. Autonomous feed-back loops; machine-learning based image analysis | Ouput data and parameters sheets | Docs and many types depending on techniques | TB |

The detailed data flow between the tasks of WP5 and between WP5 and the other WPs are shown in Fig. WP 1.7 and Fig. WP 1.8.

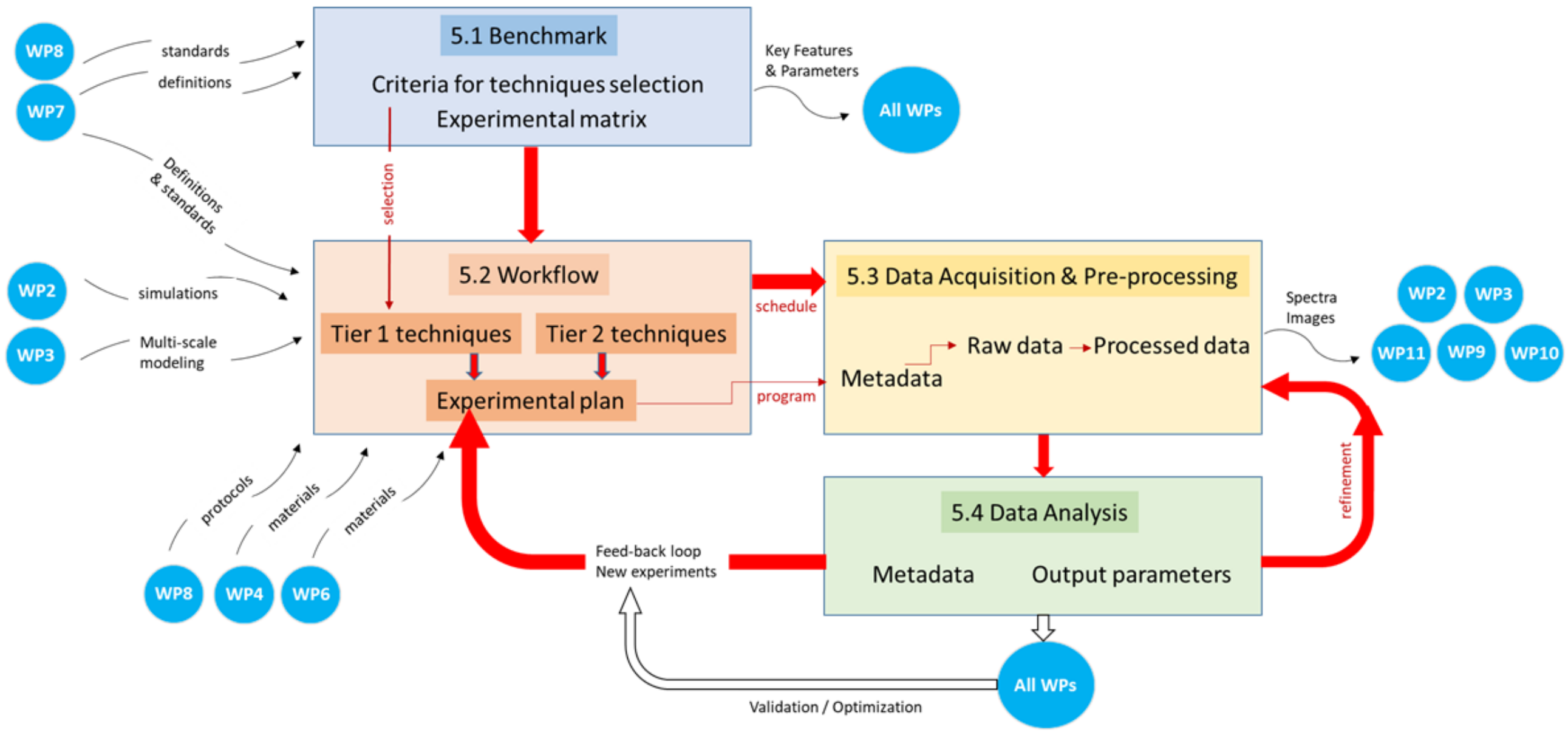

*Fig. WP 1.7. Data flow between the tasks of WP5 and relations for the other WPs.*

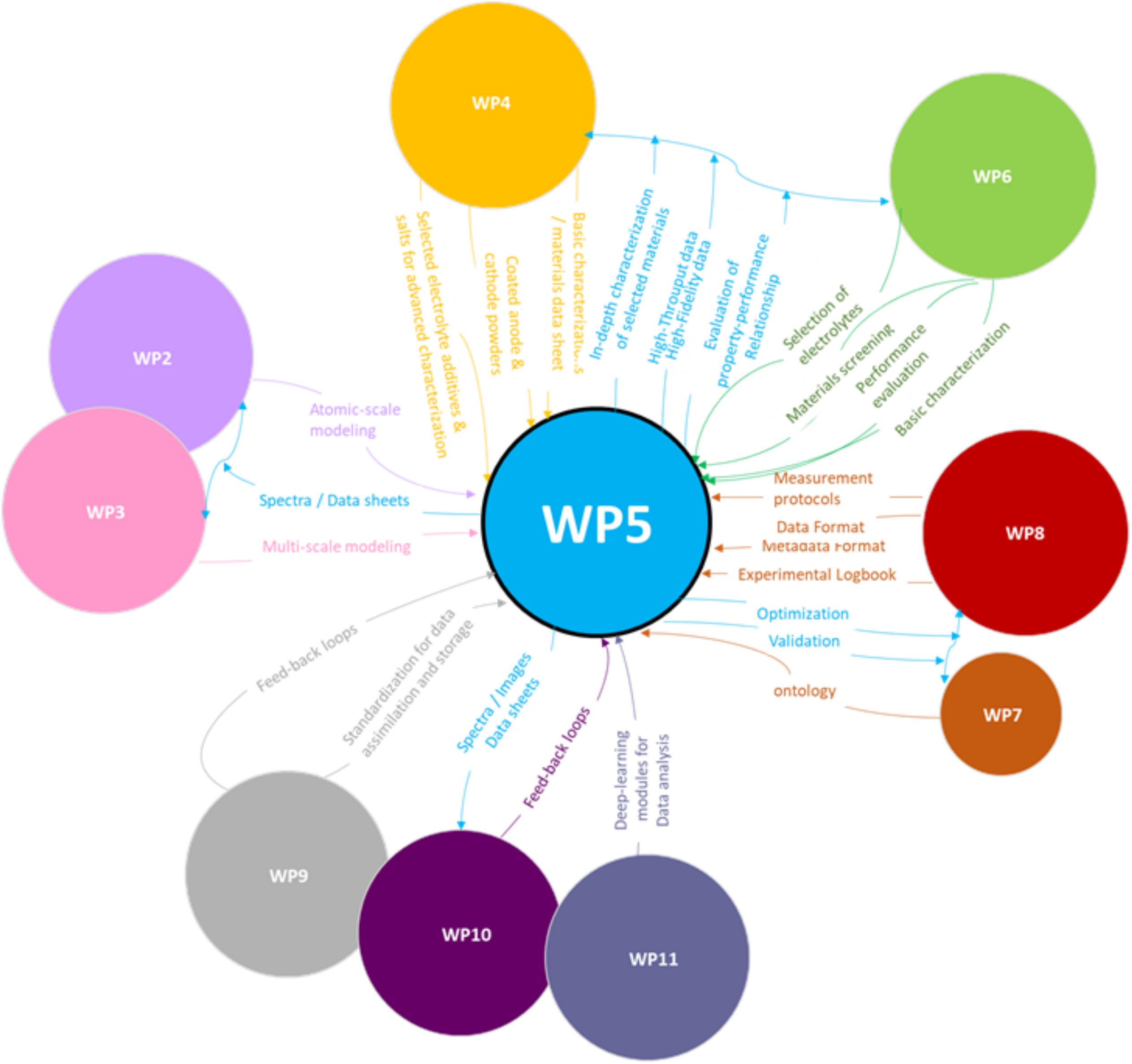

*Fig. WP 1.8. Data flow between WP5 and the WPs of BIG-MAP.*

### 1.5.4 Specify if existing data are being re-used (if any)

Relevant publicly available data collected from data repositories & literature may be reused.

### 1.5.5 Specify the origin of the data

Data will be generated in the following tasks:

Task 5.1: benchmark. Tables gathering information on techniques and observables (TUD, CEA, UU, ESRF, SOLEIL, ILL, CNRS, CTH, UOXF, UCAM, CSIC, NIC, UMI, SOLB, SAFT).

Task 5.2: workflow. Timeline, organization, site coordination for the experiments (CEA, UU, DTU, TUD, ESRF, SOLEIL, ILL, CNRS, CTH).

Task 5.3: data acquisition and visualization. Data taken in-lab and at Large Scale Facilities by the various partners. Images, videos, displays, photos of equipment. Design of cells and technical specifications **(**ESRF, CEA, UU, TUD, SOLEIL, ILL, CNRS, CTH, UOXF, UCAM, CSIC, UMI, SOLB, SAFT)**.** Tier 1 and Tier 2 techniques will be employed (as defined in Task 5.1 and 5.2).

Task 5.4: data analysis. Algorithms and codes developed. Intermediate and final output parameters (ILL, CEA, UU, DTU, TUD, ESRF, SOLEIL, CNRS, CTH, UOXF, UCAM, CSIC, UMI, SOLB, SAFT).

### 1.5.6 State the expected size of the data (if known)

Data size will depend on the extent and the nature of the data that are created during the project and made available. We may expect:

- Big data sets from Large Scale Facilities experiments (potentially, >10 TB per tomography experiment)
- Big data sets from operando measurements (potentially, large amount of time- and space-resolved data)
- Big data sets from microscopy experiments (>1 TB / FIB-SEM)
- Standard data sets from in-lab and ex situ characterizations, specific Large Scale Facility experiments

### 1.5.7 Outline the data utility: to whom will it be useful

The data created might be useful to the following groups:

- BIG-MAP consortium. In particular, modelling WPs, Standards & Protocols, Ontology, Infrastructure and Interoperability.
- European Commission and European Agencies
- Stakeholders involved in the field of Li-ion batteries and materials
- Scientific community
- EU national bodies
- General public

## 1.6 WP6 – HTS and Testing

### 1.6.1 Purpose of data collection/generation

The scope of WP6 is to establish a chemistry-neutral, high-throughput formulation-characterization-performance evaluation chain across time and length scales with a focus set on advanced liquid electrolyte formulations and compatible electrode materials with proven capabilities for lithium-based batteries. Essential characterization on electrolyte, electrode and cell level, as well as characterization of electrolyte/electrode interfaces and interphases will be performed along the entire materials lifecycle, including preselected physicochemical, electrochemical, structural and analytical characteristics /performance evaluation (results will be complemented with the ones obtained in WP4, WP5 and WP10). The high-throughput screening approach will serve through a tiered screening pipeline, accelerating the identification of hit/lead electrolyte candidates for selected cell chemistry application(s) hand in hand with generation of abundant pertinent interphase/interface/device data across the entire lifetime of the battery enabling generation of interfaces. Single instrument active learning feedback loops can be built in close collaboration with WP4 and WP10. The abundance of generated in-depth data will be used to identify vital input parameters relevant for the development and validation of computational models. Simultaneously, modelling results will be validated against experimentally obtained results (strong correlation and synergy with WP2 and WP3).The interoperability of the experimental data (raw, processed and analyzed) will be ensured by applying the ontology developed and defined within WP7 and the standards for data and metadata defined in WP8. The generated output data will be specified in the form defined by WP9 and labelled in terms of fidelity level. Obtained experimental data will be transferred to WP11 through WP10 being in charge of handling, storage and AI-training.

### 1.6.2 Explain the relation to the objectives of the project

A straightforward correlation is given to BIG-MAP's Specific Objective 1: "Develop the scientific and technological building blocks and modes for accelerated battery discovery" relating more closely to the sub-objective: "Optimized discovery process for advanced electrolyte and electrode materials using HTS guided by advanced hybrid AI-models" with the following key performance indicator: "Reduction in the time/resources needed to identify and demonstrate new materials and/or additives through HTS of electrolyte formulations, and documenting improved interface performance (factor of 2 or more)". In order to establish HTS SEI Demonstrator using integrated high throughput electrochemistry and ex situ high throughput spectroscopy to optimized electrolyte formulations and materials as a key demonstrator within the WP6, following aspects will be considered:

- Effective integration of individual and synergistic experimental tasks towards accelerated identification of targeted battery materials/chemistry
- Development and validation of a methodology for an effective, accelerated experimental workflows
- Demonstration of applicability/transferability of experimental data relevant for modelling and AI-training

### 1.6.3 Specify the types and formats of data generated/collected

Data generated/collected in the frame of WP6 will be mostly text-based and can be supplied in different text-based formats depending on individual requirements (.json, .csv, .xml). The experimental data comprise 4 main categories: raw data (1D, 2D, 3D data sets), processed data (corrected, normalized, averaged), analysed data (e.g. fitted) and output data (e.g. parameters from fits).

Types and formats of the data that will be generated by WP6 are listed in Table WP 1.13.

*Table WP 1.13. Types and formats of the data that will be generated by WP6*

| Datatype | Description | Data sets | Type | Format | Size |
|---|---|---|---|---|---|
| Methodology data | | | | | |
| Experimental logbooks | Conditions of realization of an experiment including sample description, technical set-up, type of measurement and output parameters, localization of raw data | Filling the on-line tool | Sheet | .doc | MB |
| Experimental matrix | Classification of characterization techniques, criteria & indicators | Inputs from all partners and compilation of information using matrix template | Table | .doc | KB |
| Competence matrix | Summary of technical and experimental capabilities; mappings of competence organized according to criteria (type of information, cells availability, etc.) | Inputs from all partners and compilation of information using mappings | Table/ graphical mapping | .doc | KB-MB |
| Level organization | Description of techniques grouped in levels: electrolyte, electrode and cell | Competence matrix | Table/ graphical mapping | .doc | KB-MB |
| Selection of Tier1 and Tier 2 techniques | Identification of key techniques based on competence and experimental matrixes | Compilation of available capabilities and selection according to other WPs targets | Table | .doc | KB |
| Experimental workflow | Tier 1 and Tier 2 workflows describing the implementation of the experiments (timeline, sites, interconnections, cells) | Analysis of the level and tier techniques | Table/ graphical mapping | .doc | KB-MB |
| WP6 workflow interfacing with other WPs | Connections of WP6 with other WPs to develop the experimental program | Feed-back loops | Table/ graphical mapping | .doc | KB-MB |
| Data acquisition methods | New modes of acquiring high throughput data | Feed-back loops with modelling and AI<br>Depends on the techniques, to be adapted for each level | Table/graphical mapping | .doc | KB-MB |
| Data analysis methods | Codes and algorithms for data analysis; interfaces for data representation | All partners' internal tools, mutualisation of best practices<br>Depends on the techniques, to be adapted for each level | Many types depending on data | .py<br>or<br>others | MB |
| Data sheets elaboration | Transforming analysed data into data transferable to other WPs | Depends on the techniques, to be adapted for each level, guidelines to users to be able to extract, plot, represent the output parameters | Sheets | .doc | KB-MB |
| Cell harmonization | List of available cells classified by level, benchmark, development of new compatible cells | Identification of limits and needs for measurements (technique dependent); selection of key cell designs | Summary of cell designs | .doc | KB |
| Characterization data for electrolytes, electrodes and cells | | | | | |
| Electrolyte level | | | | | |
| Electrochemical data | Conductivity mechanisms and data on the internal resistance of electrolytes | Electrochemical impedance spectroscopy (EIS) | Spectra/table | .csv,<br>ASCII<br>.json<br>.csv<br>xls; opj | MB-GB |
| | Redox properties of electrolytes and electrodes | Linear sweep voltammetry (LSV); cyclic voltammetry (CV) | Graph/table | .csv,<br>ASCII | |
| Physicochemical data | Residual water content | Karl-Fischer titration | Table | .csv,<br>ASCII | KB - MB |
| | Viscosity | Rheology/EIS | Table | .csv,<br>ASCII | KB - MB |
| | Transference number | EIS | Table | .csv, ASCII | KB - MB |
| Electrode level | | | | | |
| Electrochemical data | Conductivity mechanisms and data on the resistance of electrodes | EIS | Spectra/Table | .json .z; ASCII | MB-GB |
| | Transport properties of $Li^+$ in active materials | GITT | Graph/table | .xlsx; ASCII | KB - MB |
| | Ionic and electronic conductivity of artificial SEI films ($\sigma_{e-}$, $\sigma_{Li+}$, $\varepsilon_r$), possibly as a function of T ($E_a$) | EIS; constant polarization on metal insulator metal (MIM) stacks | Spectra/table | .mpt .txt | MB-GB |
| Spectroscopic data | Local chemical environment | FT IR/Raman | Spectra/table | .json | MB |
| Structural data | Elemental composition | EDAX | Table | ASCII | KB - MB |
| | Electrode morphology | SEM | Images | .tiff | MB |
| | Crystallinity and composition of electrode materials | XRD | Diffractogram | .raw, .uxd, ASCII | MB |
| | Crystal structure of the cathode and the anode before and after cycling | (Grazing Angle) X-Ray Diffraction | XRD patterns, tables | .xy | MB |
| Chemical data | (More or less resolved) mapping of the elemental composition | EDS/EDX; XPS; ToF-SIMS (depth profiling); LIBS; RBS | Spectra/table | .xy(z) .txt | MB |
| | Local chemical environment at the surface or slightly below the surface before and after cycling | XPS (+'depth profiling') | Spectra/table | .xy .txt | MB |
| Geometric data | Actual thickness measurement on selected locations before and possibly after cycling | FIB/STEM | Images | .tiff | MB |
| | Thickness distribution of artificial SEI prepared by combinatorial deposition on the 4'' wafer substrate | Stylus profilometry (direct measurement or extrapolated using calibration) | Tables | .xyz | MB |

| Cell level | | | | | |
|---|---|---|---|---|---|
| Electrochemical data | Overall cell resistance and individual electrode resistances | EIS | Spectra | .z; ASCII | MB-GB |
| | Redox properties of electrolytes and electrodes | LSV, CV | Graph/table | .csv; ASCII | MB-GB |
| | Cycle life (formation, fading)<br>Rate capability<br>Charge/Discharge profiles for each electrode (EL CELL)<br>Aging studies under different experimental conditions | Galvanostatic cycling | Graph/table | .csv; ASCII .json .db, .xlsx; .xls, .opj .txt | MB-GB |
| | Ionic and electronic conductivity of single component artificial CEI films | EIS | Spectra/table | .csv | MB-GB |
| | Electrochemical behaviour of coated electrodes in half-cell setup during operation | Galvanostatic cycling; CV | Graph/table | .mpt | MB-GB |

WP6 will collect data to the WPs listed in Table WP6.2.

*Table WP 1.14. WP6 will collect data to the following WPs*

| WP | What | To be used for | Suggested | | |
|---|---|---|---|---|---|
| | | | type | format | size |
| WP2 | Structures, Energy-related data, wave functions and electronic properties, AIMD trajectories Atomic trajectories and associated transport, spectral etc. properties, simulated experiments | Cross-analysis of experimental data using the modelling predictions<br><br>Design new experiments to validate the modelling predictions | Parameters data, spectra | .docs, .csv | KB MB |
| WP3 | Predictions of structures and morphology (SEI, CEI electrolyte); electronic structure and reactions of molecules; atomic structures, vibrational properties trajectories and representative | Cross-analysis of experimental data using the modelling predictions Design new experiments to validate the modelling predictions | Parameters data, spectra | .docs, .csv | KB MB |
| | configurations, connectivity prediction of electrochemical observables and reaction/transport heterogeneities | | | | |
| WP4 | Initial preselection of electrolyte components, coated anode and cathode powders; basic characterizations | Design experiments for advanced characterization | Sample sheets, data sheets | Docs | KB |
| | | | Samples for characterization/tests | g-kg | |
| WP5 | In-depth characterization of interphases (SEI, CEI) High-fidelity and high-throughput data | Evaluation of property-performance relationship Suggest next optimal screening & formulations | Output data, parameters sheets | Docs | KB |
| WP8 | Standardization for experimental protocols and logbooks, metadata and data formats | Perform experiments according to standards, generate metadata and data according to formats | Logbooks, metadata data | Docs, Technique dependent types | KB |
| WP9 | Standardization for data assimilation and storage (specified form definition, labelling in terms of fidelity level) | Analyse and store data according to standards | Data sheets | Docs | KB |
| WP10 | Complementary high-throughput electrochemical and spectroscopic data Automated data analysis framework Experiment suggestion | Interphase optimization through characterization and ML-orchestrated analysis Evaluation of structure-property-performance relationship FOM extraction Experiment planning | Output data Parameters sheets | Docs .json | KB-MB |
| WP11 | Prediction of electrochemical observables | Cross-analysis of experimental data using the modelling predictions Design new experiments to validate the modelling predictions | Predicted property and structure/composition | Docs Data files | KB-MB |

WP6 will deliver data to the WPs listed in Table WP 1.15.

*Table WP 1.15. Data delivery from WP6 to other WPs*

| WP | What | Usable for | Suggested | | |
|---|---|---|---|---|---|
| | | | type | format | size |
| WP2 | Parameters from electrochemical and spectroscopic characterizations | Validate and refine models, design new experiments | Data analysis sheets | Parameter sheets | KB |
| WP3 | Parameters/raw data from electrochemical and spectroscopic characterizations | Validate and refine models, design new experiments | Data analysis sheets | Parameter sheets | KB |
| WP4 | Selection of hit/lead electrolyte candidates<br>High-throughput electrochemical and spectroscopic data | Evaluation of structure-property-performance relationship.<br>Suggest next optimal screening & formulations | Output data, parameters sheets | Docs | KB |
| WP5 | Selection of hit/lead electrolyte candidates<br>High-throughput electrochemical and spectroscopic data | In-depth characterization of selected electrolytes<br>High-fidelity and high-throughput characterization | Output data, parameters sheets | Docs | KB - MB |
| WP7 | Domain knowledge | Ontology design | Text, excel, images | .pdf, .doc, .xls, .ppt, .png | MB -GB |
| WP8 | Logbooks and metadata for all experiments, best practices | Evaluation/optimization of standards and protocols; optimization of formats | Sheets | Docs | KB |
| WP9 | High throughput experimental data (electrochemical and spectroscopic properties and performance) evolution | Development of software infrastructure.<br>Autonomous feed-back loops through WP2 | Output data, parameters sheets | Docs | KB - MB |
| WP10 | High throughput experimental data (electrochemical and spectroscopic properties and performance) evolution<br>Communication protocol<br>Data to run autonomous feedback loop | Autonomous feed-back loops.<br>Multi-scale modelling by coupling AI with multi-fidelity data. | Output data, parameters sheets | Docs | KB - MB |
| WP11 | High throughput experimental data (electrochemical and spectroscopic properties and performance)<br>Data to run autonomous feedback loop evolution | Autonomous feed-back loops.<br>Multi-scale modelling by coupling AI with multi-fidelity data. | Output data, parameters sheets | Docs, .csv, .json | KB - MB |

The access to data as well as transfer to other WPs is dependent on the approach (conventional or high throughput) and techniques used. Individual data sets from these experiments will be in the order of up to 1 GB, depending on the method related approach: classic or high throughput screening. The flow of the data in WP6 is shown in Fig. WP 1.9 and Fig. WP 1.10.

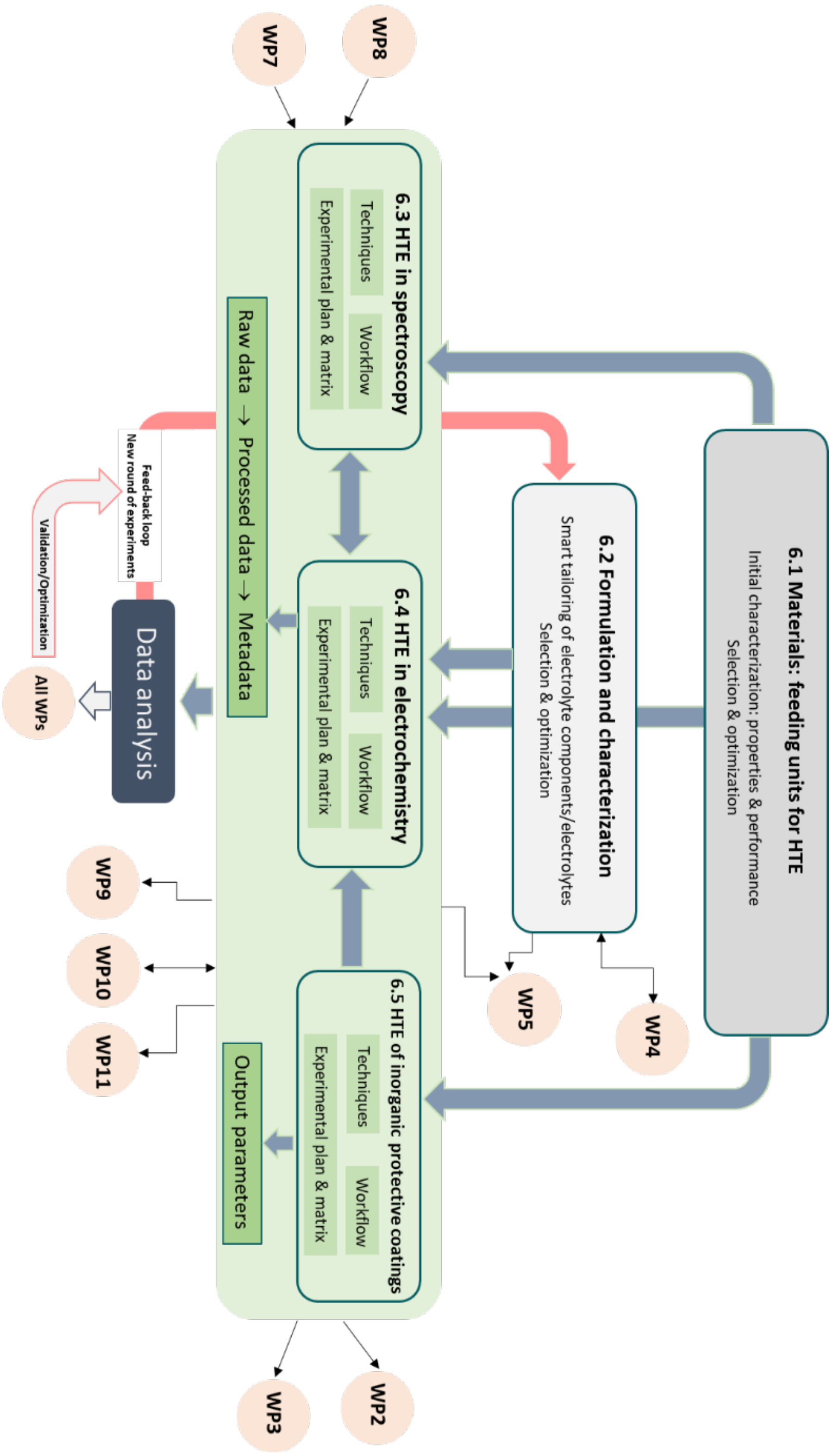

*Fig. WP 1.9. Data flow within WP6.*

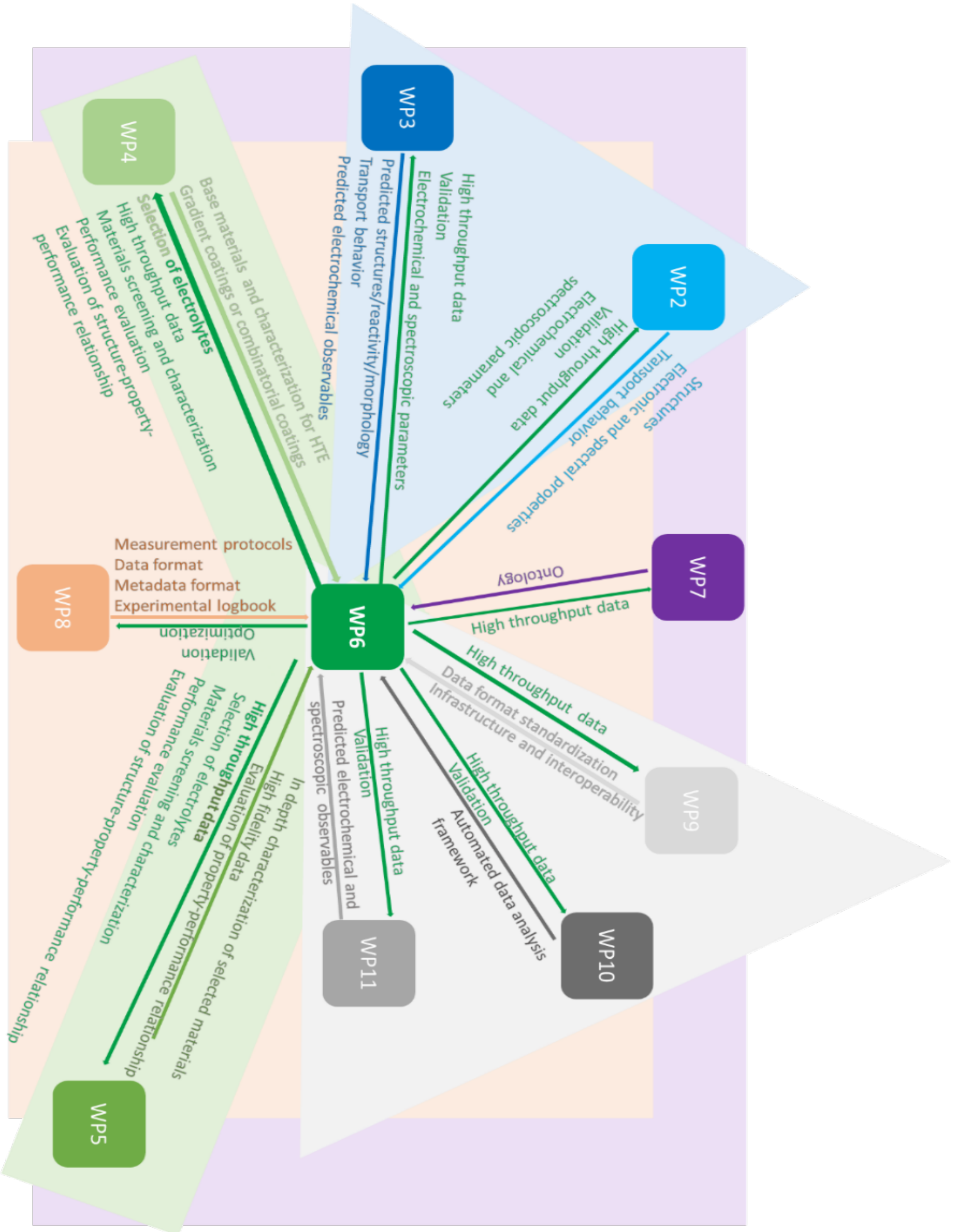

*Fig. WP 1.10. Data and information flow in WP6.*

### 1.6.4 Specify if existing data are being re-used (if any)

Relevant preliminary and publicly available data collected from data repositories & literature may be reused.

### 1.6.5 Specify the origin of the data

Data will be generated in the following tasks:

Task 6.1: Feeding units for High-Throughput Experimentation (HTE) (BASF, SOLB, UMI, CID, NVOLT, SAFT, KIT, FZJ, CEA, Fraunhofer): materials synthesis oriented, elemental analysis, EIS, GITT, SEM, XRD, EDAX, FTIR/Raman, initial galvanostatic cycling experiments

Task 6.2: Formulation and characterization (WUT, CID, KIT, FZJ, CEA, Fraunhofer): synthesis-oriented (electrolyte components), Karl-Fisher titration, LSV, CV, EIS, initial galvanostatic cycling experiments

Task 6.3: HTE in spectroscopy (KIT, WUT, FZJ, BASF, CID): SDC, XPS, SEM, FTIR/Raman, AES, ATR/DRIFTS

Task 6.4: HTE in electrochemistry (FZJ, WUT, SOLB, BASF, CID, KIT, CEA, Fraunhofer): Karl-Fisher titration, rheology related, EIS, LSV, CV, galvanostatic cycling experiments (different cell setups and test procedures)

Task 6.5: HTE of inorganic protective coatings (CEA, WUT, FZJ, Fraunhofer, CID, KIT): EIS, EDS/EDX, XPS, ToF-SIMS, LIBS, RBS, SEM, FIB/STEM, (grazing angle) XRD, stylus profilometry

### 1.6.6 State the expected size of the data (if known)

The size of the data is strongly dependent on the extent and the nature of the data generated during the project and made available:

- Large data sets from HTS galvanostatic cycling experiments ( >100 MB per cell chemistry; >1 TB over the ramp-up phase)
- Large data sets from scanning droplet cell and inorganic protective coatings related experiments (>100 MB per cell chemistry; >1 TB over the ramp-up phase)
- Standard data sets from conductivity, impedance and potentiostatic cycling experiments

### 1.6.7 Outline the data utility: to whom will it be useful

The data created might be useful to the following groups:

- BIG-MAP Consortium: especially WP2 - WP5, WP8, WP10, WP11
- European Commission and European Agencies
- Stakeholders involved in the field of Li-ion batteries and materials
- Scientific community
- EU national bodies

## 1.7 WP7 – Battery Interface Ontology

### 1.7.1 Purpose of data collection/generation

The purpose of WP7 is to develop a common representational system for describing battery interfaces enabling interoperability within and between the other work packages. The ontology is intended to be consumed by both humans and machines. Hence, in addition to formal relations between concepts, it is essential that the concepts are annotated with human descriptions in natural language (English). An ontology is a data model that represents knowledge as a set of concepts within a domain and the relationships between the concepts. By creating a standardized representation of a system, including its constituent concepts with properties and relations, an ontology provides a means not only for classifying data but also for inferring associations. Put simply, the ontology defines some basic concepts and relationships between them from which it is possible to gain new knowledge. One practical benefit of this is that linking an ontology with rich machine-processable semantic descriptions to a database enables AI-based tools to accelerate the discovery of new appropriate materials.

### 1.7.2 Explain the relation to the objectives of the project

The ontology will be used to support the interoperability of data across multiple scales, techniques, and domains in the battery discovery process. Applying the ontology will help BIG-MAP create a database utilizing a descriptive, formalized scheme to improve the reproducibility of experiments and simulations. From D7.1: BIG-MAP links together the main European actors contributing to the development of new battery materials. This include both experimental and theoretical actors working at many different scales, from electronic via atomistic and discrete mesoscopic to continuum scale. To support the efficient reuse and interoperability of research data between all these actors, it is essential to have a common way to describe the real world entities constituting batteries and their interfaces, their associated physical properties, and how these properties are measured and/or modelled. This common representational framework should be easy to read and understand by humans. At the same time, it must be highly formalized and accurate to be consumed by machines supporting flexible automated workflows that connect different domains and provide rich content to descriptors used for artificial intelligence. Another application of such a common representational framework is to enable semantic search across multiple databases. The battery interface ontology will serve exactly this purpose of being a highly formalized common language for describing battery interfaces. BattINFO will rely on other WPs to provide domain knowledge that will be incorporated into the ontology (e.g. by filling out questionnaires, tables, process diagrams, etc.). BattINFO can then be utilized by users and AI platforms in BIG-MAP to identify data links and support interoperability.

<u>First Use Case Definitions</u>
To facilitate the early development of the ontology, a proof-of-concept use case will be defined. Having a first concrete use case will help to ensure that the ontology is robust and stable, as well as allow the development team to test the links to other WPs. The proof-of-concept use case is defined as a result of discussions with representatives from WPs 2, 3, 8, 9, and 11. The first step of establishing the strategy for ontology development will be the discussion of the use case with other WPs, especially with WP2, which is seen as starting point. Also, the WP9 will define use cases, which must be ontologized.

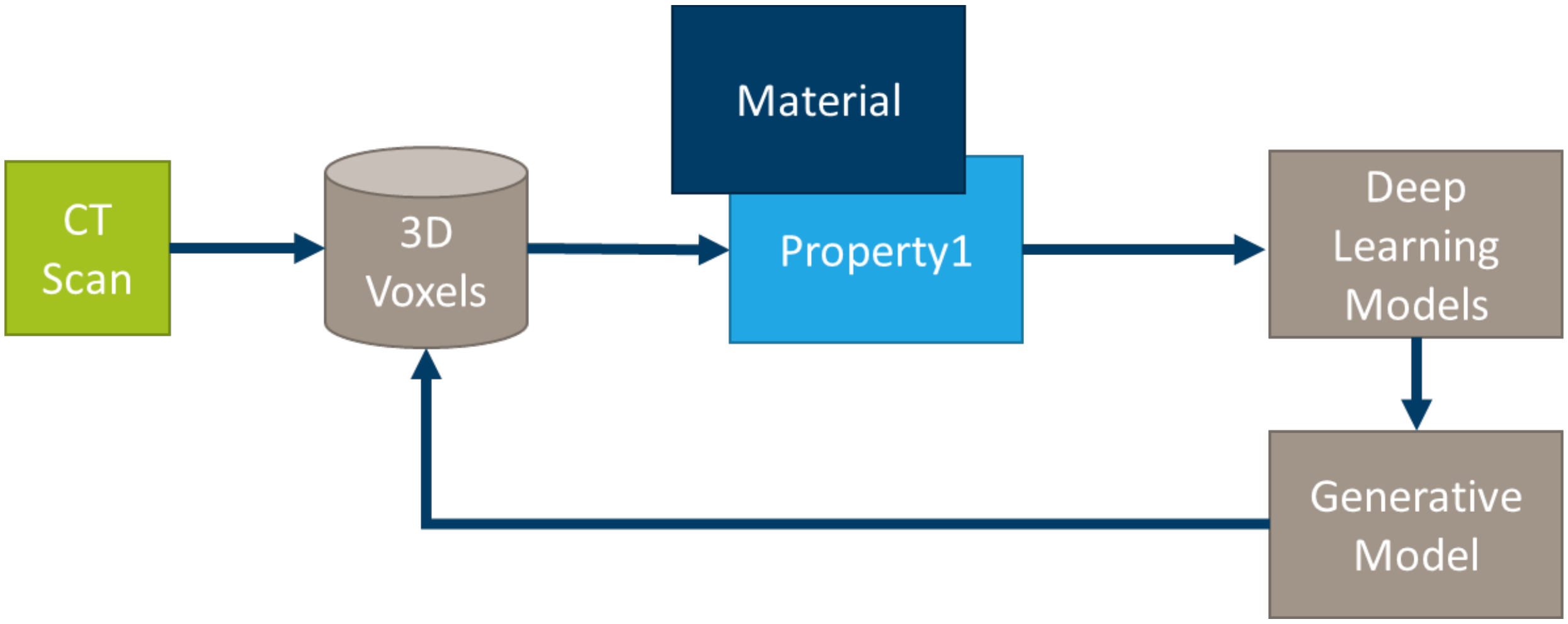

*Fig. WP 1.11. Schematic of the proposed proof-of-concept use case for the battery ontology. The ontology will facilitate access for deep learning models in WP11 to access 3D voxel data from CT scans of battery materials. The deep learning model can then be used to inform a generative model to generate virtual 3D voxel data of the same material.*

Fig. WP 1.11 shows a schematic of the proposed proof-of-concept use case. Three-dimensional voxel data of battery material structures – whether it is generated experimentally from CT or FIB-SEM characterization or virtually through stochastic microstructure models - is (i) essential to both battery model development and experimentally optimized material design, (ii) available as already existing datasets, and (iii) very large in size and somewhat difficult to manage. As such, it presents a good opportunity as a first use case for the ontology.

WP11 currently has plans to implement two classes of models: deep learning models and generative models. The deep learning models aim to fit data from different scales to observe properties using explainable AI models. The development of the deep learning models requires for instance data from battery material microstructures in a standardized format. Similarly, the generative models aim to generate data at multiple scales. The generative model must first learn from data that is physically correct (e.g. real microstructures).

<u>Other Use Cases</u>
Other early use cases that could help develop the BattINFO and contribute to work in BIG-MAP include using the ontology to structure the development of an online electronic lab notebook. BattINFO will also support model-based material property calculations in WP2 and WP3, such as calculating potentials from DFT. Electrochemical data also offers a promising first use case for the ontology. For example, linking CV data with ab-initio molecular dynamics simulations to feed neural networks with the goal of training potentials. Further development of the concepts for early use cases will be refined with the leaders of the relevant WPs, including WP2, 3, 8, 9, 10, and 11.

### 1.7.3 **Specify the types and formats of data generated/collected**

The ontology will comprise files in the Web Ontology Language (OWL) written in rdf/xml format.

Types and formats of the data that will be generated by WP7 are listed in Table WP 1.16.

*Table WP 1.16. Types and formats of the data that will be generated by WP7*

| Datatype | Description | Data sets | Type | Format | Size |
|---|---|---|---|---|---|
| Battery object information | Semantic descriptions of the objects to be included in the battery taxonomy and ontology | Textual data listing the Model, Description, Parts, Properties, and Observation Procedure (e.g. the template shown in Table WP 1.19) | Excel tables | .xlsx | KB to MB |
| Battery process information | Process diagrams mapping the causality of the main processes to be considered in the ontology | Process flow diagrams (e.g. the template shown in Fig. WP 1.14) | PowerPoint Slides | .pptx | MB |
| Battery ontology | An ontology describing a generic battery | Ontology | Protégé files | .owl, .ttl | MB |
| Battery interface ontology | An ontology describing battery interfaces | Ontology | Protégé files | .owl, .ttl | MB |
| Use case Studies | Use cases showing how the ontology can be applied to support specific activities in BIG-MAP | Report | Document | .pdf, .ttl | MB |

WP7 will collect data to the following WPs (Table WP 1.17).

*Table WP 1.17. WP7 will collect data to the following WPs*

| WP | What | To be used for | Suggested | | |
|---|---|---|---|---|---|
| | | | type | format | size |
| WP2-6 | Domain knowledge on battery active materials, coating materials, characterization of interfaces, characterization of bulk and surface reactions in electrochemical systems | Ontology design | Text, excel, images | .pdf, .doc, .xls, .ppt, .png | MB-GB |
| WP8 | Online lab notebook | Ontology design | Text, excel, images | .pdf, .doc, .xls, .ppt, .png | MB-GB |
| WP8 | Measurement protocols | Ontology design | Text, excel, images | .pdf, .doc, .xls, .ppt, .png | MB-GB |
| WP8 | Metadata standards | Ontology design | Text, excel, images | .pdf, .doc, .xls, .ppt, .png | MB-GB |
| WP9 | Materials cloud archive | Submission of demonstrators: ontology-compliant data | JSON, OWL, XML | .json, .xml, .owl, .ttl | MB-GB |
| WP9 | BIG-MAP AppStore | Open-access upload of ontologies, standards and protocols | Text, JSON, JSON-LD, XML | .txt, .json, .xml | KB |

WP7 will deliver data to the WPs listed in Table WP7.3.

*Table WP 1.18. WP7 will deliver data to the following WPs*

| WP | What | Usable for | Suggested | | |
|---|---|---|---|---|---|
| | | | type | format | size |
| WP8 | Guidelines and metadata regarding the ontology for data | The received ontology will be used to sort the list of metadata necessary to ensure the completion of the metadata defined in WP8 | text | .txt | KB, MB |
| WP8 | Battery and battery interface ontology | Establishing online lab book, other standards and protocols | Protégé file | .owl, .ttl | KB-MB |
| WP9 | Battery and battery interface ontology | Dissemination through the BIG-MAP AppStore | Protégé file | .owl, .ttl | KB-MB |
| WP10 & 11 | Battery and battery interface ontology | Supporting a machine learning model | Protégé file | .owl, .ttl | KB-MB |

All WPs will receive the ontology through WP8 and WP9. The detailed data flow between the tasks of WP7 and between WP7 and the other WPs are shown in Fig. WP 1.12 and Fig. WP 1.13.

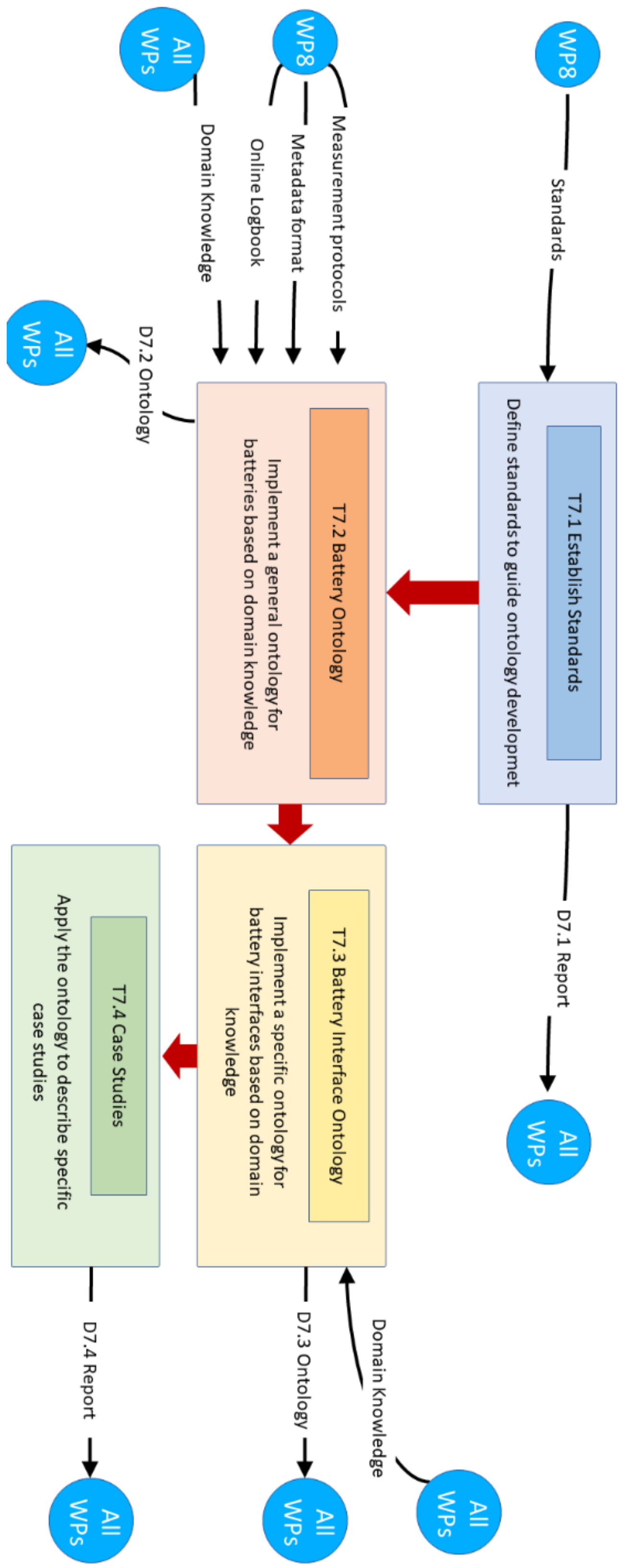

*Fig. WP 1.12. Data flow between the tasks of WP7 and relations for the other WPs.*

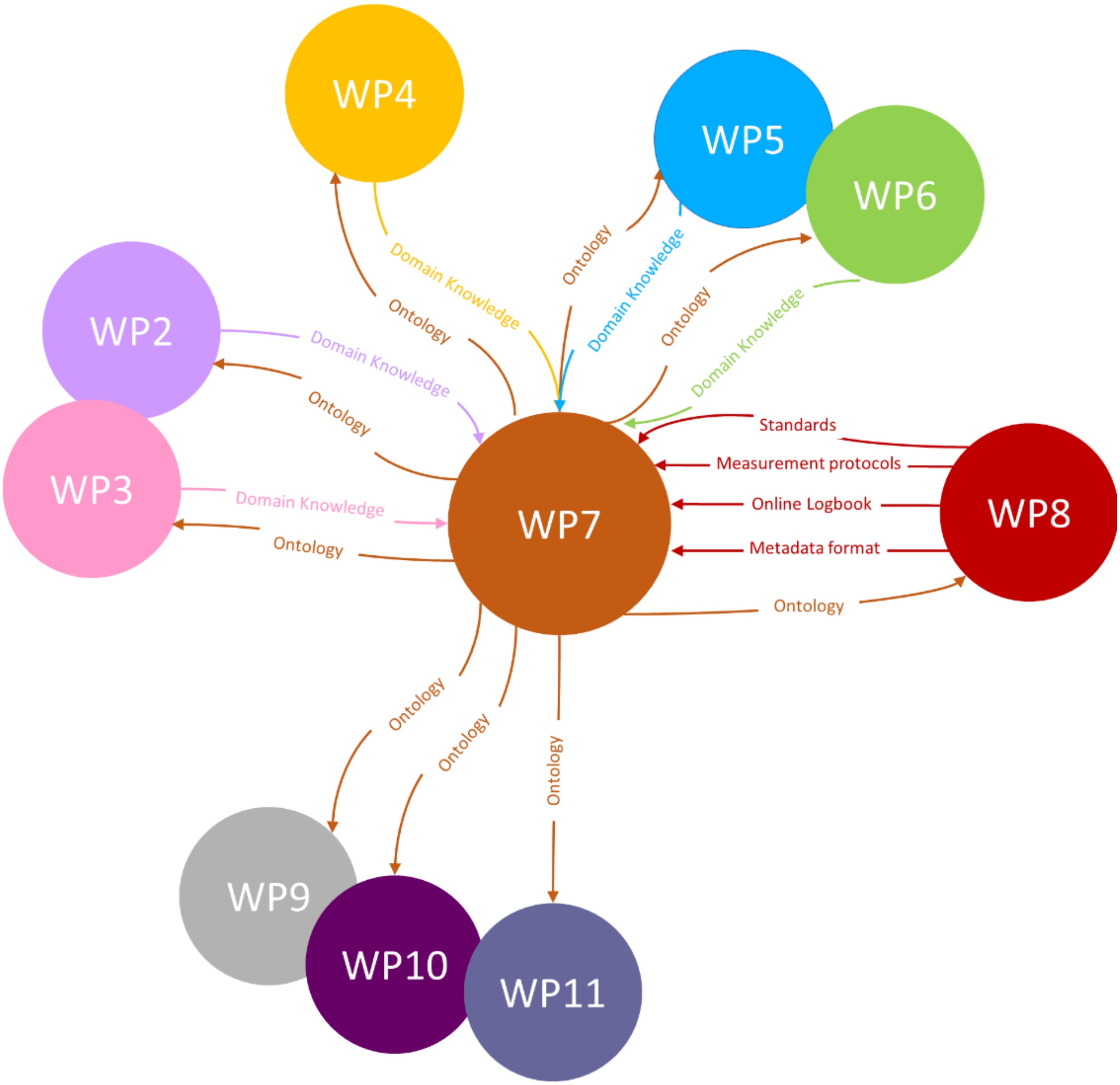

*Fig. WP 1.13. Data flow between WP7 and the WPs of BIG-MAP.*

Table WP 1.19 shows a template example for the battery object development table (Source: E. Ghedini, OntoTrans (GA 862136)).

*Table WP 1.19. Template example for the battery object development table*

| Model (of material object) | Description | Part | Property | | | Observation Procedure | |
|---|---|---|---|---|---|---|---|
| | | | Name | Units | Symbolic Represen-tation | Description of the Semiotic Process | Interpreter |
| Object 1 | (natural language description) | list of the sub-parts comprising the object | Property 1 | (SI unit if it is a quantity, or a reference to the scale used) | (e.g. vector, scalar, RGB colour scale, names list) | (how the observation occurs) (e.g. ISO procedure) | (who is assigning the property value to the object) (e.g. customer, measurement device) |
| | | | Property 2 | | | | |
| | | | Property 3 | | | | |

Fig. WP 1.14 shows the template for building a process flow diagram in the ontology. Such flow diagrams identify which material objects are participants in the process and describe the flow in terms of causality.

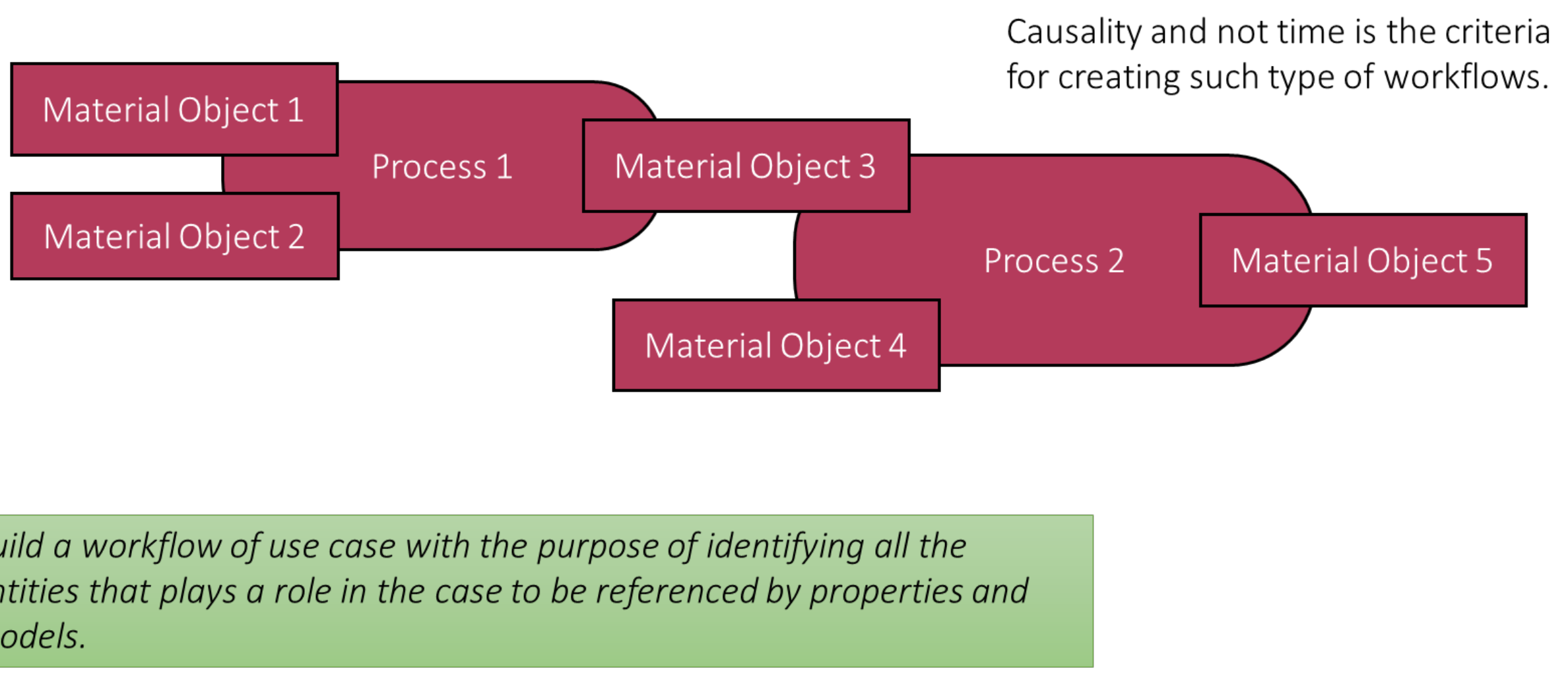

*Fig. WP 1.14. Example of process flow diagram (Source: E. Ghedini, OntoTrans (GA 862136)).*

### 1.7.4 Specify if existing data are being re-used (if any)

The Battery Interface Ontology reuses existing ontologies. It is a domain ontology based on the European Materials & Modelling Ontology (EMMO), which is a multi-disciplinary effort maintained and coordinated by the European Materials Modelling Council. EMMO itself uses a few general standards like Dublin Core (for annotating the ontology itself) and SKOS. In addition, BattINFO will, when relevant, import and reuse other EMMO domain ontologies, like the crystallography and atomistic domain ontologies.

### 1.7.5 Specify the origin of the data

BattINFO is declared within BIG-MAP based on domain knowledge (batteries, experiments, modelling, IA, ontologies, etc.) brought in by partners.

Task 7.1: Establish ontology standards to support interoperability and dissemination (SINTEF, WUT, DTU, EPFL): discussions with domain experts, excel tables, flow diagrams, text documents.

Task 7.2: Design a general battery ontology (SINTEF, WUT, DTU, EPFL, CSIC, NIC): discussions with domain experts, excel tables, flow diagrams, text documents, implementation in OWL via Protégé.

Task 7.3: Design the battery interface ontology (SINTEF, WUT, DTU, EPFL, CSIC, NIC): discussions with domain experts, excel tables, flow diagrams, text documents, implementation in OWL via Protégé.

Task 7.4: Implement the ontology to describe specific case studies (SINTEF, WUT, DTU, EPFL, CSIC, NIC): discussions with domain experts, excel tables, flow diagrams, text documents, implementation in OWL via Protégé.

### 1.7.6 State the expected size of the data (if known)

About one or a few MB in total.

### 1.7.7 Outline the data utility: to whom will it be useful

Aims at providing a common foundation for semantic interoperability for battery discovery, including experiments and autonomous synthesis robotics, physical modelling and AI within Battery 2030 and related projects and initiatives. Hence, large long-term value.

## 1.8 WP8 – Standardization and Protocols

### 1.8.1 Purpose of data collection/generation

The goal of WP8 is to define standards that should be met across the whole BIG-MAP project regarding data acquisition and reporting as well as to set protocols for the data acquisition and battery performances evaluation. The generated protocols and standards should then be implemented by the WPs dealing with experimental data (WP4, WP6, WP10, WP11) as well as simulations/modelling data (WP2, WP3, WP11) in order to ensure reproducibility, transferability as well as traceability within the project and across the different partners. These standards will be established in the form of metadata (text files) as well as graphical user interface (web-based software).

The interoperability of the standards and protocols will be ensured by applying the ontology developed and defined in WP7. The generated output data will be specified in the form defined by WP9.

### 1.8.2 Explain the relation to the objectives of the project

The work in WP8 relates to BIG-MAP's Specific Objective 2: "deliver cross-cutting initiatives to ensure implementation and use of project data and results across the battery discovery value chain" and especially the sub-objective: "Creation of a common European web-platform for materials and battery testing with community-wide accepted standards and protocols".

- The key demonstrator of WP8 is the development of an open European platform with BIG-MAP standards and testing protocols for battery materials
- Provide protocols that will pave the way for the implementation of a European platform for battery testing
- Provide lists of metadata for collecting and sharing experimental data and modelling/simulations data.

### 1.8.3 Specify the types and formats of data generated/collected

The types and formats generated and collected in WP8 will be mostly centered around lists of metadata (text files) that are necessary to collect for WPs generating data (WP2, WP3, WP4, WP5, WP6, WP10, WP11) so that WPs dedicated to building the Battery Interface Genome (BIG) can communicate comprehensible results to the WPs dedicated to develop the accelerated materials discovery platform (WP1, WP7, WP9 and WP11). The lists of metadata, which will be guidelines for the different WPs, will cover the minimum set of information that must be collected when acquiring data of any type in order to be shared across the project.

In addition to these lists of metadata, software and graphical user interfaces (GUI) will be developed within WP8 to help rationalizing the storage, the exchange and the analysis for electrochemical data generated in WP4, WP6 and WP10, as well as internally in WP8, and battery materials characterization generated in WP5. Additionally, standardized electrochemical results (.txt) and testing protocols will be generated in WP8 to be shared across the project and to serve as standards for any improvements regarding electrochemical performances obtained by the development realized in the different experimental WPs. To do so, WP8 will collect guidelines and metadata, i.e. an ontology, from WP7 as well as use the infrastructure and the guidelines regarding interoperability developed in WP9.

Types and formats of the data that will be generated by WP8 are listed in Table WP8.1.

*Table WP 1.20. Types and formats of the data that will be generated by WP8*

| Datatype | Description | Data sets | Type | Format | Size |
|---|---|---|---|---|---|
| Lists of Metadata | The lists of metadata will cover the minimum required set of metadata that must be recorded when acquiring data | Internal discussions with the WP leaders and key PIs | Table and text files | .txt | KB-MB |
| Electrochemical results | Battery full cell performances | Galvanostatic charge and discharge, constant potential steps, all together forming testing protocols | Table, texts files readable by the different battery testing apparatus | .txt, .csv | GB |
| Graphical User Interfaces (GUI) | The GUI will be online tools given to the BIG-MAP community for analyzing the electrochemical data and providing the testing protocols, for requesting samples, as well as for tracking data generated across the different WP | Developed internally using already existing web-design tools | GUIs will consist in web interfaces that will be accessible for every BIG-MAP partners | .exe, .html, python, java, css | MB |

WP8 will collect data to the WPs listed in Table WP 1.21.

*Table WP 1.21. WP8 will collect data to the following WPs*

| WP | What | To be used for | Suggested | | |
|---|---|---|---|---|---|
| | | | type | format | size |
| WP7 | Guidelines and metadata regarding the ontology for data | The received ontology will be used to the list of metadata necessary to ensure the completion of the metadata defined in WP8 | Text | .txt | KB, MB |
| WP8 | Battery & battery interface Ontology | Establishing online laboratory notebooks, other standards and protocols | Protégé file | .owl | KB-MB |
| WP9 | Infrastructure/software as well as guidelines for ensuring the interoperability | The online laboratory notebooks as well as the protocols developed in WP9 will be made available to the whole BIG-Map community using the infrastructure developed in WP9 | Web interface | .html or java | MB, GB |

WP8 will deliver data to the WPs listed in Table WP 1.22.

*Table WP 1.22. Data delivery from WP8 to other WPs.*

| WP | What | Usable for | Suggested type | Suggested format | Suggested size |
|---|---|---|---|---|---|
| WP2, WP3 | List of metadata that must be filled to identify the key information for modelling/simulations data | Reporting and sharing results generating in WP2 and WP3 | Text and web interface | .json | KB, MB |
| WP5 | List of metadata to be filled when carrying out characterizations for battery materials | Reporting and sharing results generating in WP5 | Text and web interface | .json | KB, MB |
| WP4, WP6 and WP10 | List of metadata to be filled when testing battery performances. Standardized electrochemical data to be used as references for electrodes tested within the project. Graphical user interface for electrochemical performances evaluation and to track results generated. | Reporting and sharing results generated in WP4, WP6 and WP10. Standardize the analysis of electrochemical data generated in WP4, WP6 and WP10 and share them to WP11. Certify the performances of battery chemistries developed and tested in WP4, WP6 and WP10. | Text and web interface | .json | KB, MB |
| WP8 | Online Lab Notebook, Measurement Protocols, Metadata Standards | Ontology design | Text, excel, images | .pdf, .doc, .xls, .ppt, .png | MB-GB |
| WP11 | Graphical user interface in which every data recorded within the project will be given a DOI and location where it can be found to be used by WP11 | Finding and tracing data received by WP11, and ensuring that critical metadata regarding data collection and acquisition is properly passed to WP11. | Text and web interface | .json | MB-GB |

The data flow for WP8 is shown in Fig. WP 1.15 and Fig. WP 1.16.

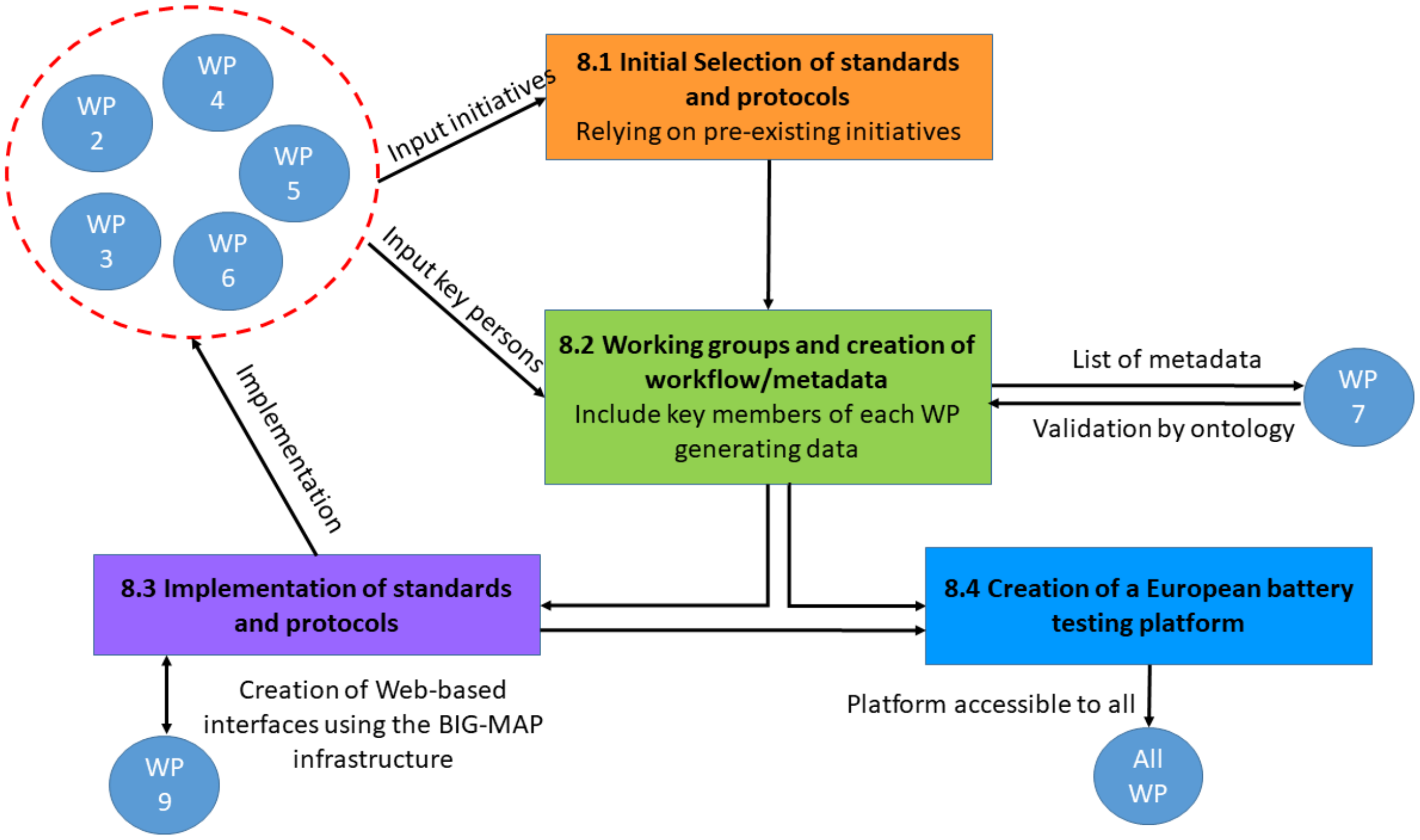

*Fig. WP 1.15. Data flow within WP8.*

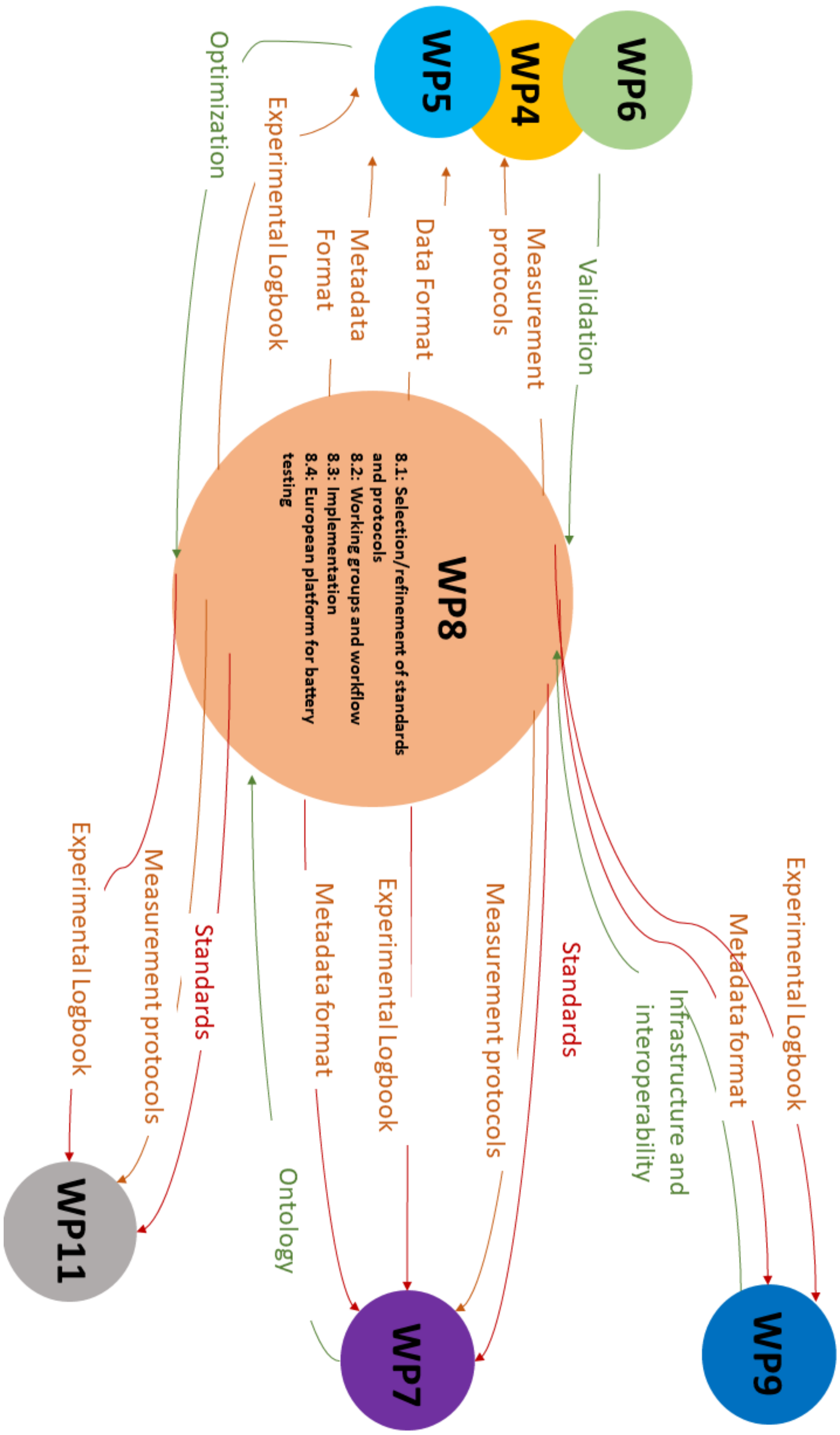

*Fig. WP 1.16. Data flow between WP8 and other relevant WPs.*

### 1.8.4 Specify if existing data are being re-used (if any)

Existing lists of metadata previously defined for materials characterizations or battery electrochemical tests.

### 1.8.5 Specify the origin of the data

Data will be generated in the following tasks:

Task 8.1: Selection of pre-existing protocols regarding characterizations (CNRS, SOLEIL, ESRF, ILL, UU, UCAM); selection of pre-existing protocols for battery testing and materials (UU, UCAM, UOXF, BASF)

Task 8.2: Definition of working groups/definition of list of metadata (CNRS, SOLEIL, ILL, ESRF, UCAM, UOXF, UU)

Task 8.3: Implementation of these standards and protocols, definition of an online laboratory notebook and GUI and generation of DOI (CNRS, SOLEIL, ILL, ESRF, UCAM, UOXF)

Task 8.4: Organization of European platform for battery testing using the defined protocols (CNRS, SOLEIL, UU)

### 1.8.6 State the expected size of the data (if known)

The expected size ranks from few KB for text files to the MB-GB scales for the graphical user interfaces and the MB scale for the electrochemical data

### 1.8.7 Outline the data utility: to whom will it be useful

The data generated will be useful for the following groups:

- BIG-MAP consortium: especially WP2, WP3, WP4, WP5, WP6, WP10, WP11
- European Commission and European Agencies
- Stakeholders involved in the field of Li-ion batteries and materials
- Scientific community
- EU national bodies

## 1.9 WP9 – Infrastructure and Interoperability

### 1.9.1 Purpose of data collection/generation

The purpose of WP9 is to develop the infrastructure and capabilities enabling simulations and experiments to run autonomously, as well as integrated, and communicate with each other, i.e. data storage, preservation, and sharing. This WP aims at providing the necessary tools to manage data and guarantee its interoperability between specific types of simulation and experiments. As a consequence, data from the entire project will be collected in order to develop robust workflows capable of handling operating with multi-sourced and diverse data types and modelling/simulation codes. WP9 further aims at outlining essential interoperability requirements for all types of data, which could be potentially enforced via interoperable data schemas (e.g. JCAMP).

### 1.9.2 Explain the relation to the objectives of the project

Some of the essential goals of BIG-MAP rely on the efficient exploration of large variable spaces. WP9 is tasked with the development of the infrastructure and tools to accelerate such exploration, whereby data from multiple WPs flows seamlessly between simulations and experiments, managed autonomously by robust workflows. Tasks 9.1 and 9.2 focus on developing workflows and integrating those with existing simulation codes and software packages, e.g., AiiDA (EPFL), SimStack (KIT) and ASE (DTU), while data flow will be managed by a common BIG-MAP infrastructure (Task 9.3) that will be ultimately deployed in the cloud (Task 9.4).

### 1.9.3 Specify the types and formats of data generated/collected

The developments within WP9 will be essentially delivered as workflow algorithms (Python code managing data generation/flow), apps (Jupyter Notebooks rendering interactive graphical user interfaces) and infrastructural software (code managing the deployment of the infrastructure in servers and databases). These are summarized in Table WP 1.23.

*Table WP 1.23. Types and formats of the data that will be generated by WP9*

| Data | Description | Source | Schema | Format | Size |
|---|---|---|---|---|---|
| Workflow algorithms | Python code controlling input/output operations, managing errors and queueing of simulation codes and experiments. Also interchangeable files storing algorithm hyper-parameters | Workflow developers | Python packages, configuration files | .py, .yml, .json | KB-MB |
| Workflow descriptions | High-level, human-readable descriptions of code, logic and structure of the workflows. | Workflow developers | Images, slides, word documents | .pdf, .doc, .ppt, .png | KB-MB |
| Jupyter-based Apps | Graphical user interfaces build in Python and | App developers | Python packages, Jupyter notebooks | .py, .ipynb | KB-MB |
| | rendered via interactive Jupyter Notebooks. | | | | |
| Infrastructure software | Software and configuration scripts to serve Jupyter-apps and host databases. | Infrastructure Developers | Python packages, Docker images, Kubenetes, etc, Configuration files and scripts | .py, .json, .tar | MB-GB |

In turn, WP9 collects (and delivers) a variety of data from multiple WPs – already described in their respective tables. For simplicity, we report here only some representative examples of data generated from simulations (Table WP 1.24) and experiments (Table WP 1.25).

*Table WP 1.24. Data collected from simulations*

| Data category | What does the data describe | Methods used for collecting the data | Type | Format |
|---|---|---|---|---|
| AiiDA calculations and workflows | Calculations typically deliver structural data and materials properties, while workflows detail the methodology used | AiiDA automatically records the full provenance graph as a workflow executes | - Electronic structure<br>- Molecular dynamics<br>- Quantum chemistry | - AiiDA export file (see specs below)<br>- .json file with relevant parsed inputs and outputs |
| ASE structures or calculations | Structures generated in or resulting from ASE calculations | Data is manually stored in an ASE database | As above | .sqlite (ASE database format) |
| 3DEXPERIENCE platform / Pipeline pilot raw data and calculation results | Structures, computational data from calculations carried out with Pipeline Pilot protocols, and workflows codified into Pipeline Pilot protocols | The Pipeline Pilot protocols can be tailored to store structures and computational data plus any additional meta data required | - Quantum mechanical<br>- Molecular dynamics<br>- Coarse grained molecular dynamics | The .3dx file format, but reader and writer components will be created to export data to commonly accepted formats facilitating exchange with Materials Cloud and other databases |
| Raw Quantum ESPRESSO output | Electronic structure calculations | A tarball is created from the calculation folder, including relevant inputs and outputs | Simulation | .tar.gz (An archive of input and output text and XML files) |
| Raw VASP output | As above | As above | As above | As above |
| Calculation metadata | Information about a calculation such as who carried it out and using which version of the software | Users are free to create this manually or have script to autogenerate it | A (possibly nested) dictionary of keys and values | .json |

*Table WP 1.25. Types and formats of experimental data, mostly collected from WP5*

| Data category | What does the data describe | Methods used for collecting the data | Type | Format |
|---|---|---|---|---|
| Experimental layouts and materials information | What layout for measurement areas on a materials library is used, what materials are used and where they are sources | Digital Journals JSON describing in key value description type | Text, .json | Text, pictures, .json |
| Measurement recipes | What kind and how a measurement was performed by listing all the measurement parameters and conditions | Machines require input to perform a measurement either that is stored or it is stored in the saved measurement data | As above | As above |
| Measurement results | The result of a measurement performed according to certain parameters | Machine saved the data | As above | Binary, sometimes proprietary |
| Analyses | Data derived from analyzing a measurement | A code runs over the measurement data and saves the result | As above | As above |

A general description of data flow is illustrated in Fig. WP 1.18.

### 1.9.4 Specify if existing data are being re-used (if any)

Public workflows, scripts, molecular and crystal structures from open source databases are re-used. The infrastructure developed here keeps the provenance of the data, making it possible to trace back where it was obtained from.

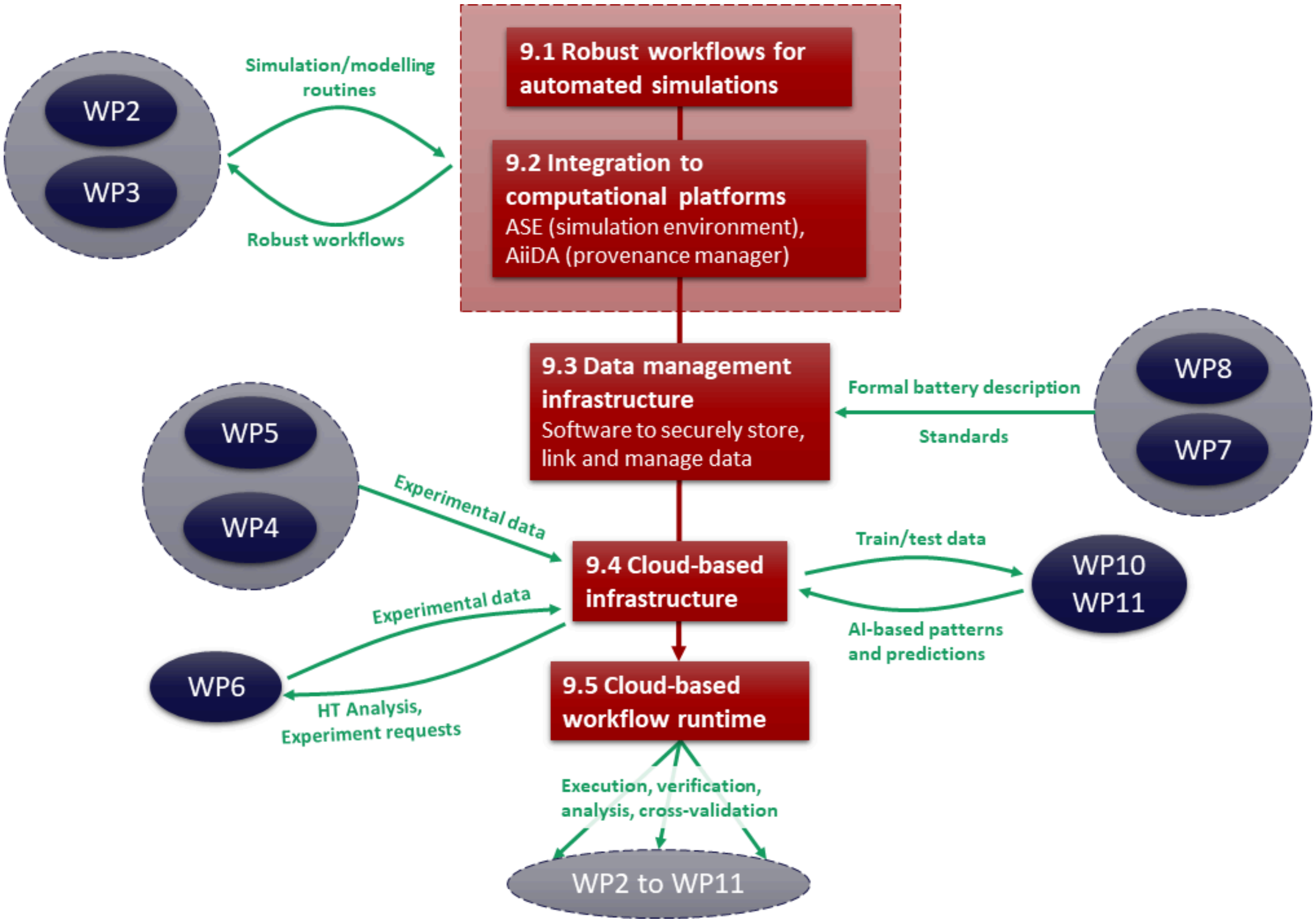

*Fig. WP 1.17 Schematic representation of the data flows between WP9 tasks and other WPs.*

### 1.9.5 Specify the origin of the data

WP2 will primarily generate atomic-scale and spectroscopic (computational) data, while WP3 is concerned with multi-scale modelling. Both WPs will be heavily reliant on infrastructure and interoperability formats from this WP. WP5 is primarily experimental in nature and will also depend on this WP for providing solutions and standards on how to couple experimental data and simulation, while WP4 and WP6 would be able to access and handshake with data and workflows produced. WP9 collects data from the whole project. In addition, open source scripts and structures are also collected. For each WP9 task, the data will originate from:

Task 9.1: The workflow codes developed by BIG-MAP partners (EPFL, CNR, UU, KIT, DTU, 3DS) will be generated as Python scripts, Python packages and user-friendly Jupyter-based apps.

Task 9.2: The integration of multiple computational platforms will be achieved via a higher-level software based on ASE and AiiDA software (EPFL, DTU, KIT, 3DS). The integration software is expected to be made available as Python code.

Task 9.3: The data-management software will enable the secure storage and retrieval of data from multiple sources, both in-house and on the cloud (KIT, CTH, DTU, EPFL, CEA, SOLB). Such software will build upon existing database management systems (e.g. SQL), authentication services (e.g. GitHub), computing-resource managers (e.g. Kubernetes) and provenance managers (AiiDA). The overarching software will be delivered as a set of Python codes, configuration files and scripts.

Task 9.4: Continuing from Task 9.3, the data management software will be deployed in the cloud to be utilized by researchers via a user-friendly Jupyter-based apps (EPFL, DTU, KIT). Such front-end/back-end ensemble will be delivered as single container image (e.g. Docker) in order to guarantee a reproducible re-deployment of the software when user traffic will demand scaling up computing resources.

Task 9.5: The key demonstrator of a simulation-experiment feedback loop (CTH, 3DS, SOLEIL, ILL, ESRF, TUD, SOLB, CEA) will implement a robust workflow (delivered as Python code), operating on data retrieved from the data-management infrastructure, controlled by the researcher form a user-friendly app (delivered as a Jupyter notebook).

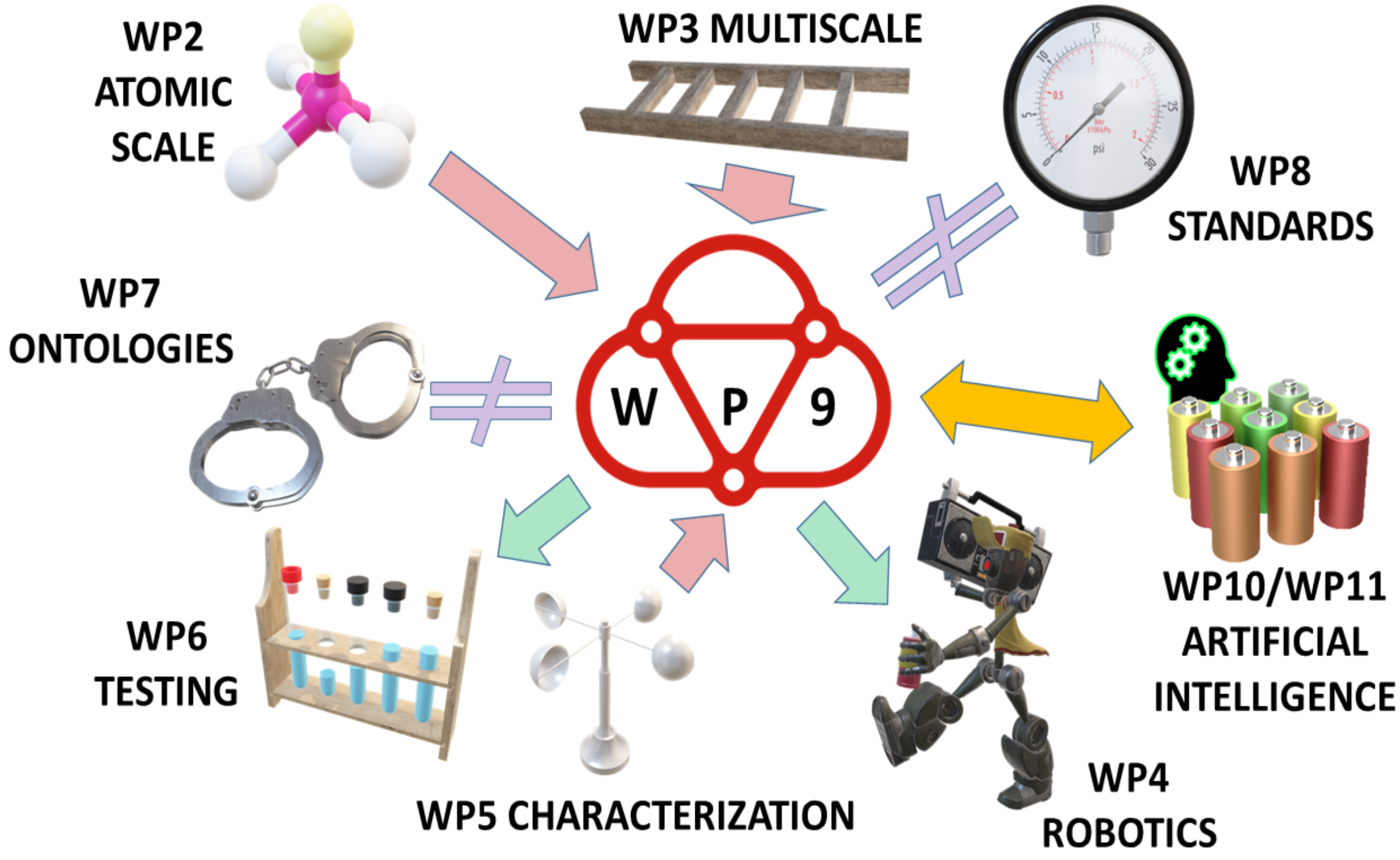

*Fig. WP 1.18. Data flow between WP9 and other WPs.*

### 1.9.6 State the expected size of the data (if known)

The expected size of the data (codes) is MB-GB.

### 1.9.7 Outline the data utility: to whom will it be useful

The *ab initio* calculations performed in this project are expected to generate large volumes of data, most of it is in the form of intermediate binary files, which can be straightforwardly regenerated from the original inputs.

As outlined in the Table above, some of the simulations in other work packages will be managed using AiiDA (*Automated Interactive Infrastructure and Database for Computational Science*), a python framework for high throughput calculations and provenance tracking (www.aiida.net). AiiDA automatically stores all information required to reproduce the result of each calculation. On the output side, AiiDA is designed to strike a balance between the cost of storage and the cost of recomputing a piece of data. For example, by default in a SCF calculation quantities like total energies, electronic band structures and log files are stored, while Kohn-Sham wave functions and charge densities are not.

Before submitting a calculation through AiiDA, all inputs (+ metadata) are automatically stored in a local database. AiiDA then transmits the necessary information to the target computer, which can be a remote supercomputer, a local cluster or the workstation of the researcher. AiiDA adds the calculation to the computer's job queue, monitors the status, and retrieves the results once the calculation is finished.

In this model, all data is generated on the target computer, while only the data intended for preservation is transferred back to the workstation of the researcher.

AiiDA records this data in the form of directed acyclic graphs. For a full specification of the data (+ metadata) stored, see the documentation of the ORM API:
https://aiida.readthedocs.io/projects/aiida-core/en/v1.3.1/reference/apidoc/aiida.orm.html

Given that a simulation may take many hundreds (if not thousands) of CPU hours it is beneficial to the materials science community to be able to access these without having to recompute them.

The corresponding accelerated (WP10) experimental campaigns to the theoretical calculations is performed in WP4, WP5, and WP6 through the interfaces and protocols defined in WP8 and WP9. The instruction and data flow direction therefore twofold: 1) data from experiments triggered by an external source (theory or BIG) and 2) experiments to be ingested in BIG from exploratory campaigns. In both cases the entire data provenance needs to be kept intact. This means that for any data structure being reported to (regardless of intend) to this WP the entire provenance needs to be carried with the data. The objective hereby is to ensure that by keeping any and all potentially relevant data gathered during the creation process, insights and uncertainties as well as mismeasurements can be dynamically identified and altered.

Types of data may contain 1D, scalar, information such as ionic conductivity at room temperature, 2D information like a charge-discharge current voltage curve or 2D diffraction images and 3D data such as current voltage curves over time from battery degradation tests. For any dimensionality type of measurement data the entire chain of origin needs to be reported with the data. This means that, where known, all source materials, their processes, and manufacturers are documented as well as the measurement equipment, measurement recipe, the analysis procedures code version, and the analysed data as well.

There is still a need for a submission platform similar to AiiDA for the experimental sciences, however a submission shall include a sequence of events and their parameters i.e. that after loading a sample a certain procedure according to a set of parameters like -1 mA current for 30 s are applied.

## 1.10 WP10 – AI-accelerated Materials Discovery

### 1.10.1 Purpose of data collection/generation

This work package links different WPs i.e. experimental and theory focused topics. This linkage encompasses linking through transfer learning and acceleration through active learning. This WP therefore sits right in the "middle" of many data flows. WP10 therefore critically depends on properly defined and structured data. At the same time WP10 will store models (as in machine learning models) in a FAIR way as this WP generates data only as a secondary product. This WP therefore requires tailored data management for the different Tasks as the data ingestion, generation and passing is fundamentally different between tasks in WP10. In Task 10.1 the goal is to deploy an active learning framework that suggests the optimal next experiment to users via a standalone software or a web service. Data management here includes the storage of received data, model initialization parameters, model states and return data. What is generated in this task is return data and models with metadata. In Task 10.2 an automatic analysis framework will be developed and principally has the same ingestion and processing logic for data collection and generation just that here versioning of data analysis codes needs to be tracked when data analysis is not handled by a machine learning type algorithm. In Task 10.3 transfer learning will be deployed i.e. models generated will be made publicly available. The data management in this task is therefore limited to logging of requests and what data was passed and what state the models were in after each request. The Task 10.4 has a fundamentally different data ingestion and processing structure as materials and data will be passed between different partners in a sequential manner. Here the goal is to track all materials and analyses from beginning to end to ensure traceability and influence.

### 1.10.2 Explain the relation to the objectives of the project

Autonomous feedback loops as envisioned by BIG-MAP need orchestration by active learning which is deployed through this work package. To deploy active learning there is however the necessity of analysing data in an automated way which is also done by this work package. As in the on-the-fly feedback loops a lot of data is being generated this work package also offers an interface for WP11 to build BIG. As a fall back feedback loop demonstrating the interaction between theory and experiment, this work package build a materials funnel at which promising materials are screened/filtered out using different tiers/cut-offs.

This work package therefore needs to offer interfaces to other work packages so that they may accelerate their experimental or theoretical efforts by means of autonomous reasoning and active learning. In addition to this, it is paramount to analyse data automatically as only with analysed data (i.e. figures of merit) it is possible to deploy active learning. As a central cross cutting initiative, this WP will therefore deploy active learning and analysis servers offering the herein developed tools as modular plug-in service to other WPs. The acceleration by means of efficient exploration of the chemical space is therefore developed in and deployed by WP10.

### 1.10.3 Specify the types and formats of data generated/collected

Ideally, the ML models receive some pre-configured input and generate inter-relatable output. Collected data should come in formats indicating which shape (dimension) and type (i.e. kind like electrochemical or diffraction) the data is (1D, 2D, 3D, ND) and what output task and return data (i.e. task of merely reformatting data or doing more advanced tasks) is needed. Ideally, these are sent as web requests in the form of JSON complaint strings. Ideally the ingested data comes exclusively from the data analysis package such that any ingested data is a scaled FOM where the active learning only needs to figure out the shape and the appropriate model to use. In task 10.4, the data will entail information about the materials and measurement areas and samples.

Types and formats of the data that will be generated by WP10 are listed in Table WP 1.26.

*Table WP 1.26. Types and formats of the data that will be generated by WP10*

| Datatype | Description | Data sets | Type | Format | Size |
|---|---|---|---|---|---|
| Data ingested through interfaces to this WP | Data sets acquired directly from experimental/simulation sources pertaining to measurements on interfaces. | Data generated by tools developed in WP2: Electronic/atomic structures and up to nano-second timescale | Molecular structure trajectories | various | MB |
| | | Data generated by tools developed in WP3: Multi-scale simulation results up to cell level and cycle life | Phase/crystallite/ micro/macro structure | 2D/3D structure files time evolving over time | |
| | | Experimental data generated in WP5: experimental characterization of interface at different time/length scales | Spectroscopy and microscopy data, electrochemical and other cycling data | various | |
| | | Experimental data generated in WP4: High-throughput interfacial electrochemistry or spectroscopy as well as long-term cycling | All of the above | various | |
| Data analysis tools | Tools that analyse measurement data from experiments mostly in WP4 and WP5 to a format that can be used for training the BIG models and for models in active learning | Library of scripts that extract pertinent FOM and or labels from experimental data | Code, database, weights | python | MB-GB |
| Active learning tools | Tools that analyse measurement data from experiments mostly in WP4 and WP5 to a format that can be used | Library of scripts that suggest optimal net experiment to run | Code, database, weights | python | MB-GB |
| | for training the BIG models and for models in active learning | | | | |
| Funnel data | Data gathered from operating the materials funnel for materials downselection | Data describing the materials, processes and results and analyses from the materials funnel as well as the thresholding rules of how materials advance to the next stage | Code, metadata, experimental data | various | MB |

WP10 will collect data from the following WPs (Table WP 1.27).

*Table WP 1.27. WP10 will collect data from the following WPs*

| WP | What | To be used for | Suggested | | |
|---|---|---|---|---|---|
| | | | type | format | size |
| WP2 | Ab-initio data for running autonomous feedback loops in-silico and jointly with experiments | Data analysis and active learning | FOM | .json, .hdf5 or database API request | KB |
| WP2 | Electronic properties and predictions of experimental spectra to facilitate analysis in the automated data analysis package | Generating input data for training the BIG models Building blocks for the BIG-MAP models Assimilating the output of the BIG models | Structural data | .json, .hdf5 or database API request | MB |
| WP2 | Curated atomistic data | Active learning and joint in-silico/experimental feedback loops | Tarball files can be created with post-processed data and parameters of interest. | .tar.gz (with databases and reproducible data sheets) | GB |
| WP3 | Domain Knowledge Workflows Training Data | Workflow interface between WP2 and wp10, Training data for WP10 | Tarball, uncompressed raw data | .hdf5, .json | TB |
| WP4 | Data for an autonomous feedback loop with high- | Electrolyte formulation or synthesis optimization | Raw experimental data | .hdf5 or .json or .csv where applicable | MB |
| | throughput experiments | | | | |
| WP5 | Experimental data or structure evolution | Autonomous feedback loops | Raw data from diffraction or spectroscopy | .json or .hdf5 | MB |
| WP6 | Complementary high-throughput electrochemical and spectroscopic data<br>Automated data analysis framework<br>Experiment suggestion | Interphase optimization through characterization and ML-orchestrated analysis<br>Evaluation of structure-property-performance relationship<br>FOM extraction<br>Experiment planning | Output data, parameters sheets | Docs<br>.json | KB-MB |
| WP6 | High throughput experimental data (electrochemical and spectroscopic properties and performance) evolution<br>Communication protocol<br>Data to run autonomous feedback loop | Autonomous feed-back loops<br>Multi-scale modelling by coupling AI with multi-fidelity data | Output data<br>Parameters sheets | Docs | KB-MB |
| WP6 | Data from high-throughput screening | Building an acquisition function in autonomous feedback loops | Electrochemical and Spectroscopy data | .json | MB |

WP10 will deliver data to the following WPs (Table WP 1.28):

*Table WP 1.28. Data delivery from WP10 to other WPs*

| WP | What | Usable for | Suggested | | |
|---|---|---|---|---|---|
| | | | type | format | size |
| WP2 | Suggestion of the next optimal experiment | Autonomous feedback loops | Json or xml or applicable run file | .json | KB |
| WP3 | AI models for scale bridging and surrogate models | Models that may bridge different scales and | hdf5 or json | .hdf5 | TB |
| | | supply surrogate modelling | | | |
| WP4 | Suggestion of the next optimal experiment or suggested candidate molecules to be synthesized. Molecules to be used for protective coatings from WP10's materials funnel | Autonomous feedback loops | Json or xml | .json | KB |
| WP4 | Data analysis | FOM extraction | Json, csv or pointer | .json | KB |
| WP5 | Data analysis for selected methods and data for experiment suggestions in active learning runs | Data processing, Spectral deconvolution, peak fitting, Experiment planning (suggestion) | Parameters, fitted data & analysed spectra | .json | KB |
| WP6 | Complementary high-throughput electrochemical and spectroscopic data<br>Automated data analysis framework<br>Experiment suggestion | Interphase optimization through characterization and ML-orchestrated analysis<br>Evaluation of structure-property-performance relationship<br>FOM extraction<br>Experiment planning | Output data<br>Parameters sheets | Docs<br>.json | KB-MB |
| WP6 | Data analysis | FOM extraction | Json hdf5 or pointer | .json | KB |

WP10's data flows are shown in Fig. WP 10.1 and Fig. WP 10.2.

### 1.10.4 Specify if existing data are being re-used (if any)

Existing data from experiment and theory is going to be reused in 10.4 as part of the materials funnel. In all of the autonomous feedback loops data is being generated on the fly and then only reused for initialization of other feedback loops. WP11 is going to re-use data generated in this WP.

### 1.10.5 Specify the origin of the data

This work package will sit inside multiple autonomous feedback loops and will receive data from experiments (in-silico or in-physico), analyse them (10.2), suggest new experiment (10.1) and pass data to WP11 (10.3). This work package is therefore a "three way valve" connecting experiments, theory and BIG.

Task 10.1: Partners involved are KIT, ULIV, ITU, BASF NVOLT, Fraunhofer and FZJ. This task is mainly concerned with developing and deploying algorithms for active learning and will not directly ingest data as all data in this task is going to be assumed to be just "numbers" or figures of merit. The aim is however to deploy active learning to KIT for electrolyte formulation optimization, ULIV, ITU, Fraunhofer for synthesis optimization, and FZJ for electrolyte optimization.

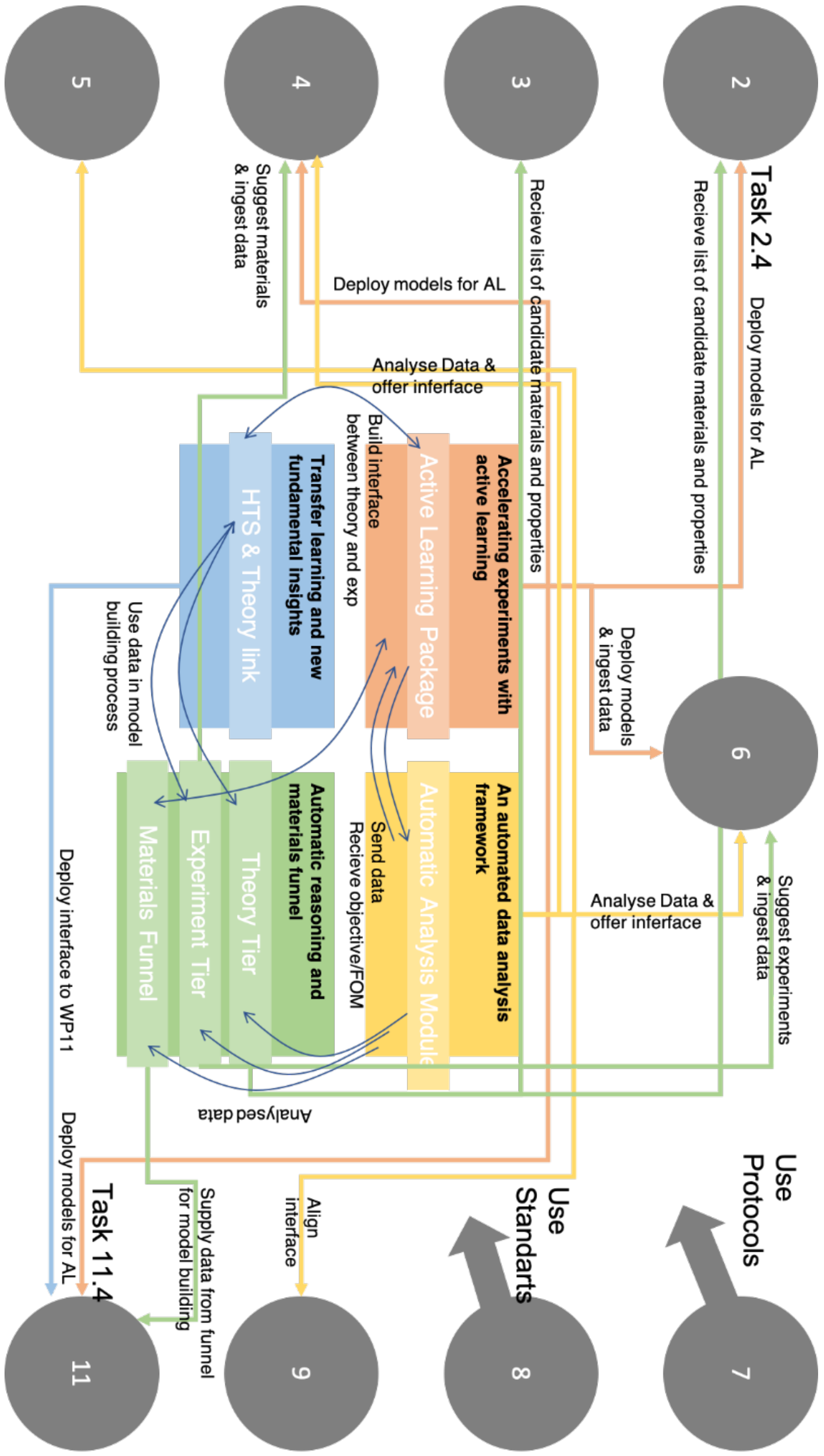

*Fig. WP 1.19. Data flow between the tasks of WP10 and their relation with other WPs.*

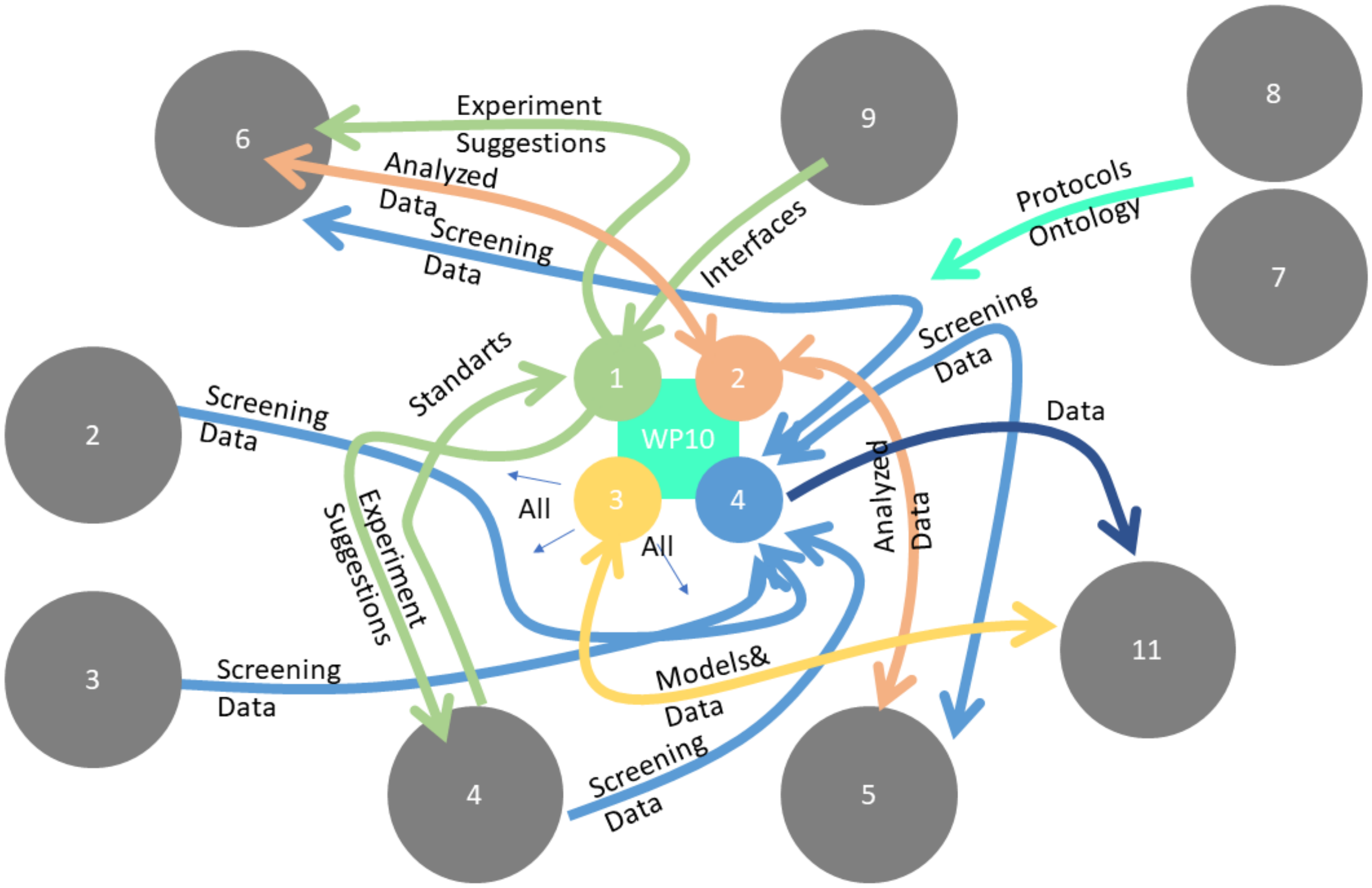

*Fig. WP 1.20. Data flow between the WP10 and other WPs.*

<u>Task 10.2</u>: This task ingests data from different work packages for analysis and for pass through to Task 10.1 Namely the plan is to ingest data from WP4, WP5, WP6 and WP9 with a focus on analysing battery cycling data from CID, FZJ, KIT as well as electrochemical data like EIS from ULIV, FZJ, CID and KIT. This will also be aligned with simulation efforts at DTU and KIT. At a later stage the ingestion of XRD data from ESRF is planned.

<u>Task 10.3</u>: This task will mainly work with WP11 on deploying the dynamic interface descriptors. This task will not work with any experimental or theoretical data per se as it only offers an interface to/from WP11.

<u>Task 10.4</u>: The automatic reasoning and materials funnel will ingest data from several WPs namely WP2 from KIT and DTU where large screening datasets are expected, data from WP3 (DTU, KIT) for scale bridging data, WP4 and WP5 regarding synthesizability (ULIV, ESRF, ILL) and quality and lastly from WP6 (FZJ, KIT) from high-throughput experimentation. The data will then be communicated with BASF and NVOLT for a potential second pass.

### 1.10.6 **State the expected size of the data (if known)**

Variable size from <10 to >10000 individual datasets each in the KB to max GB range

### 1.10.7 **Outline the data utility: to whom will it be useful**

Will be useful for WPs concerned with accelerating their experiments or simulations such as mostly WP4, WP5 and WP6. All data will be used for building BIG in WP11, here the responsibility to make data amendable for ingestion is in the Task 10.4 that is responsible to make data “ingestible”, i.e. make compatible to WP11 related needs.

## 1.11 WP11 – Battery Interface Genome

### 1.11.1 Purpose of data collection/generation

The goal is to establish predictive deep learning-based Battery Interface Genome (BIG) models for inverse design and characterization of battery interfaces and interphases. The deep-learning model shall learn to predict dynamic events occurring at the battery SEI.

Data describing the interface structure and the evolution of interphases for all length - and time scales will be utilized to train the model. The multi-scale (atomic scale to micron scale), spatiotemporal training data will be collected from WP2, WP3, WP5, WP6 and from publicly available open databases (e.g. NOMAD, Materials Cloud, Computational Materials Repository).

The interoperability of the data, both the training data and the data generated by the BIG models, will be ensured by applying the ontology developed and defined in WP7 and the standards for data and metadata defined in WP8.

The models developed in WP11 will be used in WP10 for autonomous decision making for optimal experiments to be carried out.

### 1.11.2 Explain the relation to the objectives of the project

The work relates to BIG-MAP's Specific Objective 1: "Develop the scientific and technological building blocks and modes for accelerated battery discovery" and especially the two sub-objectives:

- Multi-scale modelling by coupling AI with multi-fidelity data from simulations and experimentation
- Development of an uncertainty-guided hybrid physic and data-driven generative BIG-model to predict the spatio-temporal evolution of battery interfaces, and enable inverse materials design

This translates to the following objectives for WP11:

- Build a model for dynamic fidelity assignment for data collected from sources having different levels of errors and uncertainties.
- Development of an uncertainty-guided hybrid physic and data-driven generative BIG-model
- Predict the spatio-temporal evolution of battery interfaces
- Enable inverse materials design
- Identify dynamic interphase descriptors of the battery's SEI able to characterize the battery performance
- Define a multi-scale representation of the battery's SEI interfaces ranging from atomic scale to electrode scale
- Map uncertainty flow from data sources and throughout models' scales
- Demonstrate transferability/applicability of models to select novel battery materials/chemistry

### 1.11.3 Specify the types and formats of data generated/collected

Types and formats of the data that will be generated by WP11 are listed in Table WP 1.29.

*Table WP 1.29. Types and formats of the data that will be generated by WP11*

| Datatype | Description | Data sets | Type | Format | Size |
|---|---|---|---|---|---|
| Data collected from other WP and pre-existing external sources. Heterogeneous methods of creation<br><br>WP11 only assigns fidelity to the data in collaboration with other WP | Data sets acquired directly from experimental/simulation sources. Such data will contain information about spatio-temporal evolution of interfaces but not in a directly accessible form. Processing need to be done for joining information from multiple sources and length/time scales | Data generated by tools developed in WP2: Electronic/atomic structures and up to nanosecond timescale | Molecular structure trajectories | various | TB |
| | | Data generated by tools developed in WP3: Multi-scale simulation results up to cell level and cycle life | Phase/crystallite / micro/macro structure | 2D/3D structure files time evolving over time | |
| | | Experimental data generated in WP5: experimental characterization of interface at different time/length scales | Spectroscopy and microscopy data, electrochemical and other cycling data | various | |
| | | Data collected from external sources has been generated for other purposes and outside BIG-MAP | All of the above | various | |
| Data processing tools<br><br>Primarily contributed by WP7/9 | Tools that process data to a format that can be used for training the BIG models | Library of scripts that convert different data into a standardized hierarchical format suitable for generative deep learning | Code | Python codes in gitlab | MB |
| Curated databased | A curated database holding interface structure data at multiple length scales | Database linked to high speed access interface | DB | Hdf or .json | TB |
| Primarily contributed by WP9 | and its evolution at time scales | | | | |
| Programs that train the BIG models | Deep-learning models | Python codes built with pytorch | Code Scripts | Python codes in gitlab | MB |
| The BIG models | The trained BIG models | Generative models at multi scale Explainable models for descriptor discovery | Saved model with all parameters | Trained neural networks saved in .pth file | MB -GB |
| Output data | Data describing dynamic instances at battery interfaces by means of the developed battery ontology | Conditionally inverse designed battery interfaces | Multi-scale structure of battery interface with good property | Atomic structure in .traj file; meso and microstructur e as voxel based binary file .hdf extension | MB |

WP11 will get data from the WPs listed in Table WP 1.30.

*Table WP 1.30. WP11 will get data from the following WPs*

| WP | What | To be used for | Suggested type | format | Size |
|---|---|---|---|---|---|
| WP2 | Atomistic-scale dynamics data for interfaces, related properties and features for descriptor search | Building spatio-temporal generative model and ML based descriptor search models | Tarball files can be created with post-processed data and parameters of interest. | .tar.gz (with databases and reproducibl e- data sheets) | GB |
| WP3 | Mesoscopic models that incorporates materials specific information<br>Macroscopic P2D continuum models<br>Multi-physics models at FEM level | Model building and verification and Validation of the performance of generative model | Tarball, Uncompresse d Raw Data | .tar.gz (with databases and reproducibl e- data sheets) | TB |
| WP5 | Curated experimental data or properties (chemical environment/structure/morpholo | Generative models at microstructur e level. | Ouput data, parameters sheets | Docs and many types depending | MB-TB |
| | gy) evolution; 3D images of microstructures | Autonomous feed-back loops; machine-learning based image analysis | | on techniques | |
| WP6 | High throughput experimental data (electrochemical and spectroscopic properties and performance)<br>Data to run autonomous feedback loop evolution | Autonomous feed-back loops Multi-scale modelling by coupling AI with multi-fidelity data | Output data, Parameters sheets | Docs, .csv, json | MB-GB |
| WP7 | The battery ontology | Standards in data access | Standard | OWL | MB |
| WP8 | Graphical user interface in which every data recorded within the project will be given a DOI and location where it can be found to be used by WP11 | Finding and tracing data received by WP11, and ensuring that critical metadata regarding data collection and acquisition is properly passed to WP11. | Text and web interface | .txt, html or java | KB, MB |
| WP9 | Standardization for data assimilation and storage | Storage, collection and delivery of data | Infrastructure | Python code | MB |

WP11 will deliver data to the WPs listed in Table WP 1.31.

*Table WP 1.31. Data delivery from WP11 to other WPs*

| WP | What | Usable for | Suggested type | Suggested format | Size |
|---|---|---|---|---|---|
| WP2 | ML based descriptor search methods at atomic scale and uncertainty estimation of ML models | Optimal chemical space and interface composition exploration | Tarball files can be created with post-processed data and parameters of interest | .tar.gz (with databases and reproducible-data sheets) | MB |
| WP3 | Uncertainty propagation scheme between scales | Efficient bound on uncertainty for multiscale simulation | Github code / tarball | Python code | MB |
| WP5 | Data analysis; ML algorithms for segmentation of 3D volumes and statistical analysis of 3D microstructures | Extract quantitative parameters and their spatial distribution; design new experiments | Computer codes | Python code | MB |
| WP6 | Prediction of electrochemical observables | Cross-analysis of experimental data using the modelling predictions Design new experiments to validate the modelling predictions | Predicted property and structure/composition | Docs<br>Data files | KB-MB |
| WP10 | Latent space representation of materials and interfaces | Data analysis and active learning | Encoder algorithm | pth | KB |
| WP10 | Dynamic interface descriptors | Data analysis and active learning | parameters | npy | KB |

The data flows in the WP are shown in Figs. WP11.1 and WP11.2.

### 1.11.4 Specify if existing data are being re-used (if any)

Relevant publicly available data collected from data repositories (e.g. NOMAD, Materials Cloud, CMR etc.) will be reused.

### 1.11.5 Specify the origin of the data

The data collected/generated origin from computational work performed by the involved organization: DTU, KIT, CNRS, NIC, CSIC, CNR, UCAM, UOXF, ESRF, CID, UMI, UNIVIE and 3DS.

This work package receives data from experiments and simulations from multiple work packages and then provide guidance to those for further experiments and simulations. Within WP11, WP11.1 gathers and processes data from multiple other WPs and provides it to WP11.2/3/4. WP11.2/3/4 are the core model development activity. WP11.5 centres around verification and validation of built models and transferability testing on new battery chemistries.

Task 11.1: Partners Involved are ESRF, 3DS, CNR, CNRS, CSIC, DTU, KIT, UOXF. This Task is mainly concerned with demonstrating the integration of multi-fidelity data from simulations, experimental characterization, testing to create a unified description of chemical and spatio-temporal evolution of the battery interface in a cost effective manner. ESRF will lead the experimental fidelity activities with support from UOXF and CNRS, while CSIC will lead the computational efforts with support from 3DS, CNR, DTU and KIT.

Task 11.2: This task involves DTU, UCAM, CNRS, KIT, UOXF and focuses on identifying the multitude of descriptors or genes that encode the spatio-temporal evolution of battery interfaces and interphases. To help with the inverse design process and to make the descriptor discovery process semi-autonomous, semi-supervised deep learning will be used. DTU will lead the method development for automated feature/descriptor identification with support from KIT, while UCAM, CNRS and UOXF will perform the experimental validation and testing needed.

Task 11.3: This task is with DTU, 3DS, UNIVIE, CID, CSIC, KIT. The goal of this task is to develop a novel ML methodology for modelling time evolution of dynamic systems simultaneously at multiple time and length scales with a unified approach using hierarchical latent variable models. DTU will lead the hierarchical latent space model development with support from complementary internal activities. UNIVIE, 3DS and in particular KIT will provide link to Task 3.3 (AI enhanced models) and models will be validated experimentally by data gathered by CID.

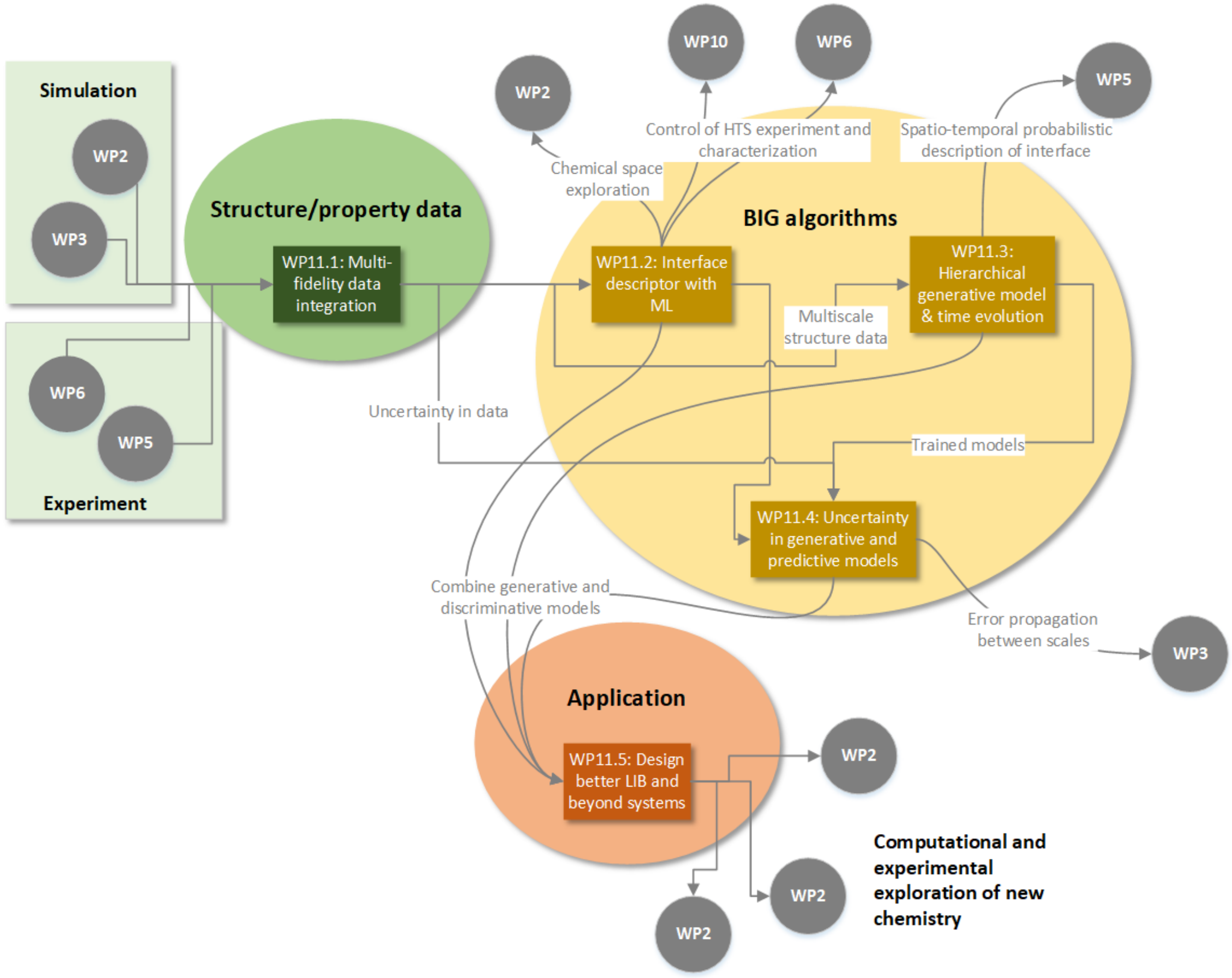

*Fig. WP 1.21. Data flow between the tasks of WP11 and their relation with other WPs.*

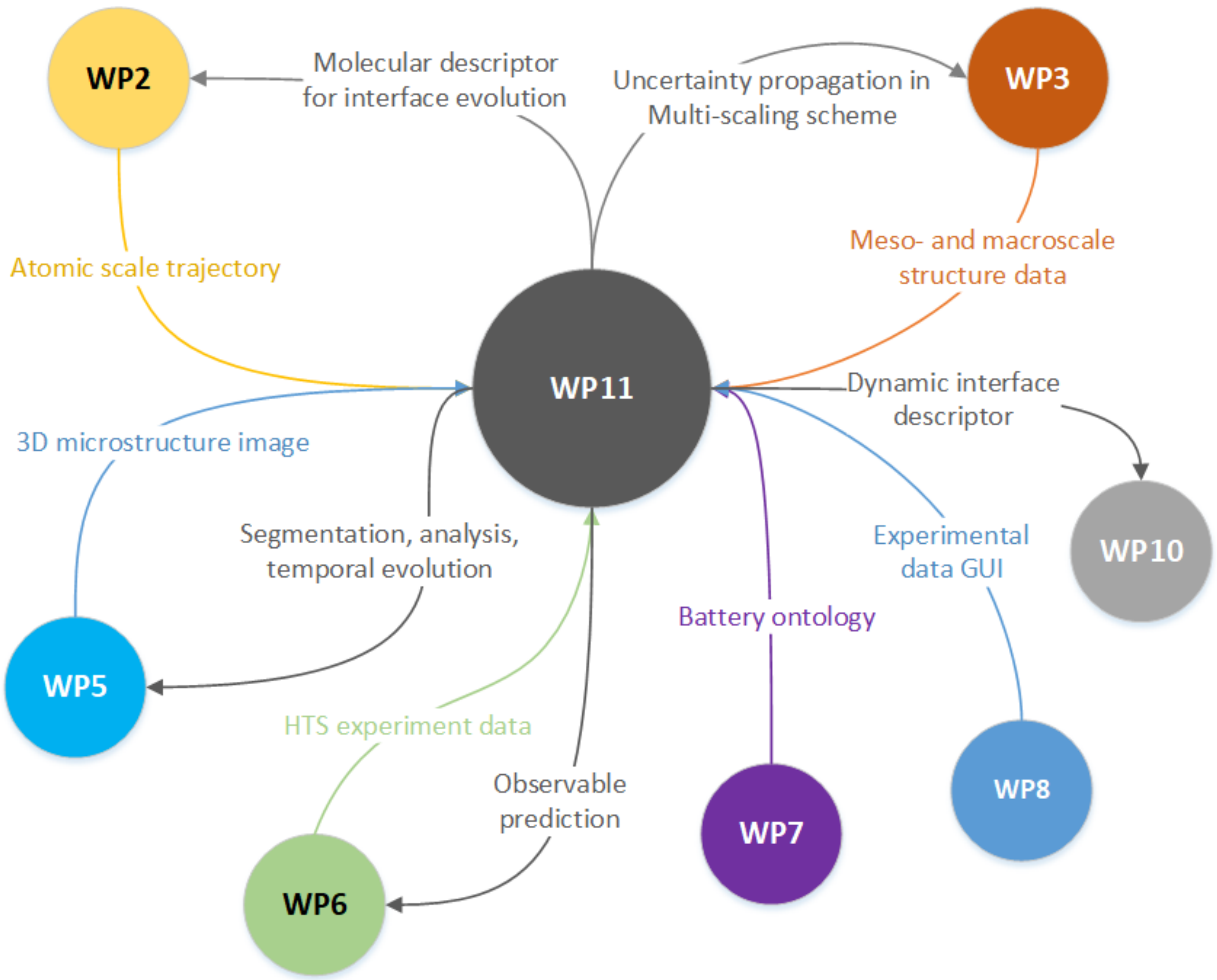

*Fig. WP 1.22. Data flow between WP11 and other WPs.*

<u>Task 11.4:</u> This task involves DTU, UNIVIE, CNR, CSIC, KIT. The goal of this task is to develop a unique framework comprising self-correcting unsupervised deep models for uncertainty-aware predictions. We will build and validate interfacial reactions and structure evolution predictor that automatically identifies and corrects uncertainty and errors in the modelled and experimental data. DTU will lead the model development and utilize synergy with the existing projects, while UNIVIE, CNR, CSIC and KIT will focus on methods for uncertainty quantification in the training data at different scales.

<u>Task 11.5:</u> Task is carried out by CID, UOXF, NIC, CNRS, CSIC, DTU, UMI. This task is dedicated to the demonstration and testing of the versatility and transferability of the developed methodologies to a novel chemistry and/or structure of active materials or electrolyte. CID, CNRS and UOXF will lead the testing activities on Na-ion, NIC and CSIC will transfer developed solutions and test them in the combination with magnesium and calcium electrodes, while DTU and UMI will focus on the general transferability of the models.

### 1.11.6 State the expected size of the data (if known)

The total size of the generated data will in the TB+ region.

### 1.11.7 Outline the data utility: to whom will it be useful

The whole battery design community but especially battery researchers focusing on developing new battery materials for the post-Li period.

# 2. FAIR Data

The guiding principles for FAIR-ification of BIG-MAP's research output are:

- H2020 guide for Open Access and Data Management
- The European Code of Conduct for Research Integrity

BIG-MAP will strive to share insights beyond state-of-the-art, but at the same time respecting the individual partners' need for restricting the access to specific data and the involved industrial partner's need for keeping data relating to their business secrets confidential.

Many of the concepts described in this FAIR section will be developed as part of the project. These will be implemented on a running basis in line with their availability. In the period up to specific concepts are ready for use, the partner will rely on local systems. As all partners are recognized for trustworthy research of high quality, the lack of dedicated concepts in the start-up phase is not regarded as a point of concern.

BIG-MAP's research output can be grouped into three categories, see Fig 2, section 1.1.1.

Upstream data (WP2, WP3, WP4, WP5 and WP6)

- Data generated by experimental methods (screening, characterization and performance testing) and by computational methods (modelling and simulation).
- Semi-automatic robotics for synthesis of battery-relevant protective coatings.

Downstream models (WP10 and WP11)

- Artificial-intelligence (AI) models for funnel-based accelerated discovery of new battery materials that form the basis for the Materials Acceleration Platform (MAP).
- Machine-learning (ML) models for inverse design of battery interfaces and interphases that form the basis for the Battery Interface Genome (BIG).

Tools (WP4, WP7, WP8 and WP9)

- Tools that are essential for achieving interoperability across the project and for feeding interoperable and aggregated upstream data efficiently into the downstream models.
- The battery-interface ontology BattINFO.
- Standards and protocols ensuring interoperability of the research output, i.e. data, code and models.
- An interoperable IT infrastructure for work in progress and research output accessible for the BIG-MAP consortium only. Interlinked with the infrastructure is a BIG-MAP's GitHub organization for shared development of computational output (e.g. software, code and scripts).

A BIG-MAP App Store has been established for easy public access to the developed computational output. The apps in the App Store link with the associated source code deposited in BIG-MAP's GitHub organization and with the data needed for running the app. This might be data deposited in public repositories and/or data stored in BIG-MAP's shared data-storage facility. This will typically be data awaiting publication or data of restricted openness. Only the apps and not the users of the apps will be granted access to data. This architecture ensures maximum exploitation of all datasets, while complying with the legal constraints attached to the use of sensitive data. The App Store's Graphic User Interface (GUI) allows searching for content via standard web browsers, and the apps APIs (Application Programming Interfaces) allow software-independent communication and usage of the apps, see Fig. 4.

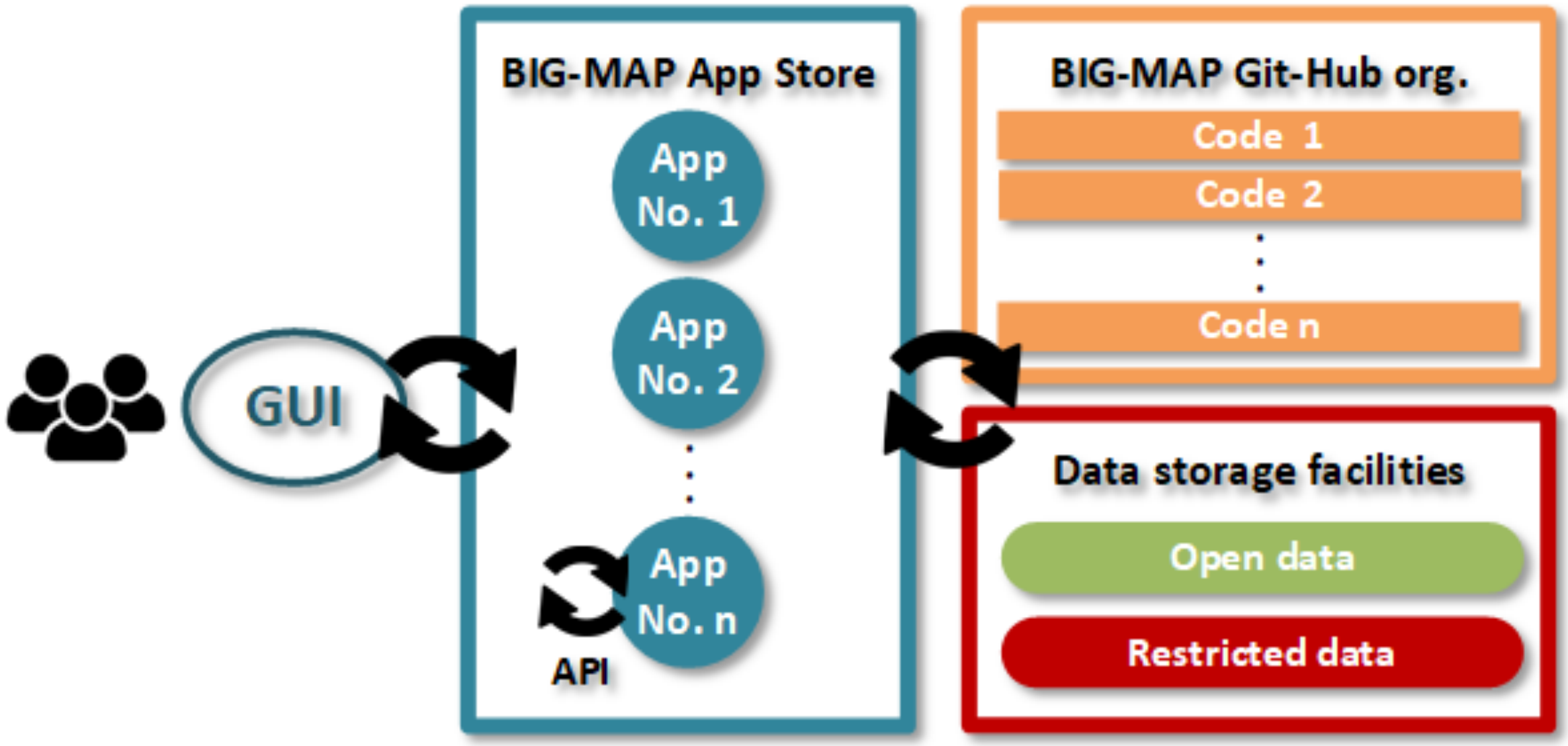

*Fig. 4. The BIG-MAP AppStore – conceptual view*

BIG-MAP's research output is divided into three levels of openness:

- Public.
- Restricted access, i.e. output to be shared within the consortium and/or with selected stakeholders, and output in progress intended for publication at a later point.
- Confidential.

Work packages 2-6 generate upstream data that is fed into the downstream models developed in WP10 and 11. These models are BIG-MAP's key products: the Materials Acceleration Platform (MAP) and the Battery Interface Genome (BIG). The tools developed in WP7-9 ensure the data's interoperability and conformability with the downstream models. The feed-back loop allows the downstream models to request new data and new materials from the upstream WPs.

BIG-MAP's strategy for FAIR-ification of the research output is described in the following subsections. The information has been organized as follows:

- The general principle applying if nothing else has been stated.
- Supplementary information relating to the general principle.
- Exceptions from the general principles.

## 2.1 Making data findable, including provisions for metadata

### 2.1.1 Discoverability of data (metadata provision)

All BIG-MAP's research output will be equipped with rich machine-readable metadata and keywords following the standard outlined in the section "Specify standards for metadata creation".

The metadata shall allow potential users to discover data via searching in repositories for research data and searching via standard web browsers. The BIG-MAP App Store has for this reason been equipped with a GUI.

### 2.1.2 Identifiability of data including standard identification mechanisms and the use of persistent unique identifiers such as Digital Object Identifiers

The assignment of identifiers will follow the rules and standards applying for the actual storage platform where the data is indexed. Experimental data generated at large-scale facilities are usually given a DOI already by the facility when generated. The same may apply for data generated in lab facilities. A procedure for alignment of identifiers assigned to data at various steps in their processing and on various platforms needs to be settled. When ready, the procedure will be included in a future update of the DMP.

Supplementary information

- BattINFO, which is metadata itself, will be made discoverable via a unique URL that redirects either directly to the development Git repository (e.g. GitHub) or to the App Store.
- When commercially licensed input data are used, e.g. in simulations, appropriate identifiers preserving the data provenance shall be used.

### 2.1.3 Naming conventions

The naming convention for data shall comply with the BattINFO ontology and the developed standards and protocols. The BattINFO ontology itself will follow the naming conventions defined in EMMO.

### 2.1.4 Search keywords

Data and metadata shall be equipped with search keywords. Search keywords shall as far as possible be machine-readable. Keywords complying with the BattINFO ontology shall always be included. The ontology-based keywords can be supplemented with keywords referring to, for instance, the developed standards and protocols, if relevant.

### 2.1.5 Versioning

Versioning of the research output shall be implemented whenever feasible. The use of a GitHub organization for computational output will have full versioning of all content via GitHub's inherent procedures for automatic versioning. For output developed and stored elsewhere, the following shall be reported: the date the research output was generated and a number giving a clear indication of version chronology (semantic versioning, calendar versioning).

Supplementary information

- All AI models and code will be timestamped, and the metadata explaining what, how and by which data the models have been trained will be hashed.
- For ML models, all model steps in active learning sessions will be saved.

### 2.1.6 Metadata standards and description of the metadata to be created and how they will be created

BattINFO shall form the basis for a metadata standard able to give a precise and unequivocal description of BIG-MAP's research output. A commonly accepted metadata standard for the electrochemical data (an essential share of the upstream data) does not currently exist. The metadata standard developed by NFFA (Nanoscience Foundries and Fine Analysis) applying for nanoscience, might serve as a source of inspiration for the electrochemical part of BATTInfo.

The BIG-MAP's metadata shall as minimum hold information on:

- The specific aspect of the battery interface described by the research output.
- The method used for generating the data with reference to standard protocols and standard operating procedures, when possible.
- Information on data fidelity and provenance.

Routines for automatic creation of metadata shall be established wherever possible and relevant. The automatically generated metadata will be supplemented by manually added metadata when needed for arriving at a complete description of the research output. The metadata and the data shall be stored together or shall be connected via persistent links. The metadata can be stored in separate files or integrated in the data file itself.

Supplementary information on experimentally generated data

A formal metadata format and content for each type of measurement will be developed, and equivalence of metadata coming from different experiments shall be pursued and implemented as soon as appropriate standards are available.

The metadata standard shall be designed to precisely reveal the following information:

- The materials, electrodes and components used for the battery assembly (each with a DOI)
- Processing and assembly protocols of test samples
- Data acquisitions protocols and specification of the instruments used
- Time and location for the experiment
- Type of output data generated and downstream processing protocols required for interpreting the results

Supplementary information on data generated by computational methods

Metadata data generated by computational methods shall to the greatest possible extent follow the AiiDA standard where the following metadata are automatically generated for each calculation:

- The inputs used
- The code and computer used
- The outputs generated
- The person who submitted the calculation
- The time and date the calculation was submitted.

ML models: The minimum standard for metadata describing ML models shall be

- The initialization parameters (including random values) used
- The data used for the training
- The hardware used

Training of the ML models will yield slightly different results when run on different hardware. The metadata shall enable any user to execute a model training and in principle, be able to reproduce results.

AI models: All available metadata need to be passed to the next stage that is responsible for passing on all previously generated metadata.

Supplementary information on ontology

The metadata shall comply with the following standards: European Materials Modelling Ontology (EMMO), Web Ontology Language (OWL), Dublin Core Metadata Initiative and Simple Knowledge Organization System (SKOS).

## 2.2 Making data openly accessible

### 2.2.1 Data that will be made openly available, and the rationale for keeping specific data closed

BIG-MAP's research data will be made publicly available to the largest possible extent without conflicting with:

- BIG-MAP' Grant Agreement and Consortium Agreement
- The project partners' local policies and regulation on openness of data
- The project's industrial partners' business secrets

Embargo periods may apply.

Exceptions

- Data stemming from the industry, typically input data to the models, will be kept confidential to the extent requested by the entity making the data available for the project.
- Data that might form the basis for IPR. If such data are generated, the owner of the data can decide to keep the data confidential in the period that publication of the data conflict with an application for IP protection or until such an application has been abandoned.
- Data generated by specific proprietary software will be made available to the extent possible under the specific licencing agreements set by the software-providing company.
- Open-source software and code will be made openly available to the extent possible under the standards and regulations set by the licence utilized.

Specific provisions that may restrict the openness apply for the following data

- Data used for developing the AI models will not be publicly available but available on a request basis. The AI model itself and the related code will be publicly available. By this, the potential users are given access to the data required for running the models on their own computer given they have access to the necessary framework and hardware.
- Commercially licensed crystal structures used as input for atomic-scale simulations. For such data provenance will be reported as needed to abide by the license agreements, and only appropriate identifiers will be preserved, in order to track the original crystal structure in the source database, for those who have access to it. Similarly, data about materials from supplies, which might be subject to confidentiality agreements, will be withheld to users without matching credentials.

All handling of confidential data and data of restricted openness shall follow what has been decided and agreed with the owner of the actual data.

If knowledge about the existence of specific data of restricted openness is regarded as important for the development of new battery materials and/or improved battery technologies, such data can be announced by indexing only the metadata in an open repository.

### 2.2.2 How the data will be made available

Computational output will be made available via BIG-Map's GitHub organization and via BIG-MAP's App Store, whereas other data will be made available via trusted repositories for research data.

A repository is regarded as trusted if it:

- Is recognized in the battery research community.
- Has clear terms and conditions.
- Guarantees sustainability.
- Provides persistent and unique identifiers, e.g. DOI.
- Offers standard license agreements for data sets.

### 2.2.3 Methods or software tools needed to access the data, and information about specific software needed

BIG-MAP's research data shall, to the greatest possible extent, be converted to open format (e.g. .txt and .tif) or commonly used formats before sharing and/or publication. Commonly used format might be commonly used in general (e.g. Word and Excel) or commonly used within the actual research discipline.

If specific software is required for accessing data, the software's name and version number will be included in the documentation following the data, or alternatively will the necessary software be provided. In the latter case, licence regulations may apply.

Supplementary information on data generated at large-scale facilities
Large-scale facilities have their own software to plot, visualize, pre-process and treat the raw data, which might be open depending on the beamline and type of data. Processed data are generally exported in a simpler format, easily implemented in python-type, home-made codes or software as Origin, IgorPro, Matlab, etc.

Supplementary information on ontology
BattINFO will be developed using Protégé and can be accessed from the toolchain either directly or via libraries like EMMO-Python. The entities in the ontology will be carefully documented using a combination of both formal annotations (for human and machine consumption) and informal annotations (human consumption). Up-to-date and easy browsable reference documentation in .html and .pdf formats will be generated from the ontology (e.g. using EMMO-Python or similar tools).

### 2.2.4 Preferred sites for deposition of data and associated metadata, documentation and code

The preferred sites for depositing BIG-MAP's output will be repositories that are recognized and commonly used by the battery research community, e.g. topical repositories as Inorganic Crystal Structure Database (ICSD), Materials Cloud, Materials Project<sup>8,9</sup><sup>8,9</sup><sup>8,9</sup><sup>8,9</sup><sup>8,9</sup><sup>8,9</sup>, Computational Materials Repository, and general-purpose repositories as GitHub, Zenodo, FigShare and the partners' institutional research-data repositories.

Associated metadata and documentation shall be deposited together with the output or linked to the output via persistent identifiers.

### 2.2.5 Access to date of restricted openness

Data of restricted openness that has been indexed in an open repository shall be equipped with a link to a contact person for the data or to a description of how admission to view and/or access the data can be requested.

## 2.3 Making data interoperable

### 2.3.1 Interoperability and the use of data/metadata vocabularies in order to facilitate interoperability

The following incentives shall ensure high interoperability of BIG-MAP's research output:

- The BattINFO ontology
- Standards for assigning metadata and keywords to data
- Conversion to open data formats, whenever feasible. If conversion to open formats is not possible, the software (name and version number) needed for accessing the data shall be included in the documentation
- The APIs integrated in all BIG-MAP apps

BIG-MAP builds on materials science, modelling, artificial intelligence and machine learning – all rooted in science and relating to materials for use in batteries. The BattINFO provides a standard vocabulary for the actual research domain and will facilitate interoperability within this domain.

### 2.3.2 The use of standard vocabularies and mapping to more commonly used ontologies in order to allow inter-disciplinary interoperability

BattINFO is a sub-domain of EMMO that provides the connection between the physical world, the experimental world (materials characterisation) and the simulation world (materials modelling). This tight linkage between BattINFO and EMMO, strengthens interoperability of BIG-MAP's research output across the disciples covered by EMMO.

BattINFO follows the W3C Web Ontology Language (OWL) standard that has been designed to represent rich and complex knowledge about things, groups of things, and relations between things, and BattINFO supports thereby versatile interoperability.

Supplementary information

The use of open-source Python infrastructures like AiiDA and ASE provide compatibility and interoperability with a wide spectrum of the most commonly used materials-modelling software.

## 2.4 Increased data re-use (through clarifying licenses)

### 2.4.1 Licensing in order to permit the widest reuse possible

BIG-MAP's research output shall be distributed and licensed in the least restricted manner:

- Software, code and scripts will preferably be released under an open source license, e.g., MIT, BSD or GPL.
- Research output not belonging to the abovementioned categories, will be distributed under the Creative Commons Attribution (CC-BY) 4.0 license.

Supplementary information

- AiiDA is licensed under the permissive MIT open source license, which puts very limited restrictions on reuse (namely: including the license and copyright notice).
- ASE carries the LGPLv2.1+ license which places similarly limited restrictions on use.
- The 3DEXPERIENCE Platform, BIOVIA Science Cloud and BIOVIA Pipeline Pilot are subject to a separate license agreement.

### 2.4.2 Time-frame for making data available for re-use, plus the reason for and length of possible embargo periods

BIG-MAP's open research output shall be made publicly available and reusable without unnecessary delays.

Exceptions

- Data published as supplementary information to a journal paper for which the publisher's demands an embargo period. The embargo period will be respected, but the embargo period shall be ≤ 6 months in order to comply with H2020's guidelines for Open Access & Data Management.
- Data for which the owner demands an embargo period for protecting business secrets, for evaluating the IPR potential or similar. The embargo period will be respected provided that the procedure for announcing possible IPR follows what has been set out in the Consortium Agreement.
- Data residing at large-scale facilities will be openly available for 3rd party after an embargo period set by the facility, typically 3 years.

### 2.4.3 Usability of data by third parties, in particular after the end of the project, and possible restriction on re-use of specific data

BIG-MAP's research output will have high value for the entire Battery 2030+ initiative and for the European battery society, including industry and academia. The project will work for high awareness of the data and thereby highest possible reuse in these communities. This will be done in collaboration with the Battery 2030+ initiative and BIG-MAP's stakeholders.

### 2.4.4 Data quality assurance processes

The BIG-MAP consortium has committed themselves to conduct their research trustworthy, transparent and in compliance with the European Code of Conduct for Research Integrity.

Additional incentives implemented in order to ensure high quality of the generated research output:

- An active review and exchange of output across the project.
- The highly skilled and experienced prime investigators involved in all the scientific tasks.
- Publications in preferably peer reviewed journals where the peer review process confirms the quality of the output itself and the methodology.

Supplementary information

- The development of software code will follow existing standards and good practices within computation research. The quality will also be ensured by depositing the code in public repositories and thereby make it available for crowd-reviewing.
- The servers used for development of the AI models will have basic validation for data plausibility i.e. shape and numbers but will not check whether the data makes sense.

### 2.4.5 The length of time for which the data will remain re-usable

BIG-MAP's research output will remain re-usable and relevant until new battery materials are capable of addressing the need for energy-storage capacity inherent in EU's vision net-zero $CO_2$. This means a time horizon of minimum 10 years. There will be no end date for the relevance of BattINFO, provided that it is constantly adapted to the progress in battery-interface research.

Supplementary information

- Data underpinning journal publications shall be stored and be available for an extended period after the publication date. The length of the storage period, typically 5-10 years, shall as minimum comply with local rules, i.e. the rules applying for the publishing entity.
- The AiiDA export files will be archived on Materials Cloud, which guarantees preservation of data sets for at least 10 years after deposition, irrespective of future funding for the repository.

## 2.5 Allocation of resources

### 2.5.1 Allocation of resources

The objective of BIG-MAP cannot be reached without research output of high FAIR-ness. The cost needed for making the data FAIR has therefore been integrated in the partners' budgets for running costs.

The cost of maintaining the generated research data FAIR after the project has ended, will be considered in connection with the mid-term review and at the end of the project.

### 2.5.2 The costs for making your data FAIR, and how the costs will be covered

The cost for making BIG-MAP's research output FAIR is presently unknown.

Incentives implemented in order to keep the costs down are:

- Utilization of national high-performance computing (HPC) and storage resources as well as the resources provided through BIG-MAP.
- Implementation of tools for automatic generation of metadata that comply with the FAIR standard.
- Generation of the AI models in a FAIR-compliant manner already from start.
- Building the BattINFO on EMMO and thereby reusing the significant effort invested in making EMMO FAIR. This is already covered by other EU-financed projects (EMMC-CSA, SimDome and partly also OntoTrans).

### 2.5.3 Responsibilities for data management

All partners in the project are responsible for handling the data they generate, collect or handle in compliance with

- Their own organization's rules and policies for data management and IT security
- H2020's guidelines for Open Access and Data Management
- The guidelines set down and agreed upon in this DMP.

The project's Data Management Responsible is responsible for

- Guiding the WP leaders, prime investigators and the partners in issues related to data management
- Reporting progress and possible data management challenges to Executive Board.

### 2.5.4 Cost and potential value of long-term preservation

The cost of long-term preservation is unknown at present and will be considered in connection with the project's mid-term review.

Supplementary information

- The cost depends on whether the data are stored locally, i.e. on the partners' locations, on a satellite storage platform, centrally at DTU or somewhere else.

- Usual rates for the storage are in the range 1 €/month for GBs of data to 10 €/month for TBs. This can most presumable be covered via the running costs of each organisation. Larger file sizes will presumably require other sources of funding.
- Software code can be preserved via public repositories at no extra costs.

## 2.6 Data security

### 2.6.1 Data recovery and secure storage and transfer of sensitive data

The entity storing the research output (data, models, tools) is responsible for

- Secure storage and preservation of the output
- Secure transfer of data to other entities
- Implementing secure means for efficient recovery of lost or damaged data, e.g. back-up, mirroring, snapshots, versioning and/or redundancy provided by the GPFS file system.

Any handling and storage of data shall comply with the local IT security policy, rules and regulations. If the one storing the data is not the owner of the data, the owner of the data shall ensure that the entity storing the data has trustworthy procedures for storage, preservation, transfer and recovery of data. Storage and transfer of confidential data shall be done in compliance with what have been agreed with the owner of the data.

BIG-MAP has not planned for any person-related data. Confidential data is therefore restricted to data that has been stamped confidential for IP reasons or because the data relate closely to business secrets.

<u>Supplementary information</u>

- Data generated on supercomputers will be stored on the supercomputer premises and protected by access control lists (ACLs). When data is transferred to workstations, the secure copy protocol (SCP) shall be used.
- Temporary files residing on the (super) computers, where calculations were performed, will not be preserved beyond the typical time of persistence of files in the /scratch file system, needed for restarts (this is typically in the range of 15-30 days).
- The ML model itself does not contain the data that has been used for training them. Sharing the model is therefore not consider a security issue. Offering active learning and data analysis as a web service however is a security issue that has to be dealt with as part of a release, e.g. via an access token.
- BattINFO is not considered sensitive, but merging changes to the upstream master branch requires review and strict access control. The use of Git ensures that the full history exists both at the central Git server, as well as on the local disk of all developers, providing a high level of protection against data loss.
- When data is stored in online repositories, preferably own-hosted solutions or commercial servers located within the European Union, will be chosen.

## 2.7 Ethical aspects

By signing the BIG-MAP Grant Agreement, all BIG-MAP beneficiaries have declared that they endorse the principles of research integrity as described in the European Code of Conduct for Research Integrity. The beneficiaries declare that they also follow their national and/or institutional version of the Code of Conduct.

## 2.8 Other

### 2.8.1 Related national/funder/sectorial/departmental procedures for data management

All BIG-MAP partners are responsible for handling their research data in compliance with this Data Management Plan and any updates of it.

This DMP will be reviewed and updated in connection with the project's midterm review.

# 3. List of Abbreviations

## 3.1 List of scientific abbreviations

| Abbreviation | Full name |
|---|---|
| (p)DOS | (Projected) Density of States |
| 3DS | https://www.3ds.com |
| AAS | Atomic Absorption Spectroscopy |
| ACE | Automated Content Evaluator |
| ADF | Amsterdam Density Functional |
| AES | Auger Electron Spectroscopy |
| AI | Artificial Intelligence |
| AiiDA | Automated Interactive Infrastructure and Database for Computational Science, https://www.aiida.net/ |
| AIMD | Ab-initio Molecular Dynamics |
| APDFT codes | Alchemical Perturbation Density Functional Theory |
| API | Application Programming Interface |
| ASE | Atomic Simulation Environment https://wiki.fysik.dtu.dk/ase/ |
| ATR | Attenuated Total Reflectance |
| BIG | Battery Interface Genome |
| Castep | http://www.castep.org/ |
| CEI | Cathode Electrolyte Interface |
| CGMD | Coarse-Grained Molecular Dynamics |
| CMR | Computational Materials Repository |
| CP2K | https://www.cp2k.org/ |
| Crystal | https://www.crystal.unito.it/index.php |
| CT | Computed Tomography |
| CV | Cyclic Voltammetry |
| DEM | Differential Electrochemical Mass Spectrometry |
| DFT | Density Functional Theory |
| DFTB+ | Density Functional based Tight Binding |
| DOS | (p)DOS (Projected) Density of States |
| DRIFTS | Diffuse Reflectance Infrared Fourier Transform Spectroscopy |
| EDS | Energy Dispersive Spectroscopy |
| EDX | Energy-Dispersive X-ray Spectroscopy |
| EDXA | Energy Dispersive X-ray Analysis |
| EELS | Electron Energy Loss Spectroscopy |
| EIS | Electrochemical Impedance Spectroscopy |
| EL CELL | Electrochemical Cell |
| EMMO | European Materials & Modelling Ontology |
| EQCM | Electrochemical Quartz Crystal Microbalance |
| ES | Electrochemical Spectroscopy |
| EXAFS | Extended X-ray Absorption Fine Structure |
| FEniCS | https://fenicsproject.org/ |
| FIB | Focused Ion Beam |
| FOM | Figure Of Merit |
| FTIR | Fourier Transform Infrared Spectroscopy |
| GAP | Gaussian Approximation Potentials |
| GAP | Gaussian Approximation Potentials |
| GAP | Gaussian Approximation Potential |
| GAUSSIAN | https://gaussian.com/ |
| GCMD | Grand Canonical Molecular Dynamics |
| GI XRD | Grazing Incidence X-ray diffraction |
| GITT | Galvanostatic Intermittent Titration Technique |
| GPAW | https://wiki.fysik.dtu.dk/gpaw/; |
| GROMACS | https://www.gromacs.org/; |
| GUI | Graphical User Interface |
| HR-TEM | High Resolution Transmission Electron Microscopy |
| HTS | High-Throughput Screening |
| ICP MS | Inductively Coupled Plasma Mass Spectrometry |
| ICP-OES | Inductively Coupled Plasma Optical Emission Spectroscopy |
| INS | Inelastic Neutron Scattering |
| IR | Infrared Spectroscopy |
| KMC | Kinetic Monte Carlo |
| LAMMPS | Large-scale Atomic/Molecular Massively Parallel Simulator |
| LIBS | Laser-Induced Breakdown Spectroscopy |
| LIGGGTHS | Open Source Discrete Element Method Particle Simulation Software |
| LSF | Large-Scale Facilities |
| LSV | Linear Sweep Voltammetry |
| MAP | Materials Acceleration Platform |
| MC | Monte Carlo |
| MD | Molecular Dynamics |
| MD | Molecular Dynamics |
| MIM | Metal Insulator Metal |
| ML | Machine Learning |
| MLFF | Machine Learning Force Fields |
| MOLPRO | https://www.molpro.net/ |
| MRCC | https://www.mrcc.hu/ |
| NEC | https://github.com/ghb24/NECI_STABLE |
| NEGF | Non Equilibrium Green's Functions |
| NEXAFS | Near Edge X-ray Absorption Fine Structure |
| NMR | Nuclear Magnetic Resonance |
| NOMAD | The Novel Materials Discovery Laboratory |
| NR | Neutron Resonance |
| NUMPY | A Python library |
| NWChem | https://www.nwchem-sw.org/ |
| OEMS | Online Electrochemical Mass Spectrometry |
| ORCA | https://orcaforum.kofo.mpg.de |
| ORM | Object Relational Mapping |
| P2D | Pseudo 2 Dimensional |
| P4D | Pseudo 4 Dimensional |
| PSI4 | https://psicode.org/ |
| PxD | Pseudo-X-Dimensional |
| PXRD | Powder X-Ray Diffraction |
| QChem | https://www.q-chem.com/ |
| QENS | Quasi-Elastic Neutron Scattering |
| QM | Quantum Mechanics |
| QMC | Quantum Monte Carlo |
| QML | Quantum Machine Learning |
| QUANTUM ESPRESSO | https://www.quantum-espresso.org/ |
| QuantumATK | https://www.synopsys.com/silicon/quantumatk.html |
| QUIP | https://libatoms.github.io/QUIP/ |
| RBS | Rutherford Back Scattering |
| RIXS | Resonant Inelastic X-ray Scattering |
| SANS | Small-Angle Neutron Scattering |
| SAS | Small-Angle Scattering |
| SAXS | Small-Angle X-ray Scattering |
| SCC-DFTB | Self-Consistent-Charge Density-Functional Tight-Binding method |
| SCIPY | https://www.scipy.org/ |
| SCM | Scanning Capacitance Microscopy |
| SDC | Scanning Droplet Cell |
| SEI | Solid Electrolyte Interphase |
| SEM | Scanning Electron Microscopy |
| SIESTA | https://departments.icmab.es/leem/siesta/ |
| SKOS | Simple Knowledge Organization System |
| SOAP | Smooth Overlap of Atomic Positions |
| STEM | Scanning Transmission Electron Microscopy |
| STEM-HAADF | Scanning Transmission Electron Microscopy High-Angle Annular Dark-Field imaging |
| ToF-SIMS | Time of Flight Secondary Son Mass Spectrometry |
| TranSIESTA | Software for modelling the electrical properties of nanoscale devices |
| TURBOMOLE | https://www.turbomole.org/; |
| VASP | Vienna Ab Initio Simulation Package |
| WFT | Wave Function Theory |
| XANES | X-ray Absorption Near Edge Structure |
| XAS (/) | X-Ray Absorption Spectroscopy |
| XPS | X-ray Photo-Electron Spectroscopy |
| XRD | X-ray Diffraction |
| XRR | X-ray Reflectivity |
| Yambo | https://yambo-studio.com/ |

## 3.2 List of abbreviations for partner names

| Partner abbreviation | Partner name |
|---|---|
| DTU | Technical University of Denmark |
| UU | Uppsala University |
| KIT | Karlsruhe Institute of Technology |
| CNRS | National Centre for Scientific Research |
| WWU | University of Münster |
| CEA | Commissariat a l'Energie Atomique et au Energies Alternatives |
| NIC | National Institute of Chemistry |
| SINTEF | Sintef A/S |
| PDT | Politecnico Di Torino |
| Fraunhofer | Fraunhofer-Gesellschaft zur Förderung der angewandten Forschung e.V. |
| CTH | Chalmers University of Technology |
| WUT | Warsaw University of Technology |
| CSIC | Spanish National Research Council |
| TUD | Delft University of Technology |
| CNR | National Research Council |
| EPFL | Swiss Federal Institute of Technology Lausanne |
| UCAM | University of Cambridge |
| UOXF | University of Oxford |
| UT | University of Tartu |
| ULIV | University of Liverpool |
| ESRF | European Synchrotron Radiation Facility |
| ILL | Institut Laue-Langevin |
| SOLEIL | Synchrotron Soleil |
| CID | Cidetec |
| EMIRI | Energy Materials Industrial Research Initiative |
| UMI | Umicore |
| SOLB | Solvay |
| BASF | BASF |
| NVOLT | Northvolt |
| ITU | IT University of Copenhagen |
| 3DS | Dassault Systems |
| FZJ | Forschungszentrum Jülich |
| SAFT | Saft |
| UNIVIE | University of Vienna |